\documentclass[a4paper,11pt]{article}
\pdfoutput=1

\usepackage{jcappub} 

\usepackage[T1]{fontenc} 

\usepackage{amssymb}
\usepackage{amsmath}

\usepackage[utf8]{inputenc}
\usepackage[normalem]{ulem}

\usepackage{makecell}

\newcommand{\mB}{\mathbf{B}}
\newcommand{\mC}{\mathbf{C}}
\newcommand{\mE}{\mathbf{E}}
\newcommand{\mF}{\mathbf{F}}
\newcommand{\mN}{\mathbf{N}}
\newcommand{\mI}{\mathbf{I}}

\newcommand{\mG}{\mathbf{G}}
\newcommand{\mR}{\mathbf{R}}
\newcommand{\mP}{\mathbf{P}}
\newcommand{\mS}{\mathbf{S}}
\newcommand{\maa}{\mathbf{A}}
\newcommand{\mA}{\tilde{\mathbf{S}}}
\newcommand{\mD}{\tilde{\mathbf{P}}}

\newcommand{\mAdicion}{\mathbf{A}}

\newcommand{\mY}{\mathbf{Y}}
\newcommand{\vsp}{\mathbf{s}}
\newcommand{\vsa}{\tilde{\mathbf{s}}}

\newcommand{\ellm}{\ell_{\mathrm{\max}}}
\newcommand{\ns}{N_{\mathrm{side}}}

\newcommand{\vx}{\mathbf{x}}
\newcommand{\vy}{\mathbf{y}}
\newcommand{\vc}{\mathbf{c}}
\newcommand{\vd}{\mathbf{d}}
\newcommand{\vb}{\mathbf{b}}

\newcommand{\va}{\mathbf{a}}
\newcommand{\vct}{\hat{\mathbf{c}}}

\newcommand{\mal}{\mathcal{C}_{IP}}
\newcommand{\mbl}{\mathcal{C}_{IP}'}
\newcommand{\mcl}{\mathcal{C}_P}

\newcommand{\ma}{$\mal$}
\newcommand{\mb}{$\mbl$}
\newcommand{\mc}{$\mcl$}

\newcommand{\vvec}[1]{\mathbf{#1}}
\newcommand{\expec}[1]{\langle#1\rangle}

\DeclareMathOperator{\tr}{tr}

\title{{\tt ECLIPSE:} a fast Quadratic Maximum Likelihood
estimator for CMB intensity and polarization power spectra.}

\author[a,b]{J.~D. Bilbao-Ahedo}
\author[a]{R.~B. Barreiro}
\author[a]{P. Vielva}
\author[a]{E. Mart\'{\i}nez-Gonz\'alez}
\author[a,b]{D. Herranz}

\affiliation[a]{Instituto de F\'{\i}sica de Cantabria (CSIC-Univ. Cantabria),\\ Avda. de los Castros s/n, 39005 Santander (Spain)}
\affiliation[b]{Departamento de F\'{\i}sica Moderna, Universidad de Cantabria,\\ Avda. de los Castros s/n, 39005 Santander (Spain)}

\emailAdd{bilbao@ifca.unican.es}
\emailAdd{barreiro@ifca.unican.es}
\emailAdd{vielva@ifca.unican.es}
\emailAdd{martinez@ifca.unican.es}
\emailAdd{herranz@ifca.unican.es}

\abstract{We present {\tt ECLIPSE} (Efficient Cmb poLarization and Intensity Power Spectra Estimator), an optimized implementation of the Quadratic Maximum Likelihood (QML) method for the estimation of the power spectra of the Cosmic Microwave Background (CMB). This approach allows one to reduce significantly the computational costs associated to this technique, allowing to estimate the power spectra up to higher multipoles than previous implementations. In particular, for a resolution of $\ns=64$, $\ellm=192$ and a typical Galactic mask, the number of operations can be reduced by approximately a factor of 1000 in a full analysis including intensity and polarization with respect to an efficient direct implementation of the method. In addition, if one is interested in studying only polarization, it is possible to obtain the power spectra of the E and B modes with a further reduction of computational resources without degrading the results. We also show that for experiments observing a small fraction of the sky, the Fisher matrix becomes singular and, in this case, the standard QML can not be applied. To solve this problem, we have developed a binned version of the method that is unbiased and of minimum variance. We also test the robustness of the QML estimator when the assumed fiducial model differs from that of the sky and show the performance of an iterative approach. Finally, we present a comparison of the results obtained by QML and a pseudo-$C_{\ell}$ estimator ({\tt  NaMaster}) for a next-generation satellite, showing that, as expected, QML produces significantly smaller errors at low multipoles. The {\tt ECLIPSE} fast QML code developed in this work will be made publicly available.}


\arxivnumber{2104.08528}

\begin{document}
\maketitle
\flushbottom

\section{Introduction}
\label{sec:introduction}
During the last decades, Cosmic Microwave Background (CMB) observations have provided very valuable information to put together our current picture of the Universe. In particular, among many other efforts, the Planck satellite has obtained the best full-sky CMB data in intensity and polarization over a large range of frequencies (30-857 GHz) up to date, allowing to impose constraints, in many cases at sub-percent level, over the cosmological parameters \cite{planck18_p1}.

Given that CMB experiments usually produce a huge amount of data in the form of pixelized maps (T for intensity and the Q and U Stokes parameters for polarization), a crucial step in their analysis is the compression of this information in a more tractable way. In particular, since CMB fluctuations are expected to be nearly-Gaussian, most of their statistical information is contained in the 2-point correlation function (or equivalently in the power spectrum). Therefore, the estimation of the power spectrum is a key point in order to extract all the valuable cosmological information encoded in the CMB.

Different approaches have been developed for power spectra estimation, which differ in their efficiency and computational cost. In particular, maximum-likelihood based-methods (e.g.~\cite{Bond98, Oh99}) provide optimal results in the sense that the estimator is unbiased and of minimum variance, but they are computationally very expensive and can not be implemented for high-resolution data. A particular case of this type of methods is the Quadratic Maximum Likelihood (QML), first introduced by \cite{QML_T} for intensity and extended to deal with polarization by \cite{QML_TEB} (see also \cite{Gru09,Sch12,2015ApJS..221....5G,Vanneste18}).
Another popular approach are the pseudo-$C_{\ell}$ algorithms (see e.g.~\cite{Peebles73,Wandelt01,Hivon02,Chon04,Tris05,Alonso19} and references therein), which are much faster than maximum-likelihood methods and can therefore be used at the resolutions provided by current and planned experiments. They are also unbiased and their efficiency is comparable to that of the optimal methods at high multipoles, but not at large scales.

Although the utility of the pseudo-$C_{\ell}$ methods is out of discussion, the use of estimators which are optimal at large and intermediate scales is becoming increasingly important since they are critical for the detection of the primordial CMB polarization B-mode, whose main contribution is present at those scales.
Note that having a QML method that can cover the full range of the reionization and the recombination peaks of the B-mode (even if a pseudo-spectrum method could be close to optimal in a part of this range)  will provide not only a consistent optimal estimation of the relevant multipole range of the spectra but also of the corresponding full covariance matrix.
Detection of primordial B-modes, which are parametrized by the tensor-to-scalar ratio $r$, would be a major breakthrough in Cosmology, since they are sourced by tensor perturbations, and its detection would constitute a definitive proof of the existence of a background of primordial gravitational waves, as predicted by inflationary models \cite{guth1981inflationary,linde1982new,starobinsky1982dynamics}. The best current constraint is given by $r_{0.05}$<0.044 at 95\% CL obtained combining Planck and BICEP2/Keck Array data \cite{Npipe_r} (see also \cite{Planck_inflacion18,BicepKeck18}), showing the faintness of the signal and the difficulty of its detection. Indeed, a large number of B-mode polarization experiments are currently on-going or planned, such as for example the BICEP array  \cite{hui2018bicep}, the Simons Observatory \cite{SO19}, the CMB-S4 experiment \cite{abazajian2016} or the JAXA LiteBIRD satellite  \cite{litebird18}, whose goal is to reach a sensitivity
$\sigma_r (r=0) \leq 10^{-3}$.

This work presents {\tt ECLIPSE} (Efficient Cmb poLarization and Intensity Power Spectra Estimator), an efficient implementation of the QML algorithm in {\tt FORTRAN}, that allows to compute the full power spectra with a very significant reduction of computational time, allowing to work at higher resolution than before. To illustrate the performance of the method we present results for a space-based B-mode mission and for a typical ground-based experiment. Although the QML method is well-known, there are several practical issues that should be taken into account when applying it to real data, such as the regularity of the Fisher matrix, the choice of an initial guess for the power spectra or the performance of an iterative scheme. In particular, depending on the observed sky fraction, the Fisher matrix can become singular. To solve this problem, we construct a binned version of the QML estimator to be used when the Fisher matrix is not invertible (see also \cite{Bilbao17} for a discussion on the regularity of the covariance matrix and the application of the QML method). 
Another important aspect of this estimator is that the user must provide an initial guess for the power spectra. However, we may wonder how the results are affected if this initial model differs from the true power spectra.  We present several tests of robustness of QML versus the assumed fiducial model and show that starting from significantly different initial guesses, through an iterative scheme map by map, QML drives statistically to the optimal estimator.

We also study the possibility of using the QML method to estimate only the polarization components with a consequent reduction of computational resources. A comparison with an advanced pseudo-$C_{\ell}$ algorithm ({\tt NaMaster}, \cite{Alonso19}) is also presented, showing the better efficiency of QML at large scales.

The rest of this paper is organized as follows. Section \ref{sec:ConfiguracionExperimentos} introduces the configuration of the experiments in which the method is tested, taking as reference typical orbital and sub-orbital cases. In section~\ref{sec:implementation} we introduce a description of the general method and of the {\tt ECLIPSE} implementation that reduces considerably the required computational resources and computing time. It presents also considerations and results on the estimation error obtained with the method when it is implemented for the T, Q and U data or only for the polarization components specifically. In section~\ref{sec:EstimadorBineado} we introduce a binned version of the estimator that works even in those cases where the Fisher matrix is not regular, providing optimal estimations of bandpowers instead of the complete power spectrum. In section~\ref{sec:Robustez} we analyse the robustness of the results when the assumed fiducial model differs from that of the maps, testing also the performance of an iterative scheme in different situations. Tests of performance are also presented for the binned and the only-polarization QML approaches. In section~\ref{sec:ComparacionMaster} we compare the efficiency of the QML and {\tt NaMaster} algorithms, focusing on the recovery of the BB component of the spectra at low and intermediate multipoles. Some conclusions are offered in section~\ref{sec:Conclusiones}. Finally, several appendices introduce more technical aspects of the work, including a detailed description of our efficient implementation (appendix~\ref{ap:AspectosComputacionales}), a discussion on the dependence of the power spectrum error on the observed sky fraction for different sky geometries (appendix~\ref{ap:fsky}), a smoothing function useful to construct an ansatz to iterate on QML results (appendix~\ref{ap:Suavizado}) and a simple estimator of the tensor-to-scalar ratio used to quantify the quality of some of our results (appendix~\ref{ap:r}).

\section{Instrumental configurations}
\label{sec:ConfiguracionExperimentos}

In order to test the performance of the QML method in different situations, we have mainly considered two different experimental configurations along the text. Showing results for different  cases is interesting since the very existence of the optimal estimator depends on the sky coverage of the experiment, while the noise level is reflected in the error of the estimated power spectra.

\begin{figure}
  \begin{center}
  \begin{tabular}{ccc}
  \includegraphics[scale=.30]{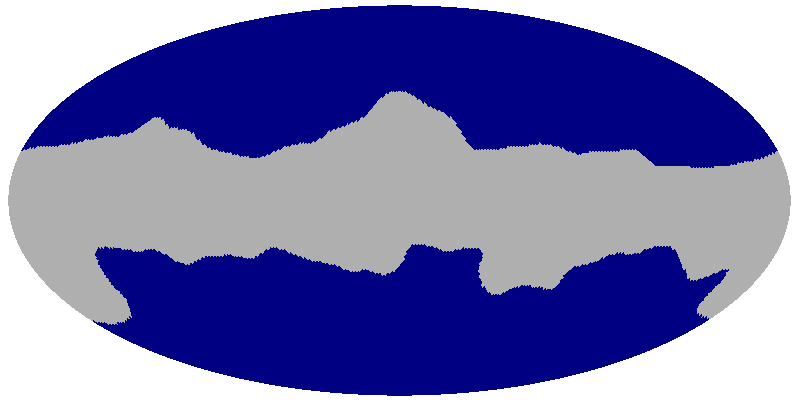} & &
   \includegraphics[scale=.30]{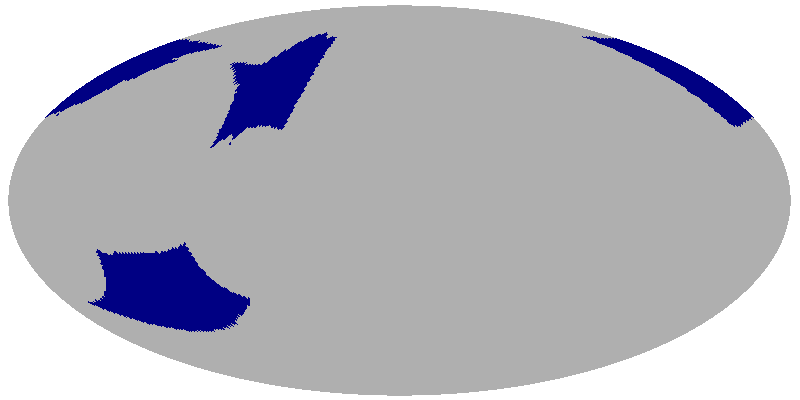} \\
  \end{tabular}	
    \caption{Sky coverage considered for the space (left) and ground-based (right) experiments, shown at a {\tt HEALPix} resolution $N_{\mathrm{side}}=64$.}
    \label{fig:Mascaras}
  \end{center}
\end{figure}

In particular, we have considered a spaceborne-like experiment, that provides a large sky coverage and a ground-based experiment that focuses on certain regions of the sky. As a typical example, we have considered the sky coverage given by the masks of figure~\ref{fig:Mascaras}  for the space (left panel) and ground-based (right panel) experiments. The first mask corresponds to the Galactic mask provided by \textit{Planck}, that allows one to use the 60\% of the sky at full resolution,\footnote{More specifically, we have used the mask file HFI\_Mask\_GalPlane-apo0\_2048\_R2.00.fits available at the \href{http://pla.esac.esa.int/pla}{\textit{Planck} Legacy Archive}.}
while the second one corresponds approximately to the three cosmological regions selected by the QUIJOTE experiment  \cite{quijote12,quijote17}. Regarding noise sensitivities, for the space mission, we have considered a noise level of 2.5 $\mu$K  $\cdot$ arcmin, similar to that expected for the JAXA LiteBIRD satellite  \cite{litebird18}, while for the ground experiment, a value of 1 $\mu$K  $\cdot$ arcmin is assumed. This sensitivity is similar to what could be obtained by the US-led CMB-S4 experiment  \cite{abazajian2016} or by a future Low Frequency Survey  \cite{LFS}. Table~\ref{tab:Experimentos} summarizes the specifications of the two selected configurations.

\begin{table}
\begin{center}
  \begin{tabular}{|lcc|} \hline
         & $f_{\rm sky}$ (\%)  &  Noise IQU (${\mu}K \cdot \mathrm{arcmin}$) \\ \hline
    Space & 59.0 & 2.5  \\ 
    Ground &  8.4 & 1.0  \\ \hline 
  \end{tabular}
  \caption{Specifications of the instrumental configurations considered along this work to test the QML estimator. The sky fraction of each mask has been obtained at a {\tt HEALPix} resolution of $N_{\rm side}=64$. A Gaussian beam of FWHM = 2.4 times the corresponding pixel size is adopted.}
  \label{tab:Experimentos}
\end{center}
\end{table}

For our tests, CMB simulations have been produced using the {\tt HEALPix} package  \cite{healpix}, for a standard cosmological model given by the best fit values of the cosmological parameters of the baseline \textit{Planck} 2018 $\Lambda$CDM model  \cite{planck18_p6}, but adding a tensor-to-scalar ratio $r=3\times 10^{-3}$.
The corresponding power spectra have been obtained using CAMB \cite{camb11}.
Along the paper, we will refer to this choice as Planck model and, except when otherwise stated, it will be the power spectra assumed in this work. A Gaussian beam of FWHM = 2.4 times the pixel size is also adopted.

\section{The QML estimator}
\label{sec:implementation}

In this section, after describing the Quadratic Maximum Likelihood (QML) estimator, we present an efficient implementation of the method ({\tt ECLIPSE}) that can reduce significantly the computational time required by this technique, allowing one to go up to higher multipoles. We also study the possibility of using a reduced version of the method, which lowers computational requirements, when one is interested in recovering only the polarization power spectra.

\subsection{General description}
\label{sec:Descripcion}

The QML is a method for obtaining an optimal estimation of the CMB power spectra and its covariance matrix from a map, which is well suited to deal with incomplete sky coverage. Assuming that the CMB fluctuations are Gaussian and isotropic, \cite{QML_T} and \cite{QML_TEB} show that given a CMB temperature map or CMB temperature and polarization maps $\vx$, it can be found an estimation of the power spectra $C_{i}$ in a two-step process:

\begin{enumerate}
  \item Starting from the pixels in the map, compute an angular quantity $y_i$ that is related to the power spectrum (see eq.~(\ref{ec:Defyi}) below).
  \item Given this quantity, define an estimator of the power spectrum (see eq.~(\ref{ec:Estimador}) below).
\end{enumerate}

Before describing the method, let us first establish some basic notation (details can be found in  \cite{QML_TEB}). In the only-temperature case, the map $\vx$ is an $N$-dimensional vector of elements $T_{i} = T(r_{i})$; in the full case, the map is a $3N$-dimensional vector of values $T_{i}$, $Q_{i}$, $U_{i}$.\footnote{Actually, the estimator can work with a different number of pixels in intensity and polarization, but for simplicity we will assume throughout this work that they are the same.} The method requires a model of the signal and of the noise, that are both introduced through the signal $\mS$ and the noise $\mN$ covariance matrices, respectively.

The statistical properties of $\vx$ are characterized by the power spectra $C_{i}$, and assuming that the signal and the noise are uncorrelated, we have
\begin{equation}
  \label{ec:DefMC}
  \mC \equiv  \langle \vx \vx^t \rangle = \mS + \mN = \sum_{i} C_{i} \mP_i + \mN.
\end{equation}
Note that in the only-temperature case the index $i$ can be directly substituted by the multipole index $\ell$, while in the full case $i$ includes $\ell$ and one of the six pairs TT, EE, BB, TE, TB, EB; thus $C_{i}  \leftrightarrow C_{\ell}^{XY}$, XY $\in \{$TT, EE, BB, TE, TB, EB$\}$. The $\mP_{i}$ matrices (see eq.~(\ref{ec:DefPi}))  connect the covariance in the harmonic space to the covariance in the pixel space. Each of them is the product of some subset of the columns of the matrix of the spherical harmonics by their transpose. Naturally, for each value of $\ell$ we have six matrices $\mP_{i} \leftrightarrow \mP_{\ell}^{XY}$ in the full case.

The first step, getting the angular quantity $y_i$  related to the anisotropies of the map but translated to the harmonics space, is achieved by
\begin{equation}
  y_i \equiv \vx^t \mE_i \vx - b_{i},
\label{ec:Defyi}
\end{equation}
where
\begin{equation}
  \label{ec:DefEi}
  \mE_i\equiv \frac{1}{2}{\mC^{-1}\mP_i\mC^{-1}}
\end{equation}
and $b_{i}$ takes into account the presence of noise
\begin{equation}
  \label{ec:PotenciaRuido}
  b_{i} = \tr \mN \mE_{i}.
\end{equation}

Arranging the sets of $C_i$ and $y_i$ in the vectors $\vc = \{C_1, C_2, \dots \}$ and $\vy = \{y_1, y_2, \dots\}$, respectively, \cite{QML_T} shows that the intermediate power coefficients $y_{i}$ are related to the power spectrum $C_i$ of the fiducial model as
\begin{equation}
  \label{ec:Mezcla}
  \expec{\vy} = \mF \vc
\end{equation}
and the covariances satisfy
\begin{equation}
  \label{ec:CovY}
  \expec{\vy \vy^t}-\expec{\vy}\expec{\vy}^t = \mF,
\end{equation}
where $\mF$ is the Fisher information matrix, which, taking into account eq.~(\ref{ec:DefMC}), can be written as:
\begin{equation}
  \label{ec:MatrizFisher}
  \mF_{ii'} = \frac{1}{2}\tr \left[ \mC^{-1} \frac{\partial\mC}{\partial C_{i}}  \mC^{-1} \frac{\partial\mC}{\partial C_{i'}} \right] = \frac{1}{2}\tr \left[ \mC^{-1} \mP_i  \mC^{-1} \mP_{i'} \right].
\end{equation}
If $\mF$ is regular, the power spectrum estimator can be defined as
\begin{equation}
  \label{ec:Estimador}
  \vct \equiv \mF^{-1} \vy.
\end{equation}
Combining this definition with eq.~(\ref{ec:Mezcla}), we get $\langle \vct \rangle = \vc$. The covariance matrix of this estimator is the inverse of the Fisher matrix
\begin{equation}
  \label{ec:CovarianzasEstimador}
  \expec{(\vct-\vc)(\vct-\vc)^t} = \mF^{-1}[\expec{\vy\vy^t}-\expec{\vy}\expec{\vy}^t]\mF^{-1}=\mF^{-1}.
\end{equation}

Therefore, the estimator is unbiased and, by the Cramer-Rao inequality, of minimum variance: QML is mathematically equivalent to the Maximum Likelihood Estimator, but, since it does not require a brutal force maximization, with a significant reduction of the computational time. Note that from eq.~(\ref{ec:DefMC}), the covariance matrix of the signal $\mS$ is computed from the fiducial model $C_i$ that the user provides. Therefore, an initial guess for the sought power spectra is required in order to compute the QML estimator. This leads naturally to the possibility of using an iterative scheme in order to update the initial fiducial model and, thus, to improve the final estimation of the power spectra. This possibility has not been fully explored in the literature, in part due to the high computational resources required by previous QML implementations. In section~\ref{sec:Robustez}, a detailed analysis of the robustness of the method versus the choice of the initial guess as well as the performance of an iterative scheme is presented.

Note that the QML estimator requires matrices $\mC$ and $\mF$ to be regular. If necessary, the covariance matrix can be regularized by adding a small amount of noise (a detailed analysis on the conditions on which $\mC$ is regular can be found in~\cite{Bilbao17}). If the Fisher matrix is singular, an optimal binned QML can be implemented as described in section~\ref{sec:EstimadorBineado}.

\subsection{Description in terms of alternative variables}

When estimating the power spectra, one usually needs to explicitly consider the effects of the instrumental beam and of the pixel window function. In addition, it is also quite common to describe the angular power spectra per logarithmic interval as $D_\ell = \ell (\ell + 1) C_\ell / 2 \pi$. Therefore, it may be convenient to implement the QML method in terms of the $D_\ell$ variables and/or including the instrumental resolution effects. This can be easily done by introducing some additional factors in the $\mP_i$ matrices of eq.~(\ref{ec:DefMC}).

Let us first denote by $B_\ell$ the beam and pixel instrumental effects, such that the harmonic coefficients (see appendix~\ref{ap:AspectosComputacionales}) of the observed signal are given by $a_{\ell m}^{\mathrm{Observed}} = B_\ell  a_{\ell m}^{\mathrm{Signal}} $ . Analogously, let us define $W_i = B_\ell^{X} B_\ell^{Y}$, that encodes these effects in the power spectra. In this way, we can write the covariance matrix of the observed (smoothed) signal as
\begin{equation}
 \mS  =  \sum_{i}  C_{i} W_i \mP_i.
\end{equation}
We can also write the previous equation in terms of the $D_i$ variables
\begin{equation}
\label{ec:ConDiBeamPixel}
 \mS  =  \sum_{i}  D_{i} \frac{2 \pi}{\ell (\ell + 1)} W_i \mP_i = \sum_{i}  D_{i}  \check{\mP}_i,
\end{equation}
where we have defined the new matrices $\check{\mP}_i$. It becomes apparent that replacing $C_i$ and $\mP_i$ by $D_i$ and $\check{\mP}_i$, respectively, in the equations of the previous section, we have an implementation of the method such that the $D_i$ quantities are estimated. Let us remark that these estimated spectra are corrected from the beam and pixel effects.

Of course, one could also easily write the equivalent expressions to estimate the power spectra in terms of the variables $C_i$ or $D_i$ and/or including the experimental beam.
Along the paper, we will use the QML in terms of different variables, as convenient, but note that this does not imply any loss of generality since all results can be straightforwardly obtained for any of the previously considered variables.

\subsection{Efficient numerical implementation}
\label{sec:efficiency}

A key element in an efficient implementation of the method is the connection between the pixel and harmonic domains, established by
\begin{equation}
\vx = \mY \va,
\label{ec:ConexionEspacios}
\end{equation}
where $\vx$ are the data in the pixel domain, $\va$ are the data in the harmonic domain, and $\mY$ is a matrix that connects both. The vectors and matrices involved in this method can be constructed in any of these domains, and so the corresponding numerical implementation. Depending on which domain they are calculated, the time and memory required can be significantly different.
In this section, we are going to outline some important steps for an efficient implementation of the QML. Unless otherwise stated, we will consider the polarization case along this discussion. For a more detailed description of these computational aspects, we refer to appendix~\ref{ap:AspectosComputacionales}.

First, we will discuss the implementation in pixel space, which follows directly from the expressions given in section~\ref{sec:Descripcion}. As we will see, some simple algebraical manipulation allows one to reduce the computational time in this case. However, the harmonic implementation, that we will outline below, is significantly more efficient.

A straightforward approach to implement the first step of the QML method, the calculation of the $y_{i} $ vector, is to compute the matrices of eq.~(\ref{ec:DefMC}), (\ref{ec:DefEi}) and (\ref{ec:MatrizFisher}) and, subsequently, the vectors  $b_i$,  $\vx^t \mE_i \vx$ and $y_{i}$. However, it is easy to show that the quantity $y_{i} $ can be obtained without the need of calculating explicitly the $\mE_{i}$ matrices, reducing significantly the number of operations. In particular, from eq.~(\ref{ec:Defyi})--(\ref{ec:PotenciaRuido}), we can write
\begin{equation}
y_{i} = \frac{1}{2} \left[ (\mC^{-1} x)^{t} \mP_{i} (\mC^{-1} x) - \tr ((\mC^{-1} \mN \mC^{-1}) \mP_{i}) \right].
\label{ec:Mejora1}
\end{equation}

For the second step of the method, we need to construct the Fisher matrix as well as its inverse. The calculation of the Fisher matrix is the part that involves the highest computational cost. From eq.~(\ref{ec:MatrizFisher}), for each element of the matrix, one needs to compute the trace of a matrix product, having a total number of elements of the order of $(6 \times (\ellm-1))^{2}/2$. Since the trace of the product of two matrices can be calculated without computing the product of the matrices (see eq.~(\ref{ec:TrazaProducto})), the number of operations can be considerably reduced, but one still has to compute --- and keep stored in the memory of the computer --- $6 \times (\ellm -1)$ products of matrices of the kind $\mC^{-1} \mP_i$. If the number of pixels in the map is large, this may require high computational resources regarding memory and CPU time (note that the number of operations to compute the product of two square matrices is of the order of the dimension of the matrices at the third power).

Alternatively, if we take into account in certain parts of the calculation the transformation from real to harmonic space given by eq.~(\ref{ec:ConexionEspacios}), it is possible to construct a significantly more efficient implementation of the QML, that we will refer as implementation in harmonic space, which is the base of our {\tt ECLIPSE} code.
The essence of the reduction in the number of operations is that, while in real space the $\mP_i$ matrices are dense, their analogues in harmonic space are sparse, with a reduced number of ones in strategic locations. To take advantage of this property, we have done a symbolic analysis of the results of the matrix operations involved and found analytical expressions of these results that can be easily implemented for a general case, reducing the number of operations tremendously. Moreover, the code can also be parallelised, further increasing the efficiency of the algorithm. We give the details of this approach in appendix~\ref{ap:AspectosComputacionales} and mention here only some of the improved aspects of the calculation.

First of all, it is possible to show that the $y_i$ quantities can be computed avoiding the calculation of the $\mP_i$ matrices. The quantities $\vx^t \mE_i \vx$ can be obtained taking the vector product $\mY^{\dag} (\mC^{-1} \vx)$ and afterwards the sum of products of subsets of its elements. To get $b_i$ one needs to compute the product $\mC^{-1} \mY$ --- a highly demanding operation that can not be avoided --- . If the noise is spatially uncorrelated, the noise matrix is diagonal and the quantities $b_i$ can be obtained through simple (non-matrix) operations with sub-blocks of elements of $\mC^{-1} \mY$. Regarding the computation of the Fisher matrix, one only has to compute the product $\mY^{\dag} (\mC^{-1} \mY)$ (actually only six square blocks out of the nine blocks of that matrix product). Once these blocks are calculated, the only work left in order to compute the Fisher matrix is collecting sub-blocks of the resultant blocks, multiply them and sum the elements of the product sub-block. This considerably reduces both the number of calculations --- orders of magnitude --- and the memory required to store the intermediate matrices. Therefore, the only highly demanding computer operations in our optimal implementation are the calculation of $\mC$, $\mC^{-1}$, $\mC^{-1} \mY$ and six blocks of $\mY^{\dag} \mC^{-1} \mY$. The rest of operations consist in taking sub-blocks of the matrices, multiplying pairs of elements and summing the resulting numbers.

To quantify better the difference in the computational cost of the direct approach and the more efficient implementation of {\tt ECLIPSE}, let us make an analysis of the number of operations involved in the calculation. For simplicity, we will focus only in the computation of the matrices needed to calculate the Fisher matrix, which is the most demanding step of the algorithm and will provide us with an approximated factor of the improvement gained with our approach.

In the straightforward implementation one has to compute the mentioned $\mC^{-1} \mP_i$ matrices multiplications. Since both matrices are of dimension $3 N_{\mathrm{pix}}$, the $6 \times (\ellm -1)$ multiplications $\mC^{-1} \mP_i$  require $6 (\ellm -1) (3 N_{\mathrm{pix}})^3$ operations. In fact, the number of operations would be even larger since, in this approach, one would need first to compute the $\mP_i$ matrices.

In our efficient implementation one has to compute instead $\mY^{\dag} \mC^{-1} \mY$. The matrix $\mY$ has $3 N_{\mathrm{pix}}$ rows and $3 L$ columns, where $L = \sum_{\ell=2}^{\ellm} (2 \ell + 1)$. Therefore, the product $\mC^{-1} \mY$ requires $3 N_{\mathrm{pix}} \times 3 N_{\mathrm{pix}} \times 3 L$ operations. A complete computation of the last product, $\mY^{\dag} (\mC^{-1} \mY)$, takes $3 L \times 3 N_{\mathrm{pix}} \times 3 L$ operations. Therefore the total number of operations is the sum, i.e.,  $27 (N_{\mathrm{pix}}^2 L + N_{\mathrm{pix}} L^2)$. In practice, the number of operations can be further reduced taking into account the structure of $\mY$ (see eq.~(\ref{ec:EstruturaY})) and the fact that one needs to compute only six out of the nine blocks of this product. This takes the number of operations needed to compute the product $\mC^{-1} \mY$ down to $15 N_{\mathrm{pix}}^2 L$ while to compute $\mY^{\dag} \mC^{-1} \mY$ one needs in this case $18 N_{\mathrm{pix}} L^2$ operations. These expressions are summarised in  table~\ref{tab:OperacionesImplementaciones}.

The specific number of required operations will depend on the values taken by the relevant parameters (resolution, number of pixels, maximum multipole) for the considered case. To have a better insight on how the different implementations scale with these parameters, let us consider the case $\ellm = 3 \ns$ and $N_{\mathrm{pix}} = 12 \ns^2$ (i.e., full sky), such that the number of operations depend only on $\ns$ through a polynomial expression. It is straightforward to show that the leading term of the direct implementation scales as $\ns^7$ while the optimal implementation goes as $\ns^6$. Moreover, this term is multiplied by a different factor that increases even further the number of operations in the direct approach. This leading term including the specific factor is given in table~\ref{tab:OperacionesImplementaciones}. For comparison, an implementation in which only polarization is computed is also shown (see section~\ref{sec:CasosTEB&EB}).

The table also shows the number of operations in the case of a map at $\ns$=64,   $\ellm = 192$ and a Galactic mask allowing for the use of $N_{\mathrm{pix}} = 29009$ pixels, corresponding to the space Configuration, for the direct and efficient approaches. As seen in the table, in order to compute the matrices needed to calculate the Fisher matrix, our efficient implementation requires around 3 orders of magnitude fewer operations than the straightforward approach for the considered case.

\begin{table}
\begin{center}
 \begin{tabular}{|ccccc|} \noalign{\hrule height 1pt}
&  \makecell[c]{ \# operations \\ (generic)} & \makecell[c]{\# operations \\  $\ellm = 192$ \\ $N_{\mathrm{pix}} = 29009$} & Ratio & \makecell[c]{ \# operations (leading term) \\ $\ellm = 3 \ns$ \\ $N_{\mathrm{pix}} = 12 \ns^2$}  \\ \noalign{\hrule height 1pt}

    Direct & $6 (\ellm -1) (3 N_{\mathrm{pix}} )^3$ &$7.55 \times 10^{17}$ & 634 & $839808 \ns^7$ \\ \hline
   \makecell[c]{Efficient \\ (T,E,B) } & $15 N_{\mathrm{pix}} ^2 L + 18 N_{\mathrm{pix}}  L^2 $  & $1.19 \times 10^{15}$ & 1  & $36936 \ns^6$  \\ \hline
    \makecell[c]{Efficient \\ (E,B) }  & $8N_{\mathrm{pix}} ^2L + 6 N_{\mathrm{pix}}  L^2$ & $4.92\times 10^{14}$ & 0.41 & $16200 \ns^6$  \\ \noalign{\hrule height 1pt}
\end{tabular}
    \caption{Number of operations to compute the matrices needed to calculate the Fisher matrix in three different approaches: the direct estimation, our efficient implementation {\tt ECLIPSE} and a case in which only polarization is computed also efficiently (see section~\ref{sec:CasosTEB&EB}). The second column corresponds to a generic case while the third one refers to the particular case with $N_{\mathrm{pix}} = 29009$ and $\ellm = 192$. The fourth column gives the ratio between the number of operations of the third column relative to our efficient approach for (T,E,B). Note that $L = 37245$ for the considered case. Finally, the last column shows the leading term of the polynomial expression that defines the number of operations in terms of the parameter $\ns$ for the case $\ellm = 3 \ns$ and full sky.}
  \label{tab:OperacionesImplementaciones}
  \end{center}
\end{table}

We can compare the performance of our code with those presented in the literature. Ref. \cite{2014MNRAS.440..957M} reports that it took roughly one day, using 16384 cores, to estimate the intensity power spectrum of 1000 maps at resolution $\ns = 64$ with a mask excluding around 20 per cent of the sky. In our case, we computed the six polarization power spectra up to $\ellm = 191$ of 1000 simulated maps at the same resolution and sky fraction in 90 minutes using 144 cores with a parallelized implementation in the Altamira\footnote{\url{https://www.res.es/en/res-sites/altamira}} supercomputer  at the Instituto de F\'\i sica de Cantabria (IFCA).

In \cite{2015ApJS..221....5G} an efficient implementation of the QML method also based in the harmonic space is described, that computes the elements of the Fisher matrix as the trace of the product of two matrices of the kind $(\mY^{\dag} \mC^{-1} \mY) \mI_i$, where  $\mI_i$ represents a sparse matrix. Although this approach is similar to ours, our implementation takes advantage of the specific details of the previously mentioned traces and products of matrices, which consequently reduces very significantly the number of operations, allowing also a parallelization of the code. In particular, the implementation from \cite{2015ApJS..221....5G} can compute  an estimation of the polarization power spectra of a map at resolution $\ns$ = 16 and $\ellm$ = 32 in 2 CPU minutes. Our parallelized code can perform a similar computation in 7 seconds in a node of the Altamira supercomputer (16 cores), and it takes 24 seconds running on a single core of a laptop. Also note that, since our code is parallelized, it can compute problems of higher dimensions. We would like to emphasize that, as far as we know, {\tt ECLIPSE} is the fastest available implementation of the QML estimator.

\subsection{Full and only-polarization implementations}\label{sec:CasosTEB&EB}

The QML estimator has been usually implemented to compute either only the temperature power spectrum or all the six possible spectra (intensity and polarization) simultaneously. However, if we ignore the information about the correlation between temperature and polarization, it is also possible to implement the QML only for the three polarization spectra, i.e., EE, BB and EB. This is interesting since it implies an important reduction of the computational requirements, allowing one to work at higher resolution. Independently of the assumed fiducial model, the QML estimator is unbiased and, therefore, these different constructions of the QML should produce, on average, the same results. However, the estimator error is only optimal if we use the correct fiducial model in the definition of the covariance matrix.

An example in which this partial estimation of the CMB angular power spectra, focused on the polarization signal, could imply a clear benefit is related to instrumental calibrations of CMB experiments. In particular,
the accurate estimation of the EB angular power spectrum can be used as a capital observable to perform the polarization angle calibration. This is recognised as one of the most important systematics to have under control for incoming high-sensitivity CMB polarization experiments. Mismatch calibrations of the polarization angle (above a few arcminutes) could induce a leakage from E-modes to B-modes that mask any possible primordial signal with  $r \lesssim 10^{-3}$. This degree of accuracy can not be obtained from astrophysical sources and, up to date, nulling the observed EB angular power spectrum (as it would be expected from the standard $\Lambda$CDM model), is a clear approach  to reach the required degree of accuracy on the polarization angle estimation~ \cite[see, for instance,][]{Minami2019}. This observable is only useful for this purpose if the EB estimation is unbiased, with optimal error bars, and up to a large multipole value.

In this section, we study the performance of the QML both when recovering the complete power spectra and also when recovering only the polarization terms,
checking that provides unbiased results and quantifying the increase in the error of the estimator when neglecting the correlation between temperature and polarization.

The structure of the covariance matrix in the temperature and polarization case is given by (for details see~\cite{QML_TEB})

\begin{equation}
  \label{ec:EstructuraMC}
  \mC =  \left(
  \begin{array}{cc}
    \mathrm{Block} \left[\begin{array}{c} TT \end{array} \right] & \mathrm{Block} \left[\begin{array}{cc} TQ & TU \end{array} \right] \\ \\
    \mathrm{Block} \left[ \begin{array}{c} QT \\ UT  \end{array} \right] & \mathrm{Block} \left[ \begin{array}{cc} QQ & QU \\ UQ & UU \end{array} \right] \\
  \end{array}
  \right).
\end{equation}

The matrix is composed of a diagonal block that accounts exclusively for temperature correlations, another diagonal block that encodes polarization correlations and two off-diagonal blocks that mix temperature and polarization. These off-diagonal blocks are related to $C_{\ell}^{TB}$, which is expected to vanish in the standard cosmological model, and to $C_{\ell}^{TE}$. Therefore, if we impose $C_{\ell}^{TE}=0$ in the fiducial model, the estimations for temperature and polarization are decoupled. Taking this into account, we have considered three different QML implementations:

\begin{enumerate}
  \item \ma: estimation of intensity and polarization spectra using a complete covariance matrix of order $3 N_{\mathrm{pix}}$ with $C_{\ell}^{TE} \neq 0$ in the fiducial model.
  \item \mb: estimation of intensity and polarization spectra using a complete covariance matrix of order $3 N_{\mathrm{pix}}$ but with $C_{\ell}^{TE} = 0$ in the fiducial model, i.e., containing zero off-diagonal blocks.
  \item \mc: a reduced version operating only on Q and U. The covariance matrix is the second diagonal block of eq.~(\ref{ec:EstructuraMC}), of order $2 N_{\mathrm{pix}}$. This version only estimates EE, BB and EB spectra.
\end{enumerate}

Let us remark that, with regard to the TT, EE, BB and EB spectra, the implementation \mb\ is equivalent to run two independent QMLs, one for intensity and one for polarization since, when $C_{\ell}^{TE} = 0$ in the fiducial model, the calculations for the estimation of intensity and polarization spectra are decoupled. Of course, the \mb\ configuration allows also the estimation of the TE and TB spectra, that would not be obtained with two independent QML estimators. Therefore, the \mb\ and \mc\ configurations should provide exactly the same results for the EE, BB and EB spectra map by map, but different from those obtained with \ma. Also, \ma\ and \mb\ provide, map by map, different estimations for TT, TE and TB.

It is clear that the lower dimensionality of the problem in the only polarization case implies a reduction of the required computational resources. Not only the covariance matrix has a smaller size, but also the number of elements in the Fisher matrix goes down from $6^2 (\ell_{\mathrm{max}} - 1)^2$ in the \ma\ and \mb\ cases to $3^2 (\ell_{\mathrm{max}} - 1)^2$ for the \mc\ configuration. In particular, we can obtain a rough estimation of the reduction of the CPU time, by referring again to
the computation of the matrix $\mY^{\dag} \mC^{-1} \mY$. In the \mc\ case, the product $\mC^{-1} \mY$ requires $8N_ {\mathrm{pix}}^2L$ operations while the computation of the blocks needed from  $\mY^{\dag} (\mC^{-1} \mY)$, only three blocks in this case, takes $6 N L^2$; thus we have a total of $8N_{\mathrm{pix}}^2L + 6 N_{\mathrm{pix}} L^2$ operations. This leads to a reduction of approximately a factor 2.3 in the number of operations with respect to the full implementation of (T,E,B) for the case considered in table~\ref{tab:OperacionesImplementaciones}. Note that this factor is only mildly dependent on the values of $\ell_{\mathrm{max}}$ and $N_{\mathrm{pix}}$ and will range approximately between 2 and 3.

We have also compared the performance of the three QML estimators on a practical example, considering 5000 CMB simulated maps for the space configuration given in table~\ref{tab:Experimentos}, at resolution  $N_{\mathrm{side}}=64$ and  $\ell_{\mathrm{max}}=128$. All the calculations were done in the Altamira supercomputer with 100 processors, with our efficient harmonic implementation, taking 32 and 18 minutes for the \ma\ and \mc\ cases, respectively, what corresponds to around a factor of 2 improvement, closer to the one found for the calculation of the matrices needed to compute the Fisher matrix. Although, given the high optimization of our code, this is not as large as one would obtain in the case of a direct implementation of the method, the \mc\ implementation still provides a significant reduction of computational time that, together with the smaller memory requirements, can be important if one wants to go to the highest possible resolution.

\begin{figure}
  \begin{center}
    \begin{tabular}{ccc}
      \includegraphics[scale=.85]{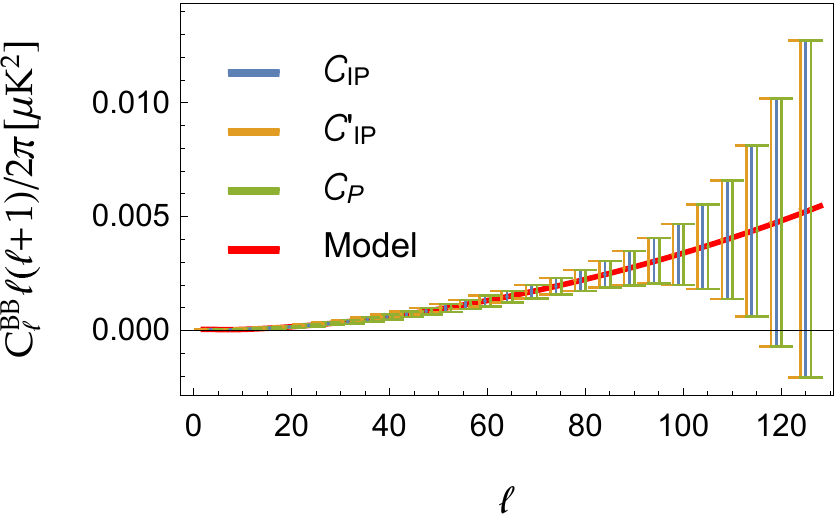} & &
            \includegraphics[scale=.85]{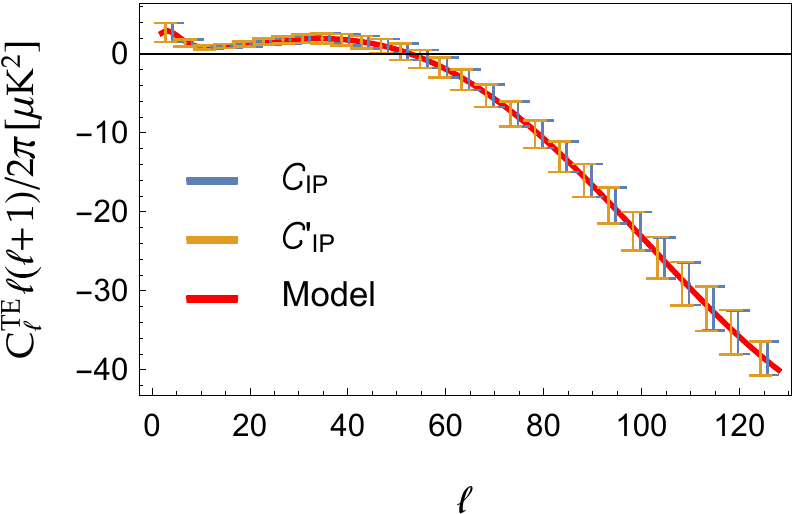} \\
    \end{tabular}
    \caption{Estimated power spectrum for BB (left) and TE (right) obtained with the different QML configurations explained in the text. For comparison the fiducial model is also plotted (red line). Error bars have been computed as the dispersion from 5000 simulations (we have tested that in the cases that the fiducial model matches the power in the maps, i.e. \ma\ and \mc, the error estimated from simulations agrees very well with that obtained from the Fisher matrix). For a better visualization, the power spectra has been binned and some of the points have been plotted with a small shift in the multipole value.}
    \label{fig:QML_TEB_1y2_EB}
  \end{center}
\end{figure}

Figure~\ref{fig:QML_TEB_1y2_EB} shows the TE power spectrum (right panel) as derived with the \ma\ and \mb\ QML estimators, and the BB power spectrum (left panel) obtained with the three different configurations. For all the different power spectra, we find that all the estimations are unbiased and that the corresponding errors are very similar for the considered configurations. We also checked that, as expected, the polarization spectra is identical for the \mb\ and \mc\ approaches.

In order to quantify the different performance of the three estimators, we have calculated the increase in the error of the \mb\ and \mc\ configurations with respect to the \ma\ reference case, which gives the best estimation since it uses the complete information in the fiducial model. In particular, for the \mb\ approach, we have calculated this increment in the error as
\begin{equation}
\sigma_{\mathrm{rel}} \left(\mbl\right) = 100 \frac{\sigma \left( D_{\ell}^{\mathrm{\mbl}} \right) - \sigma \left( D_{\ell}^{\mathrm{\mal}} \right) }{\sigma \left( D_{\ell}^{\mathrm{\mal}} \right)},
\label{ec:CalculoAumentoBarraError}
\end{equation}
where $\sigma \left( D_{\ell} \right)$ is the dispersion obtained from the corresponding 5000 simulations.

\begin{figure}
  \begin{center}
  \begin{tabular}{ccc}
      \includegraphics[scale=.85]{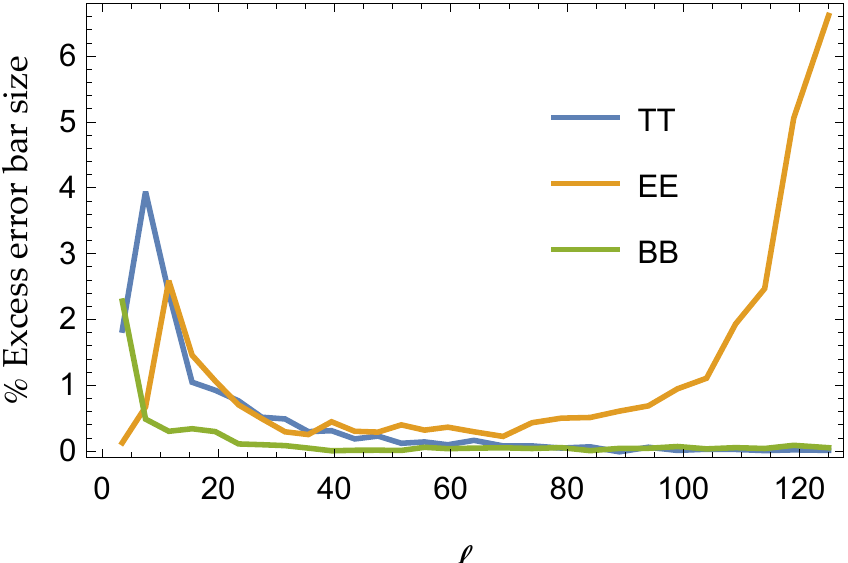} $ $
      \includegraphics[scale=.85]{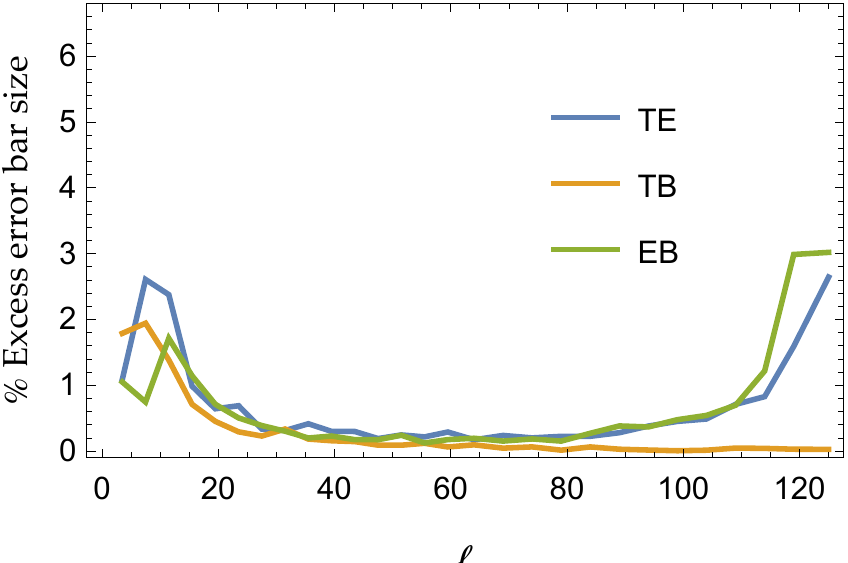}
  \end{tabular}
    \caption{Relative difference in per cent of the estimation error of the different power spectra given by the \mb\ implementation with respect to that of \ma. These quantities have been computed with eq.~(\ref{ec:CalculoAumentoBarraError}).}
    \label{fig:AumentoBarraError}
  \end{center}
\end{figure}

Figure~\ref{fig:AumentoBarraError} shows this quantity for TT, EE and BB (left panel) and for TE, TB and EB (right panel) when estimating the power spectrum with \mb\ (note that $\sigma_{\mathrm{rel}}(\mcl)$ is identical to that of \mb\ for the EE, BB and EB spectra). When excluding the information about TE in the fiducial model, we are slightly increasing the error bar in the final estimation of the different spectra. However, this increase is very moderate, showing that the method is close to optimal and that, if one is only interested in polarization spectra, the method can be implemented specifically for this case with an important reduction of computational resources. Let us remark that the reduction of computational time can be important, for instance, if one needs to repeat the process over many data sets or to iterate over the fiducial model (see section~\ref{sec:Robustez}). Another important advantage is that for a given set of computational resources, this implementation can work up to higher $\ns$ and greater $\ellm$ than the complete TEB configuration, since the memory required to store the matrices in the EB implementation is significantly lower.

\section{A binned version of the QML estimator}
\label{sec:EstimadorBineado}

It is well known that when working with an incomplete sky coverage, coupling appears between the different multipoles and the errors in the estimation of the power spectrum increase. In addition, there is a limit in the achieved angular resolution $\Delta \ell$ that depends on the size of the considered patch. As the mask grows, we are left with less information due to the pixels discarded. At some point, this leads to the Fisher matrix becoming singular. This is for instance the case when we apply the QML method to the sky coverage corresponding to our considered ground configuration (see right panel of figure~\ref{fig:Mascaras}).

As already pointed out in section~\ref{sec:implementation}, the QML method consists on two stages: (i) compute the vector $\vy$ from the anisotropies of the map that contains the coupled power of harmonic space and (ii) decouple the mixing of power making use of the inverse of the Fisher matrix (i.e. $\hat{\vc}=\mF^{-1}\vy$). If this matrix is singular, the last step cannot be completed. However, we will show that an optimal estimator of a reduced number of variables (binned spectra) can still be defined. This can be understood as an extension of the method to estimate bandpower spectra.

\subsection{Description of the estimator for binned spectrum}\label{sub:EstimadorBineado}

Let us assume that we have computed all the quantities in the harmonic space required by QML, but that we cannot invert the coupling in the $\vy$ variables (see eq.~(\ref{ec:Mezcla})) because $\mF$ is singular. From the point of view of a system of linear equations, eq.~(\ref{ec:Mezcla}) defines a problem where $\vc$ is the vector of variables to be solved. When $\mF$ is singular, the system has more variables than linearly independent equations and, therefore, is under-determined. To reduce the number of variables, we can consider a set of
bandpowers $B_{b}^{XY}$, which are linear combinations of the original $D_{\ell}^{XY}$ variables, and solve for them.

For the sake of simplicity, let us consider for the rest of the section the case of only temperature. Note that the process is the same in the full temperature and polarization case, but working with a higher dimension.

Let us take $N_{\mathrm{bins}}$, indexed by $b$, with boundaries $\ell^b_{\mathrm{low}} < \ell^b_{\mathrm{high}} = \ell^{b+1}_{\mathrm{low}}-1$, and define $L_b = \{\ell^b_{\mathrm{low}}, \ell^b_{\mathrm{low}}+1, \ldots, \ell^b_{\mathrm{high}} \}$ the set of values of $\ell$ corresponding to the bin $b$. To define the bandpower of a given bin, one usually calculates the weighted mean of the multipoles in the bin. If this is the case, the variables $B_b$ will be the mean power of the multipoles in each one of the bins weighted by their theoretical errors \cite{Eis99} taking into account cosmic variance, noise and sky fraction (see Appendix~\ref{sec:fsky} for a discussion on the validity of this error with regard to the sky fraction).
In this case, we also have to compute the mean value of the $\ell \in L_b$ to find the position of the value representative of the bin, $\ell^*_b$.
Therefore,
assuming that the fiducial is described in terms of the variables $D_\ell$ --- and therefore, that we are implementing QML to get the estimation in terms of these variables --- we can define the bandpowers of the binned fiducial as
\begin{equation}
  B_b = \frac{\sum_{\ell \in L_b} \frac{D_{\ell}}{(\Delta D_{\ell})^2}} {\sum_{\ell \in L_b}  \frac{1}{(\Delta D_{\ell})^2}}
  \label{ec:DefMediasBin}
\end{equation}
where $\Delta D_{\ell}$ corresponds to the theoretical error of the power spectrum. The position of the representative value of the bin is given by
\begin{equation}
  \ell_b^{*} =  \frac{\sum_{\ell \in L_b} \frac{\ell}{(\Delta D_{\ell})^2}} {\sum_{\ell \in L_b}  \frac{1}{(\Delta D_{\ell})^2}}.
  \label{ec:DefMediasBinPosicion}
\end{equation}
Let us note that other choices for $B_b$ are possible. For example, one could pick the $D_{\ell}$ of the fiducial corresponding to the central value of $\ell$ in the bin ($B_b = D_{\ell_c}$) or use an unweighted mean of the values of the multipoles. As we will see, the binned QML constructed consistently with these different definitions will provide as output the considered binned power. Whichever the choice for the variables $B_b$, once they are fixed, the next step is to define a set of factors $f_\ell^b$ making use of the information in the fiducial model, $D_\ell$ as

\begin{equation}
  \label{ec:DefinicionCoeficientesf}
  f_{\ell}^{b} = D_{\ell}/B_{b}.
\end{equation}

Introducing the factors $f_\ell^b$ and the variables $B_b$ in eq.~(\ref{ec:Mezcla}) we get (assuming again that QML is implemented in terms of the variables $D_\ell$)
\begin{eqnarray}
  \label{ec:SistemaRectangular}
  \nonumber \expec{y_{\ell}} & = & \sum_{\ell'} F_{\ell\ell'} D_{\ell'} =
  \sum_{b'} \sum_{\ell' \in L_b'} F_{\ell\ell'} f^{b'}_{\ell'} B_{b'} \\
  &  = & \sum_{b'} B_{b'} \sum_{\ell' \in L_b'} F_{\ell\ell'} f^{b'}_{\ell'}.
\end{eqnarray}
Now this system of linear equations has $\ellm-1$ equations and $N_{\mathrm{bins}}$ variables $B_b$, thus it contains more equations than variables. To reduce the number of equations accordingly, we can simply combine them linearly.
Although, as it will be shown later, the most efficient estimator can be obtained by a particular combination of the equations, for the purpose of illustration, let us combine them now just summing together the $\ell^b_{\mathrm{high}} - \ell^b_{\mathrm{low}}+1$ equations of each bin. This leads to $N_{\mathrm{bins}}$ equations
\begin{eqnarray}
  \sum_{\ell \in L_{b}} \expec{y_{\ell}} & = &\sum_{\ell \in L_{b}} \sum_{b'}B_{b'} \sum_{\ell' \in L_{b'}} F_{\ell \ell'} f^{b'}_{\ell'} \nonumber \\
  & = & \sum_{b'} B_{b'} \sum_{\ell \in L_{b}} \sum_{\ell' \in L_{b'}} F_{\ell \ell'} f^{b'}_{\ell'}  \nonumber \\
  & = & \sum_{b'} B_{b'} G_{bb'},
  \label{ec:SumaEcuaciones}
\end{eqnarray}
where in the last step we have defined the matrix $\mG$. Defining $z_{b} \equiv \sum_{\ell \in L_{b}} y_{\ell}$, we get
\begin{equation}
  \label{ec:FinalBines}
  \expec{z_{b}} = \sum_{b'} G_{bb'} B_{b'},
\end{equation}
or, arranging $B_{b}$ and $z_{b}$, respectively, in the vectors $\vvec{b}$ and $\vvec{z}$, we can write the previous equation in matrix notation as
\begin{equation}
  \label{ec:MezclaBin}
  \expec{\vvec{z}} = \mG \vvec{b}.
\end{equation}
Note that the matrix $\mG$ is square. Since we can reduce the number of variables as needed to make it regular, we can take as our estimator of the power in the bin
\begin{equation}
  \label{ec:EstimadorPotenciaBin}
  \hat{\vvec{b}} \equiv \mG^{-1} \vvec{z},
\end{equation}
that by eq.~(\ref{ec:MezclaBin}) and (\ref{ec:EstimadorPotenciaBin}) is unbiased, i.e.,
$\expec{\hat{\vvec{b}}} = \vvec{b}$.

Expressions~(\ref{ec:MezclaBin}) and (\ref{ec:EstimadorPotenciaBin}) are analogous for the binned case to equations (\ref{ec:Mezcla}) and (\ref{ec:Estimador}), respectively. It is important to note that the matrix $\mG$ is not symmetric, thus it is not the Fisher matrix expressed in terms of the variables $B_b$.

\subsection{Covariance matrix for the binned estimator}

To determine the covariance matrix of the binned estimator, let us define the rectangular matrix $\mAdicion$ of dimensions $N_{\mathrm{bins}} \times (\ell_{max}-1)$ as
\begin{equation}
  \mAdicion_{b\ell}  = \left \lbrace  \begin{array}{ll}
  1 & {\rm if} \quad \ell \in L_{b} \\
  0 &  {\rm otherwise} \\
  \end{array} \right.,
  \label{ec:DefinicionS}
\end{equation}
such that for a given bin (row) it has non-null values only for those multipoles (columns) belonging to the considered bin $b$. Using this definition, the sum $z_{b} = \sum_{\ell \in L_{b}} y_{\ell}$ can be expressed as $\vvec{z} = \mAdicion  \bf{\vy}$. Therefore, combining this expression with eq.~(\ref{ec:Mezcla}) and replacing the vector $\vc$ by a vector $\vd$ that contains the variables $D_{\ell}$, we have
\begin{equation}
  \label{ec:SumaMatricial}
  \expec{\vvec{z}} = \mAdicion  \expec{\bf{y}} = \mAdicion  \mF \vd.
\end{equation}
Let us also define a matrix $\mR$ of dimensions $(\ell_{max}-1) \times N_{\mathrm{bins}}$ to transform the set $\{B_{b}\}$ into the set $\{D_{\ell}\}$
\begin{equation}
  R_{\ell b} = \left \lbrace \begin{array}{ll} f_{\ell}^b & {\rm if} \quad \ell \in L_{b} \\
  0 & {\rm otherwise} \\ \end{array} \right. ,
  \label{ec:DefinicionMatrizR}
\end{equation}
so $\vd = \mR \vvec{b}$.
Expression (\ref{ec:SumaMatricial}) then becomes
\begin{equation}
  \label{ec:FinalCero}
  \expec{\vvec{z}} = \mAdicion  \expec{\bf{y}} = \mAdicion  \mF \mR \vvec{b}.
\end{equation}
The matrix $\mAdicion  \mF \mR$ is a square matrix. By choosing an appropriate binning, this matrix is regular and therefore we can compute the estimator
\begin{equation}
  \label{ec:Final}
  \hat{\vvec{b}}  \equiv [\mAdicion  \mF \mR]^{-1} \mAdicion  \bf{y},
\end{equation}
which is the same of eq.~(\ref{ec:EstimadorPotenciaBin}), and is unbiased. Finally, we can find the covariance matrix for the estimator
\begin{eqnarray}
  \label{ec:FinMCBineado}
  \expec{(\hat{\vb} - \vb)(\hat{\vb}-\vb)^t}  & = &
  [\mAdicion  \mF  \mR ]^{-1} \mAdicion  [\expec{\vy \vy^t}-\expec{\vy}\expec{\vy}^t] ([\mAdicion  \mF \mR]^{-1} \mAdicion )^t \nonumber \\
  &  = & [\mAdicion  \, \mF \, \mR]^{-1} \mAdicion  \mF \mAdicion ^t [\mR^t \, \mF \, \mAdicion ^t]^{-1}.
\end{eqnarray}

\subsection{Fisher matrix of the binned spectrum}

In order to calculate the Fisher matrix corresponding to the $B_b$ variables, let us first write the covariance matrix in terms of the binned power spectrum $B_{b}$ and the factors $f^{b}_{\ell}$
\begin{equation}
\label{ec:CSoloBines}
  \mC = \sum_{\ell}  D_{\ell} \check{\mP}_{\ell} + \mN = \sum_{b} \sum_{\ell \in L_b} f^{b}_{\ell} B_{b} \check{\mP}_{\ell} + \mN = \sum_{b}  B_{b}  \sum_{\ell \in L_b} f^{b}_{\ell} \check{\mP}_{\ell} + \mN.
\end{equation}
The Fisher matrix expressed in terms of the variables $B_b$ is then given by\footnote{Since we are binning in terms of the variables $D_\ell$, in the next expression we write explicitly the matrices $\mP$ as $\check{\mP}$. The Fisher matrix should also be written as $\check{\mF}$ in that expression and the rest of the section, but we will keep the notation $\mF$ for simplicity.}
\begin{eqnarray}
  \label{ec:MatrizDeFisher}
 \mF^B_{bb'} & = & \frac{1}{2}\tr\left[
  \mC^{-1} \frac{\partial\mC}{\partial B_b}
  \mC^{-1} \frac{\partial\mC}{\partial B_{b'}}\right] \nonumber \\
  & = & \frac{1}{2}\tr\left[\mC^{-1}  \left(\sum_{\ell \in L_b} f_{\ell}^b \check{\mP}_{\ell}\right) \mC^{-1} \left(\sum_{\ell' \in L_{b'}} f_{\ell'}^{b'} \check{\mP}_{\ell'}\right)\right] \nonumber \\
  & = & \sum_{\ell \in L_b} \sum_{\ell' \in L_{b'}}  f_{\ell}^b  f_{\ell'}^{b'} \mF_{\ell\ell'} =  \mR^t \mF \mR,
\end{eqnarray}
where in the last step we have used the definition of matrix $\mR$.
To avoid confusion, it is important to note that the matrix  $\mF$ in the previous equation is the Fisher matrix corresponding to the power spectrum $D_{\ell}$, while $\mF^{B}$ is the Fisher matrix corresponding to the binned quantities $B_b$. For the covariance of the estimator to be minimum, the last term of (\ref{ec:FinMCBineado}) has to be equal to the inverse of $\mF^B$, thus at first sight the estimator $\hat{\vb}$ does not have minimum variance.

\subsection{Method of minimum variance}

We recall that in the previous sections, we have simply added sets of linear equations eq.~(\ref{ec:SumaEcuaciones}) to reduce the dimensionality of the problem. However, as we have seen, this leads to a non-optimal estimator. Therefore, we have to look for the appropriate way of combining the equations (i.e., to determine the appropriate $\mAdicion $) in order to obtain an estimator with minimum variance. In particular, if we choose $\mAdicion  \equiv \mR^{t}$, we find
\begin{equation}
  \label{ec:eqMCMinimaVarianza}
  \expec{(\hat{\vb} - \vb)(\hat{\vb}-\vb)^t}  = [\mR^t \, \mF \, \mR]^{-1} \mR^t \mF \mR [\mR^t \, \mF \, \mR]^{-1} = [\mR^t \, \mF \, \mR]^{-1},
\end{equation}
where the last expression is the inverse of $\mF^B$. Therefore, we find the optimal estimator for the binned case to be
\begin{equation}
  \label{ec:FinalEstimador}
  \hat{\vvec{b}}  = [\mR^t \mF \mR]^{-1} \mR^t \vy.
\end{equation}

\bigskip
\bigskip
\bigskip

In summary, to construct the optimal and unbiased binned version of QML, one first has to compute $\mF$ and $\vy$ as in the standard case. The next step is to define the set of bins as well as to obtain the value of the considered bandpowers from the full set of $D_{\ell}$ (usually as a weighted mean of the power spectrum in the bin). The binned QML will provide as output estimations of these bandpowers. Once this is established, one needs to compute accordingly the $f_\ell^b$ values using the information of the fiducial model, which will allow us to construct the $\mR$ matrix (given by eq.~(\ref{ec:DefinicionMatrizR})). Using this matrix, it is straightforward to estimate the binned power spectrum according to eq.~(\ref{ec:FinalEstimador}). Finally, the Fisher matrix of the binned estimation is given by eq.~(\ref{ec:MatrizDeFisher}).

\begin{figure}
\begin{center}
 \includegraphics[scale=.9]{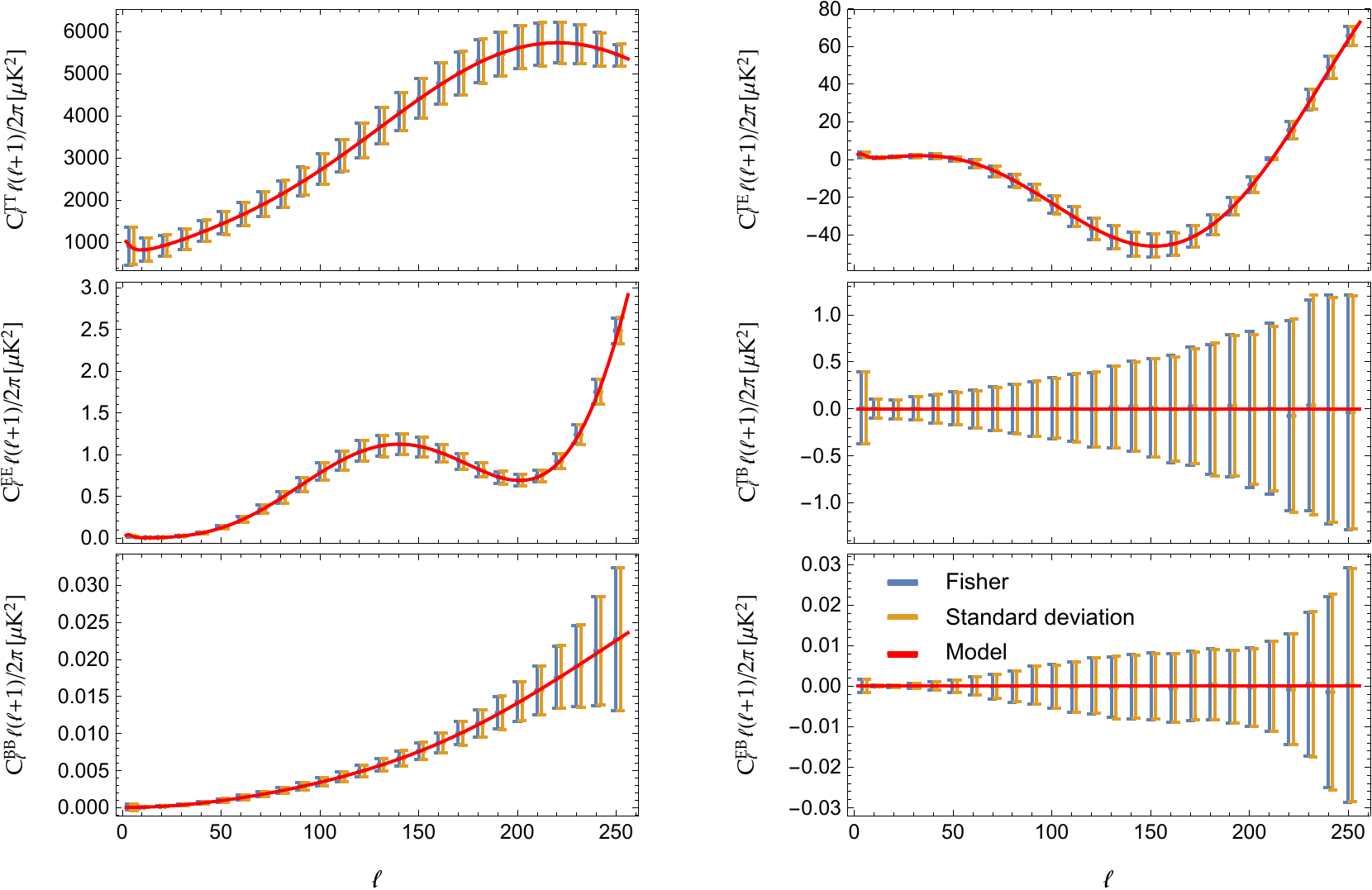}
\end{center}
  \caption{Unbiased binned power spectrum estimation in a case when the sky coverage is such that the Fisher matrix becomes singular. Ground experiment at resolution $N_{\mathrm{side}} = 128$. The red line shows the model in the simulated maps used as fiducial; the orange error bars, the dispersion on the estimated power spectra obtained from the 1000 simulations; the blue error bars, the error estimated from eq.~(\ref{ec:eqMCMinimaVarianza}).}
  \label{fig:GraficaBinesQuijote}
\end{figure}

Figure~\ref{fig:GraficaBinesQuijote} shows the results of the application of this method to an estimation of the mean power spectrum and its corresponding error using 1000 simulated maps at resolution $N_{\mathrm{side}} = 128$ for the ground configuration given in table~\ref{tab:Experimentos}, up to $\ellm=256$. The fist bin runs from 2 to 5; the second, from 6 to 10; and the rest of the bins are of length ten. As seen, the agreement between the estimated binned spectrum and the underlying true model is very good and we also find a very good match between the errors estimated from simulations and from the Fisher matrix. Therefore, this confirms that the method is unbiased and of minimum variance. The values of $B_{b}$ (and consequently the values of $f_{\ell}^{b}$) and $\ell_{b}^{*}$ were calculated from the fiducial model as the mean of the values in the bins weighted by their theoretical error (according to eq. (\ref{ec:DefMediasBin})).

\section{Discussion on the performance of QML}\label{sec:Robustez}

As shown in section~\ref{sec:Descripcion}, the QML method is, under the considered assumptions, unbiased and of minimum variance. However, this requires the use of the correct fiducial model, what is in general unknown. In practice, it is expected that small deviations of the fiducial model with respect to the true underlying model produces still unbiased results, although with a sub-optimal error. In this case, one may use an iterative scheme, such that the initial fiducial model is updated taking into account the output of the QML. Therefore, it is interesting to test this approach and to check how the estimator and its error depend on the choice of the fiducial model.

So far, we have shown results for the QML method using a fiducial model that perfectly matches that of the simulated maps, with the exception of the calculation of  $C_{\ell}^{TE}$ in the \mb\ implementation, in which $C_{\ell}^{TE}=0$ is assumed for the fiducial model. In this section, we will study the robustness of the results when the fiducial model differs from that assumed for the simulations in different cases, also checking the convergence of an iterative approach. We will also show the performance of the binned method and a comparison of the results for the full and only-polarization approaches described in section~\ref{sec:CasosTEB&EB}.

\subsection{Robustness of QML with respect to the assumed fiducial model}
\label{sec:robustness}

\begin{figure}
  \begin{center}
    \includegraphics[scale=1.3]{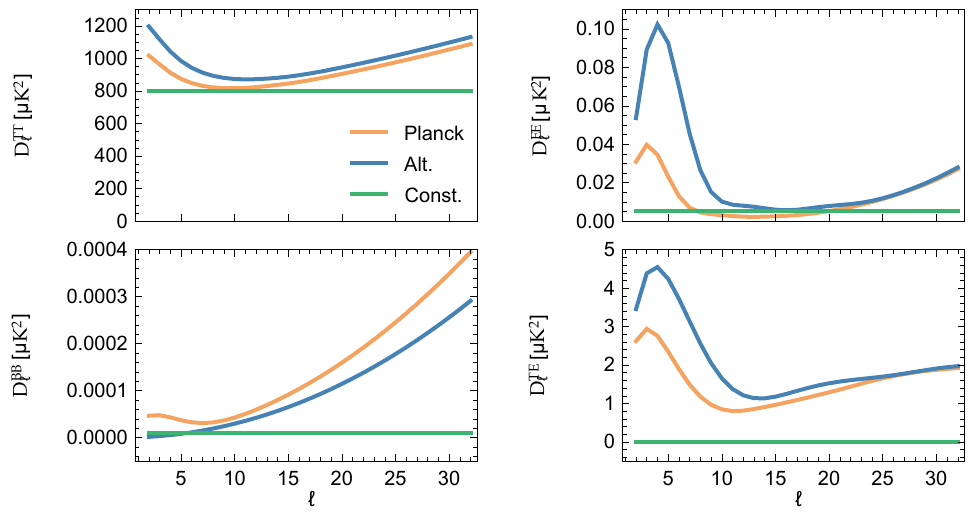}
  \end{center}
  \caption{Power spectra considered to test the robustness of the QML method versus the assumed fiducial model.}
  \label{fig:TresFiducial}
\end{figure}

In order to test the robustness of the power spectra estimated by QML versus the initial assumed model, we have generated $N_\mathrm{sim}=10000$ simulations in the space configuration using our Planck model (i.e. the Planck best-fit $\Lambda$CMD model but adding $r$=0.003) for a resolution of $\ns = 16$ and $\ellm = 32$. We have estimated the power spectra of each simulation assuming three different power spectra: the Planck model (i.e., the correct underlying model), an alternative $\Lambda$CDM model (with a larger scalar amplitude than the previous Planck model and $r=0$)
and a constant value for each of the six components. The three models are given in figure~\ref{fig:TresFiducial}. Note that we have assumed a null TB and EB spectra for the three cases.

\begin{figure}
    \centering
    \includegraphics[scale=.95]{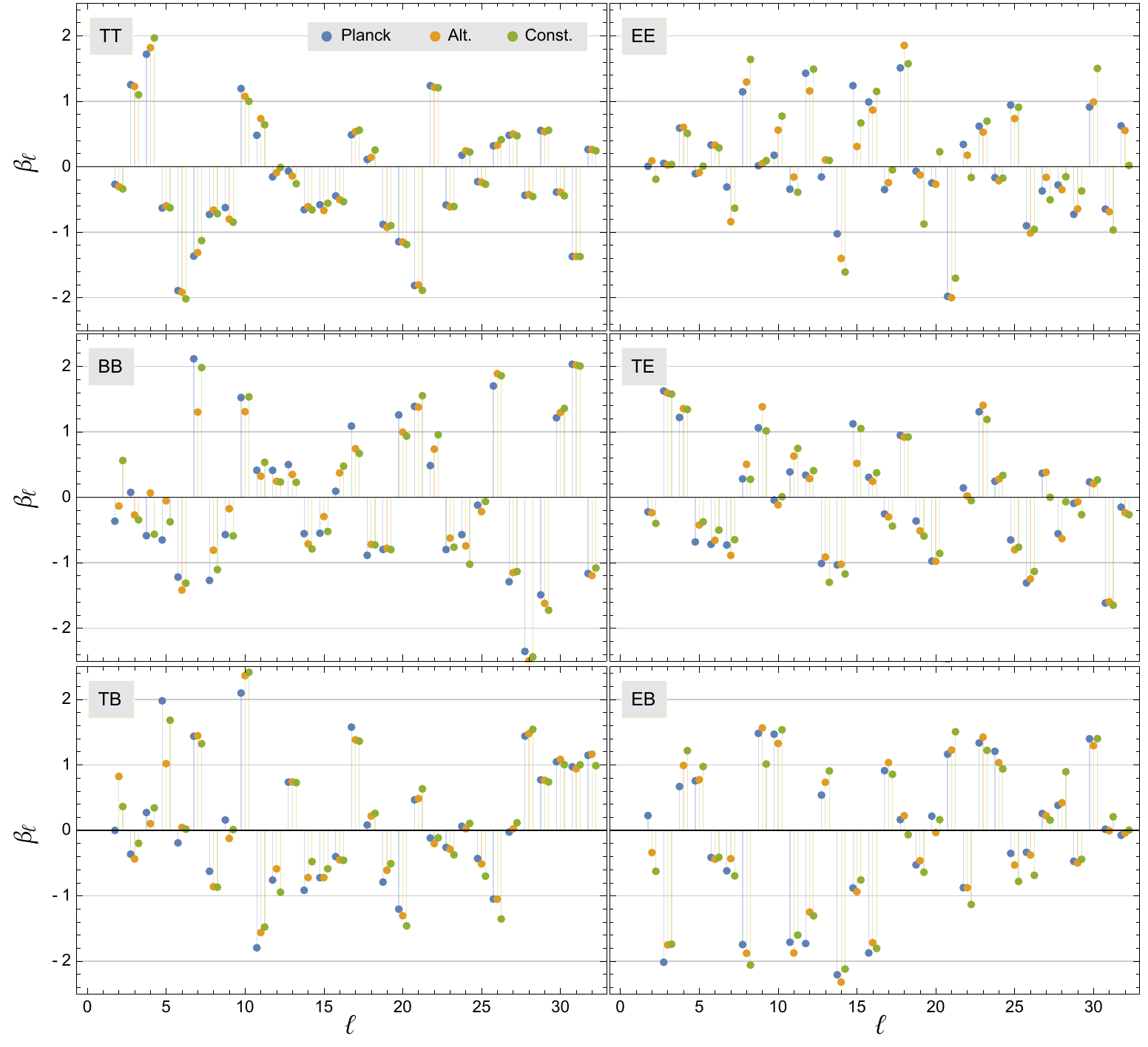}
    \caption{Relative bias on the power spectra estimated with QML when starting with the Planck (blue), alternative $\Lambda$CDM (orange) and constant models (green) obtained using 10000 simulations generated with the Planck model.}
    \label{fig:SesgoSinBinear}
\end{figure}

We find that the estimated power spectra averaged over the simulations follow the Planck model closely, for the three considered cases, i.e, the method is basically unbiased even if the power spectrum assumed to calculate the QML estimates differs, even greatly, from the true underlying model. To quantify this result, we have calculated the relative bias $\beta_{\ell}$ between the true and estimated spectra (averaged over simulations) with respect to the estimated error on the mean average of the power spectrum, i.e.
\begin{equation}
\beta_{\ell}^{\mathrm{Altern.}}=\frac{\langle D^{\mathrm{Altern.}}_\ell \rangle-  D^{\mathrm{Planck}}_\ell }{\sigma^{\mathrm{Altern.}}_\ell/\sqrt{N_\mathrm{sim}}}
\end{equation}
when assuming the alternative $\Lambda$CDM model and analogously for the Planck and constant spectra. Figure~\ref{fig:SesgoSinBinear} gives the relative bias for the three considered fiducial models, showing that there are not significant outliers for any multipole or spectra component, confirming that the method is unbiased.

Regarding the errors of the estimated power spectra, as expected, we find that they are generally sub-optimal when starting with a wrong fiducial model. This has been quantified by looking at the ratio between the errors $\sigma_\ell$ at each multipole obtained from simulations with a wrong fiducial (alternative or constant) versus those of the Planck fiducial, i.e.
\begin{equation}
    \eta_{\ell}^{\mathrm{Altern.}}=\frac{\sigma_{\ell}^{\mathrm{Altern.}}}{\sigma_{\ell}^{Planck}}
\end{equation}
when assuming the alternative $\Lambda$CDM model and analogously for a constant spectrum.

In particular, table~\ref{tab:BarrasRelativas} shows the maximum value of $\eta_{\ell}$ for the alternative and constant cases (top) and its average value over multipoles (bottom) for the six components of the power spectra. The maximum difference is found for the alternative case and the BB  spectra corresponding to a multipole $\ell=2$, with a ratio of 1.79. The TB and EB spectra are also affected, while for TT, the errors increase only slightly. Regarding the mean ratio, again the alternative case for the BB spectra gives the largest errors (1.08) when compared to the case when the correct fiducial model is used.
\begin{table}
    \centering
    \begin{tabular}{|lcccccc|} \hline
        Fiducial & TT & EE & BB & TE & TB & EB \\ \hline \hline
         \multicolumn{7}{|c|}{Maximum $\eta_{\ell}$} \\ \hline
        Alternative & 1.01 & 1.11 & 1.79 & 1.04 & 1.37 & 1.39 \\
        Constant  & 1.02 & 1.11 & 1.30 & 1.05 & 1.14 & 1.14 \\ \hline \hline
         \multicolumn{7}{|c|}{$\left <\eta_{\ell}\right>$} \\ \hline
          Alternative &  1.00 & 1.02 & 1.08 &  1.01 & 1.04 &  1.05  \\
        Constant    &  1.01 & 1.07 &  1.04 & 1.03 &  1.02 & 1.06 \\ \hline
    \end{tabular}
    \caption{Top: maximum value of $\eta_{\ell}$, the ratio between the errors obtained when the assumed fiducial is the alternative or the constant model versus those obtained when the fiducial is Planck. Bottom: average value of $\eta_{\ell}$.}
    \label{tab:BarrasRelativas}
\end{table}
In summary, these results show that the QML estimate is unbiased versus the choice of a (reasonable) fiducial model but sub-optimal with regard to its errors. This leads naturally to the possibility of using an iterative scheme, where the assumed fiducial model is updated taking into account the output of the QML estimator. In the next subsections we will check the validity of this approach.

We may wonder if the binned QML will also be robust versus the choice of the fiducial model. As shown in section~\ref{sec:EstimadorBineado}, we recall that the binned QML makes used of the information of the fiducial model not only to evaluate the matrix
$\mC$ (as in the unbinned case) but also to construct the $\{f_{\ell}^b\}$ set which is needed to reduce the dimensionality of the Fisher matrix in order to make it regular.

To test the performance of the binned estimator, we have applied it to the same 10000 simulations considering the three different fiducial models again and using the bins limits
\begin{equation}
 \ell^b_{\mathrm{high}} = \{4, 8, 12, 16, 20, 24, 28, 32\}.
\end{equation}

\begin{figure}
    \centering
    \includegraphics[scale=0.95]{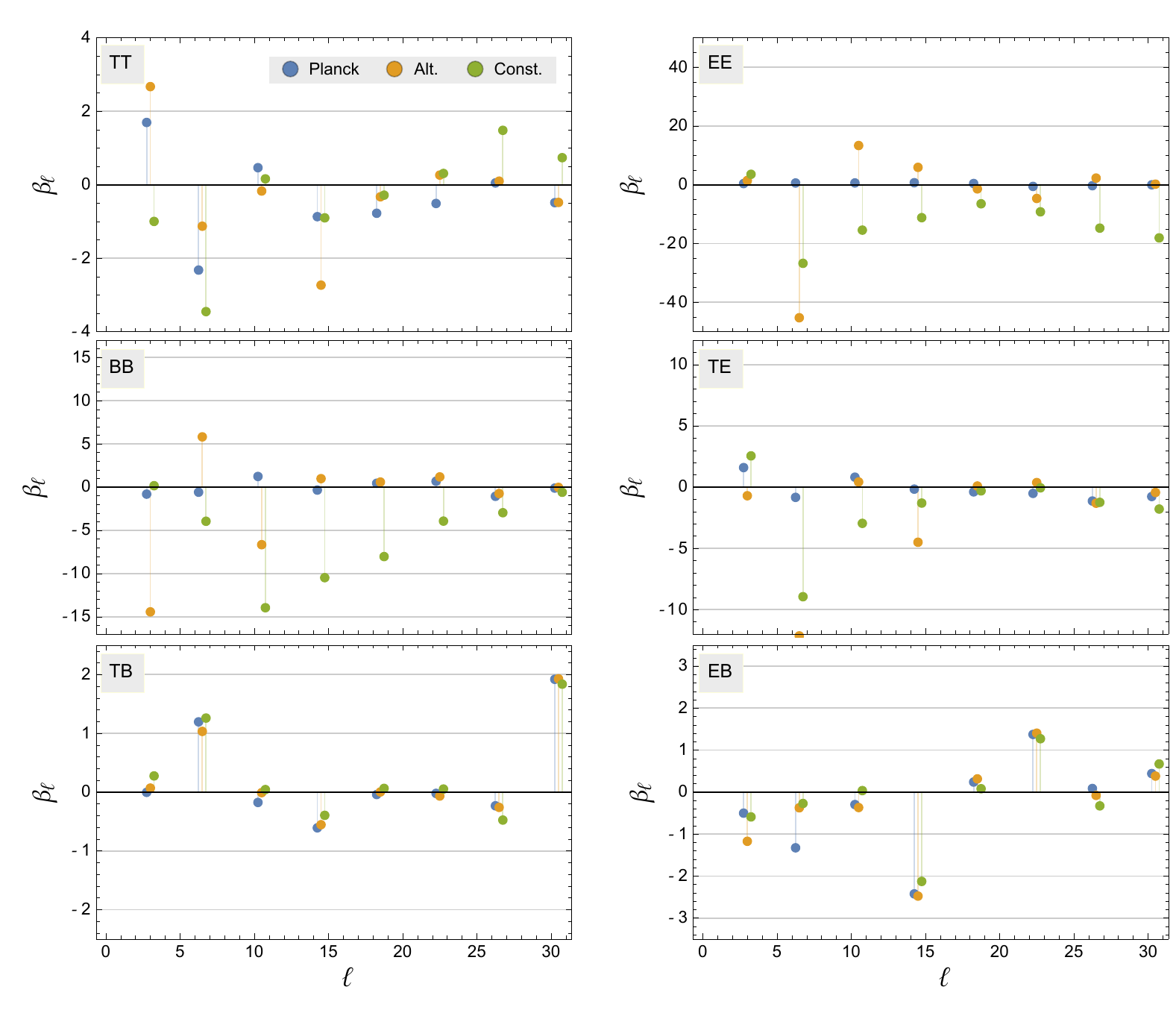}
    \caption{Relative bias on the power spectra estimated with the binned QML when assuming the Planck (blue), alternative (orange) and constant (green) models obtained using 10000 simulations generated with the Planck model.}
    \label{fig:SesgoBineado}
\end{figure}

Figure~\ref{fig:SesgoBineado} shows the relative bias $\beta_{\ell}$ for the different assumed fiducial models when estimating the power spectra with the binned QML. In this case, we find that the method is unbiased only when starting with the correct fiducial model (Planck), while significant biases are found when assuming the constant or alternative models. This reflects the fact that the binned QML is more sensitive to the choice of the initial power spectra. This could be understood since this quantity appears in a different way in the binned estimator. In particular, for the standard QML the estimated spectrum is given by eq.~(\ref{ec:Estimador}), therefore the fiducial model enters twice through the inverse of the covariance matrix in the  matrices $\mE_i$ (eq.~(\ref{ec:DefEi})) and $\mF$ (eq.~(\ref{ec:MatrizFisher})). 
The estimation itself is given by the multiplication of the vector $\vy$ and the inverse of the Fisher matrix, being
its effect somehow partially compensated.  However, in the binned QML estimation the fiducial enters also through the matrix $\mR$ (eq.~(\ref{ec:FinalEstimador})) through the factors $\{f_{\ell}^b\}$. This matrix is used twice to reduce the size of the Fisher matrix and only once for the same process for the vector $\vy$ and, therefore, the effect of the fiducial is more unbalanced. We may wonder if the situation improves if we do not include the information about the fiducial model in the binning (i.e. using $f_\ell^b=1$). However,
we note that this is actually equivalent to assume a constant fiducial model for the binning step. Therefore, when including constant weights in the binning, we are actually considering a extreme case for the fiducial and, as one would expect, this leads in general to larger biases.
Therefore, once a (reasonable) fiducial model is assumed, it is convenient to include this information in all the steps.

\begin{figure}
  \begin{center}
  \begin{tabular}{ccc}
   \includegraphics[scale=.77]{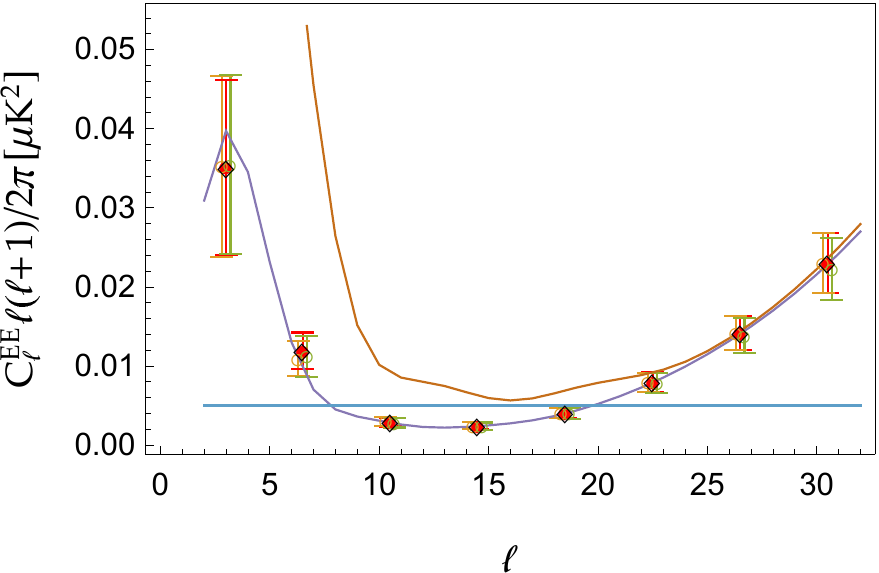} & &
     \includegraphics[scale=.77]{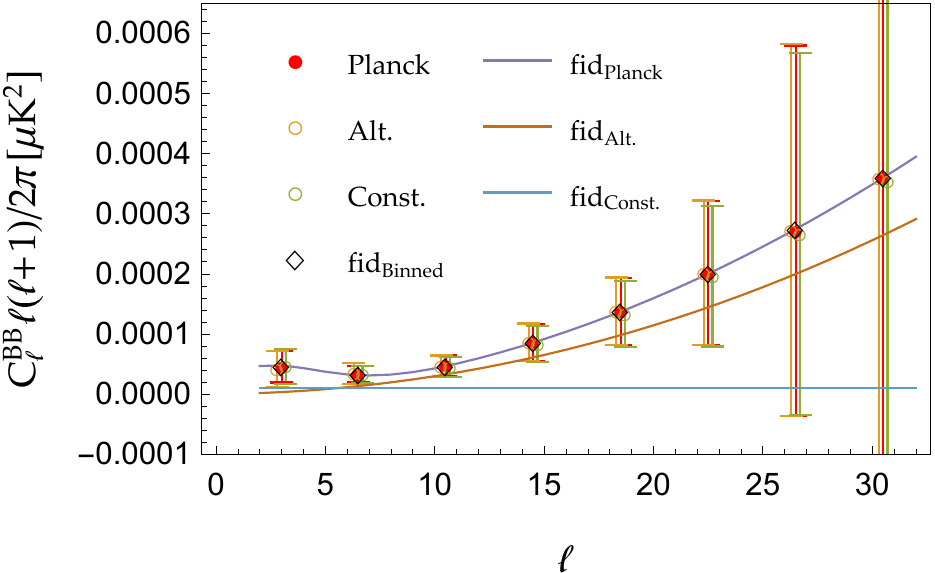}
  \end{tabular}
    \caption{The mean and dispersion of the power spectra for EE (left) and BB (right) estimated with the binned QML, averaging over 10000 simulations and starting with three different fiducial models are shown. The used fiducial models (labelled as fid$_\mathrm{Planck}$, fid$_\mathrm{Alt.}$ y  fid$_\mathrm{Const.}$) are also given as solid lines. For comparison, the binned spectra obtained from the Planck model (labelled as fid$_\mathrm{Binned}$), i.e. the model used to generate the simulations, is also given.}
    \label{fig:QMLBineadoTresFiducial}
 \end{center}
\end{figure}

We should also note that the biases found are well within the error of the estimated power spectra for a single realization and, therefore, in practice, they are relatively small. This can be better appreciated in figure~\ref{fig:QMLBineadoTresFiducial}, which, as an illustration, shows that the three estimations for the EE (left) and BB (right) spectra are
actually quite similar independently of the initial guess. However, the small error in the average power spectra allows to detect the presence of these biases. One can also appreciate that, when assuming a wrong fiducial spectrum, the output binned QML moves from it towards the correct model. This is a clear indication that the initial fiducial model does not reflect the true underlying spectra and, therefore, some kind of iterating scheme is recommended. This is discussed in more detail for the binned estimator in subsection~\ref{sec:robust_r}.

Regarding the increase of the estimation error, table~\ref{tab:BarrasRelativasBineado} shows the maximum and mean value of $\eta_{\ell}$ for the binned QML. We find a similar behavior to that of the standard QML, although with lower ratios, especially for BB whose maximum ratio is found to be 1.25 (when starting with the alternative fiducial model).

\begin{table}
    \centering
    \begin{tabular}{|lcccccc|} \hline
        Fiducial & TT & EE & BB & TE & TB & EB \\ \hline \hline
         \multicolumn{7}{|c|}{Maximum $\eta_{\ell}$} \\ \hline
        Alternative & 1.01 & 1.11 & 1.25 & 1.05 & 1.18 & 1.25 \\
        Constant  & 1.01 & 1.12 & 1.08 & 1.07 & 1.07 & 1.30 \\ \hline \hline
         \multicolumn{7}{|c|}{$\left<\eta_{\ell}\right>$} \\ \hline
          Alternative &  1.00 & 1.02 & 1.06 &  1.01 & 1.04 &  1.07  \\
        Constant    &  1.00 & 1.06 &  1.00 & 1.04 &  1.02 & 1.08 \\ \hline
    \end{tabular}
    \caption{Top: maximum value of $\eta_{\ell}$, the ratio between the errors obtained when the assumed fiducial is alternative or constant versus those obtained when the fiducial is Planck. Bottom: average value of $\eta_{\ell}$. Results are obtained for the binned QML.}
    \label{tab:BarrasRelativasBineado}
\end{table}

\subsection{Iterative QML}
\label{sec:iterativeQML}
To test if an iterative process would lead to an unbiased and minimum variance QML estimator, independently of the initially assumed fiducial model, we have carried out a further test using 500 simulations with the same characteristics as those of the previous subsection. In particular, we have estimated the power spectra with three different initial power spectra (as before, Planck, alternative $\Lambda$CDM and constant spectra) for each of the simulations. After this first estimation, we have iteratively estimated the spectra another four times modifying the assumed fiducial model taking into account the output of QML for the previous step. Therefore, in this test, we have applied QML a total of $3 \times 500 \times 5$ times. In this subsection, we have considered only the standard (unbinned) QML, while the performance of the binned QML will be explored for a specific case in the next subsection.

When iterating, one could simply use as the updated fiducial model, the values of the power spectra directly estimated with QML. However, we have tested that this may lead to failures in the method, due for instance to the fact that some multipoles are estimated as negative (especially for BB), leading to singular covariance matrices. Therefore, it is convenient to use a smoothed version of the output spectra as a guess for the next iterative step. Details about how this smoothing has been implemented are given in appendix~\ref{ap:Suavizado}. Even with this approach, some instabilities can be present and it has not been possible to complete the full iterative process for all the simulations and the three different initial fiducial models. Note that one could complete the full process for all the simulations by tuning the parameters used in the smoothing. However, we have seen that this does not affect our results and, for simplicity, we have just discarded those simulations that have failed at some step.
Therefore, we present results for a total of 453 simulations, for which the full process has been completed without further tuning.

To test the performance of the iterative QML, we have studied the convergence of the method independently of the initial fiducial model, the consistency of the results for the three considered initial spectra and the evolution of the estimation errors with iterations.

\begin{figure}
  \begin{center}
  \begin{tabular}{ccc}
  	\includegraphics[scale=.8]{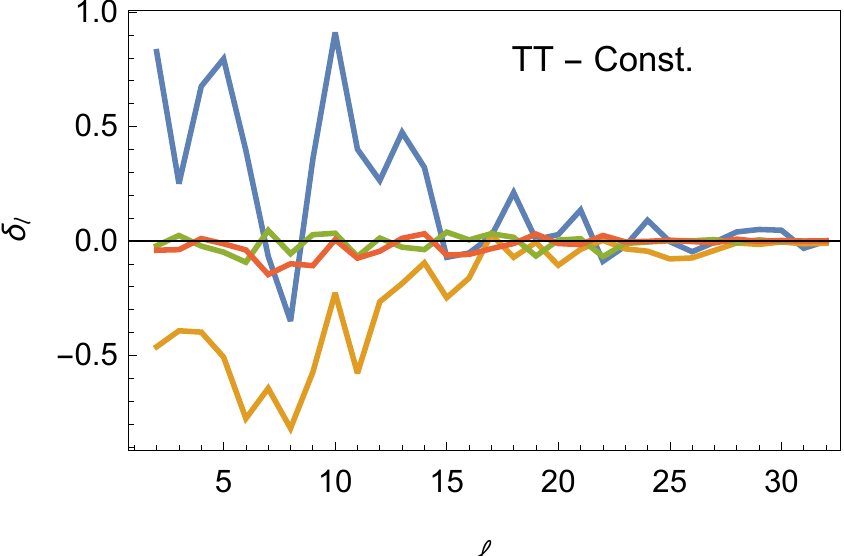}  & &
  	\includegraphics[scale=.8]{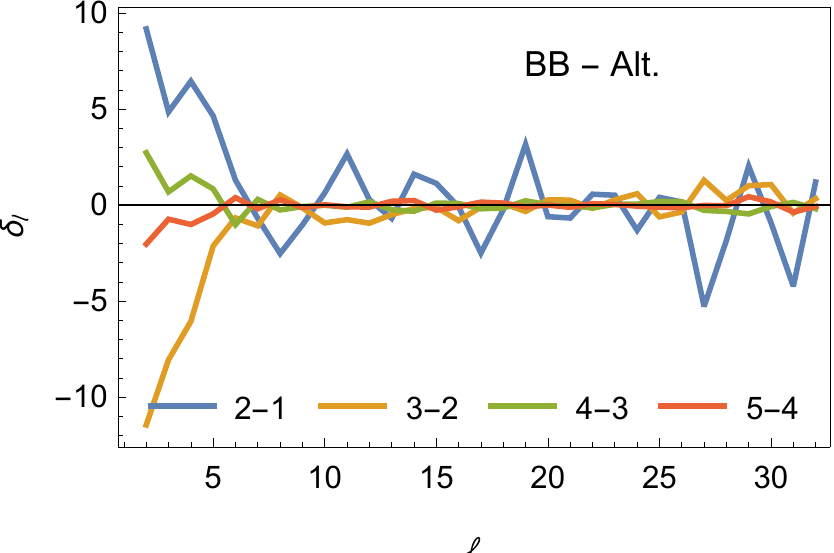}
  \end{tabular}
    \caption{Convergence of the iterations for TT when starting with the constant spectra (left) and BB when starting with the alternative model (right). The different colours indicate the relative difference (in percentage) between two consecutive steps averaged over simulations with respect to the value estimated with the correct fiducial model.}
    \label{fig:ReducionDiferencias}
  \end{center}
\end{figure}

Regarding the convergence of the method, we find that after around five steps (i.e., the initial QML plus four iterations), the results are already quite stable. In particular, we have studied the evolution of the convergence by looking at the relative difference between two consecutive steps with respect to the spectra estimated with the correct fiducial model (in this case without iterating), i.e.
\begin{equation}
\delta_{\ell}=\frac{\langle D^{\mathrm{j+1,Altern.}}_{\ell} - D^{\mathrm{j,Altern.}}_{\ell}\rangle}{\langle D^{\mathrm{Planck}}_{\ell} \rangle} \times 100
\end{equation}
when starting with the alternative $\Lambda$CDM model and analogously for the constant model. Note that $j$ corresponds to the step in the iterative process and that averages are obtained over simulations. As an example, figure~\ref{fig:ReducionDiferencias} shows this quantity for two cases: TT spectrum starting with the constant model (left) and BB spectrum starting with the alternative model (right). As one would expect if the method converges, these differences decrease when advancing in the number of iterations. A similar behaviour is found for the other considered cases.

\begin{figure}
  \begin{center}
  \begin{tabular}{ccc}
  	\includegraphics[scale=.8]{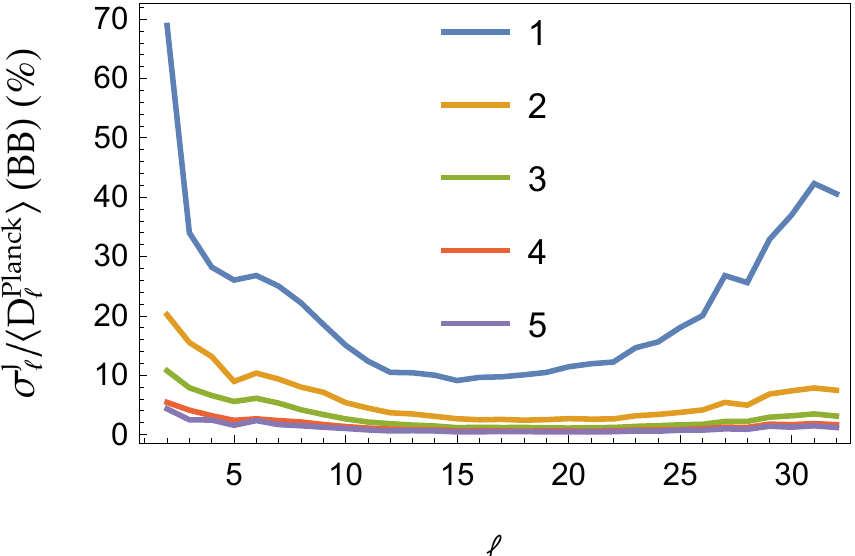} & &
  	\includegraphics[scale=.8]{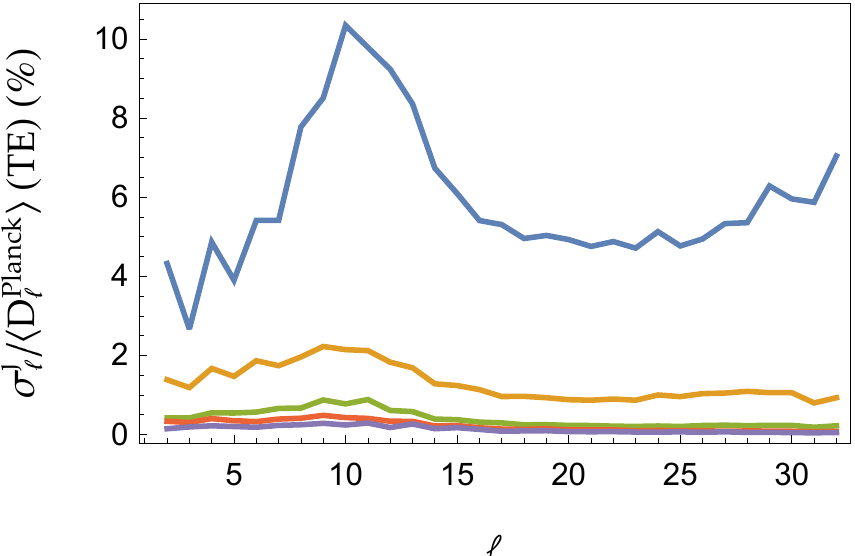} \\
  \end{tabular}
    \caption{Ratio (in percentage) of the average dispersion obtained over the three different estimates of the spectra over the average power spectra obtained with the Planck fiducial (without iterating) for each iterative step $j=1,5$ (indicated by different colours) for BB (left) and TE (right). }
    \label{fig:EvolucionSigmaTresComienzos}
  \end{center}
\end{figure}

We have also tested if the results obtained with the QML by the three different starting fiducial models converge to the same values as the iterations progress. To quantify this point, we have calculated for each simulation and for each iteration the dispersion between the three estimations obtained with the three different starting models (Planck, alternative and constant) at each multipole. Therefore we have the function
\begin{equation}
   \label{ec:SigmaDiferencias}
    \sigma_{\ell,i}^\mathrm{j} = \mathrm{Dispersion} \{ D_{\ell,{\mathrm{i}}}^{\mathrm{j,Planck}}, D_{\ell,{\mathrm{i}}}^{\mathrm{j,Alt.}}, D_{\ell,{\mathrm{i}}}^{\mathrm{j,Const.}} \}
\end{equation}
for each multipole $\ell$, simulation $i$ and step $j$. Figure~\ref{fig:EvolucionSigmaTresComienzos} shows the ratio (in percentage) of this quantity averaged over simulations relative to the estimated spectra obtained when starting with the Planck model (without iterating), also averaged over simulations, for BB (left) and TE (right). As seen, the dispersion between estimates is significantly reduced when increasing the number of iterations, showing that the iterative QML leads to very similar results, not only on average but also simulation by simulation, independently of the chosen initial spectra. Similar conclusions are reached for the other spectra.

We have also checked that, independently of the starting fiducial model, the errors in the estimated power spectra at the end of the iterative process are very similar to those obtained for the optimal case (i.e., using the correct fiducial model and not iterating). This is expected since we have seen that the iterative QML converges basically to the same result for the three considered cases.
Note that these conclusions also hold for the case in which we start with the correct fiducial model, showing that the process is stable and that there is not danger in iterating even when one is already in the right initial point.

\subsection{Robustness of QML with respect to the assumed tensor-to-scalar ratio}
\label{sec:robust_r}

In the previous subsections, we have considered the robustness of the QML versus the initial choice of the fiducial model as well as the performance of an iterative approach for different initial generic spectra. However, future experiments will focus on the estimation of the tensor-to-scalar ratio and, therefore, we think it is interesting to study specifically the sensitivity of the method to a wrong initial value of $r$. Given that many of these experiments observe only a small fraction of the sky, we will also consider the binned version of the QML. This also allows us to check the consistency between both approaches and whether information could be missed when using the binned estimator.

In this section, the iterative approach will also be considered, although we will use a different method to provide an initial guess for the fiducial model. We will assume that all cosmological parameters are known, except for the tensor-to-scalar ratio $r$, so we will consider three different initial fiducial models for QML that differ in the value of $r$. Then, rather than smoothing the output spectra of the previous step, we will estimate $r$ from the QML spectra and use it to construct the guess spectra for the next iteration. The estimator used for $r$ is described in detail in appendix~\ref{ap:r}.

In particular, we have carried out the following procedure:

\begin{enumerate}
\item We simulate one map (including CMB and noise) with the specifications of the space case (given in table~\ref{tab:Experimentos}) and with a tensor-to-scalar ratio $r_{true}=0.003$. The map is generated at $\ns=16$, with $\ellm=32$ and smoothed with a Gaussian beam of 8.79 degrees of full-width half maximum.

\item We apply the unbinned and the binned QML to estimate the power spectrum, starting with a wrong fiducial model (i.e. $r_0 \ne r_{true}$). More specifically, we consider two cases: $r_0=0$ and $r_0=0.03$. For comparison, the case $r_0 = r_{true}$ is also considered. For the binned estimator, we have used 8 bins with $\ell^{b}_{\mathrm{high}}$ given by \{4, 8, 12, 16, 20, 24, 28, 32\}.

\item An estimator $\hat{r}$  of the tensor-to-scalar ratio is obtained as explained in appendix~\ref{ap:r}.

\item We update the fiducial model using the estimated value $\hat{r}$ and apply again the unbinned and binned QML. A total of five iterative steps are performed.

\end{enumerate}
The full process is repeated for 200 simulations.

\begin{figure}
\centering
\begin{tabular}{ccc}
\includegraphics[scale=0.72]{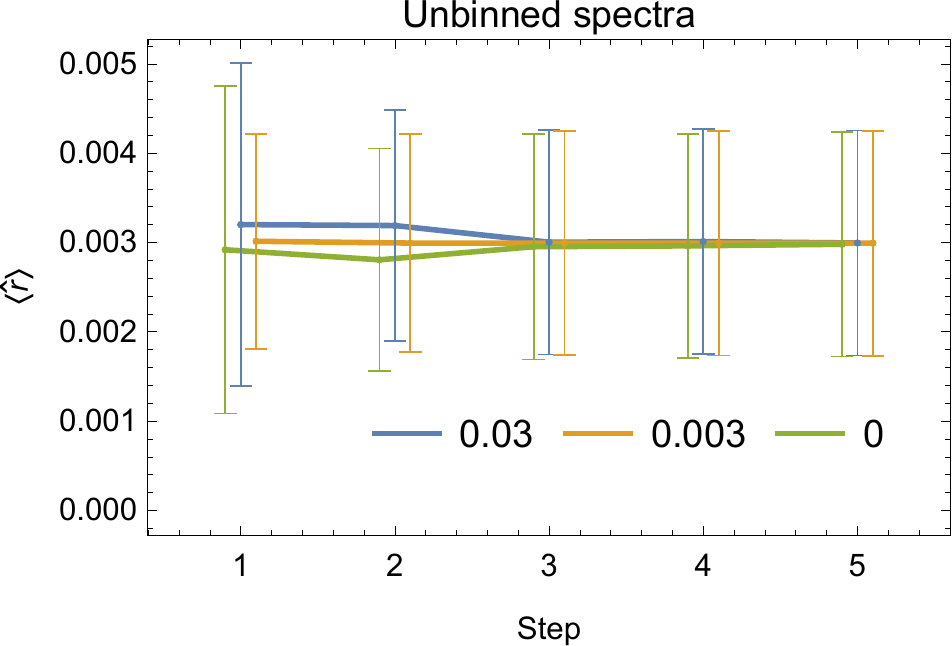} & &
\includegraphics[scale=0.72]{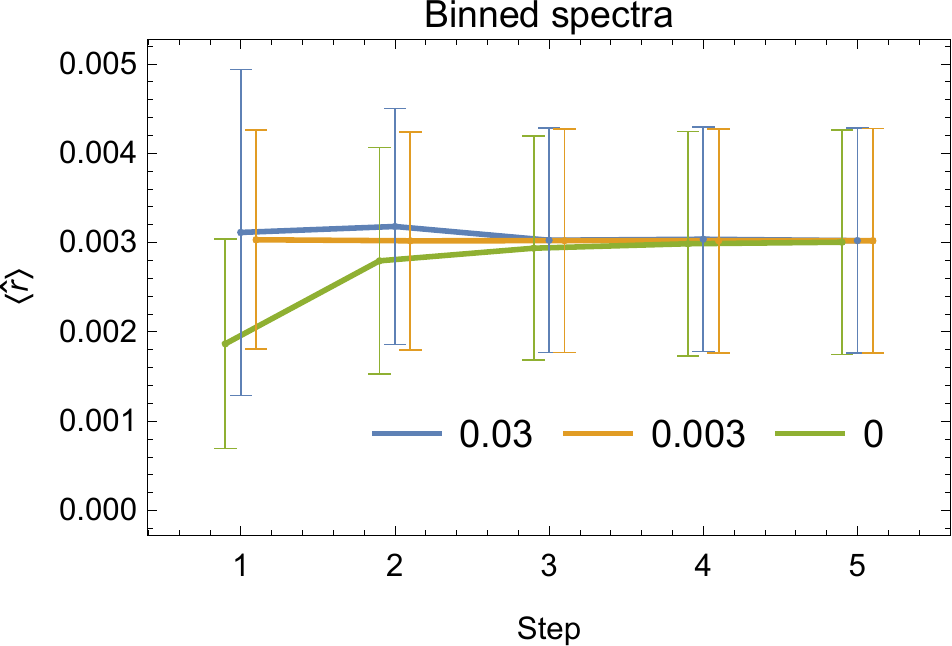}\\
\end{tabular}
\caption{Left: evolution through the steps on each one of the five iterations with starting points $r_0$ = 0.03, 0.003 and 0, of the mean and the standard deviation over 200 simulated maps of $\hat{r}$ estimated from the QML power spectrum. Right: $\hat{r}$ obtained from the same maps in the same conditions than in the left part, but from a binned spectra.}
\label{fig:EvolucionREstimado}
\end{figure}

Figure~\ref{fig:EvolucionREstimado} shows the progression with the number of iterations of the mean value and dispersion of $\hat{r}$ obtained over the simulations, for the three values of $r_0$ and for the unbinned (left) and the binned (right) QML estimator. For the standard (unbinned) estimator,
the results indicate that when starting with a wrong fiducial, even without any iteration, the estimator is close to unbiased although not of minimum variance. For the binned estimator, this is also the case when starting with a fiducial model with $r_0=0.03$. However, for $r_0=0$ without iterating, we find that the mean of the estimated values of $r$ is around $1\sigma$ below the true value. Since that for the standard QML case we do not find that deviation (left panel), this indicates again that the binned QML is more sensitive to the choice of the initial fiducial model, at least in some cases, which is consistent with the results found in section~\ref{sec:robustness}.  However, we see that $ \left< \hat{r} \right> $ converges rapidly with the iterations to the true value, independently of the starting point, for both the binned and unbinned QML. Also, the error of the estimation of $r$ decreases, becoming quite stable after around four steps.

\begin{figure}
  \begin{center}

    \includegraphics[scale=0.65]{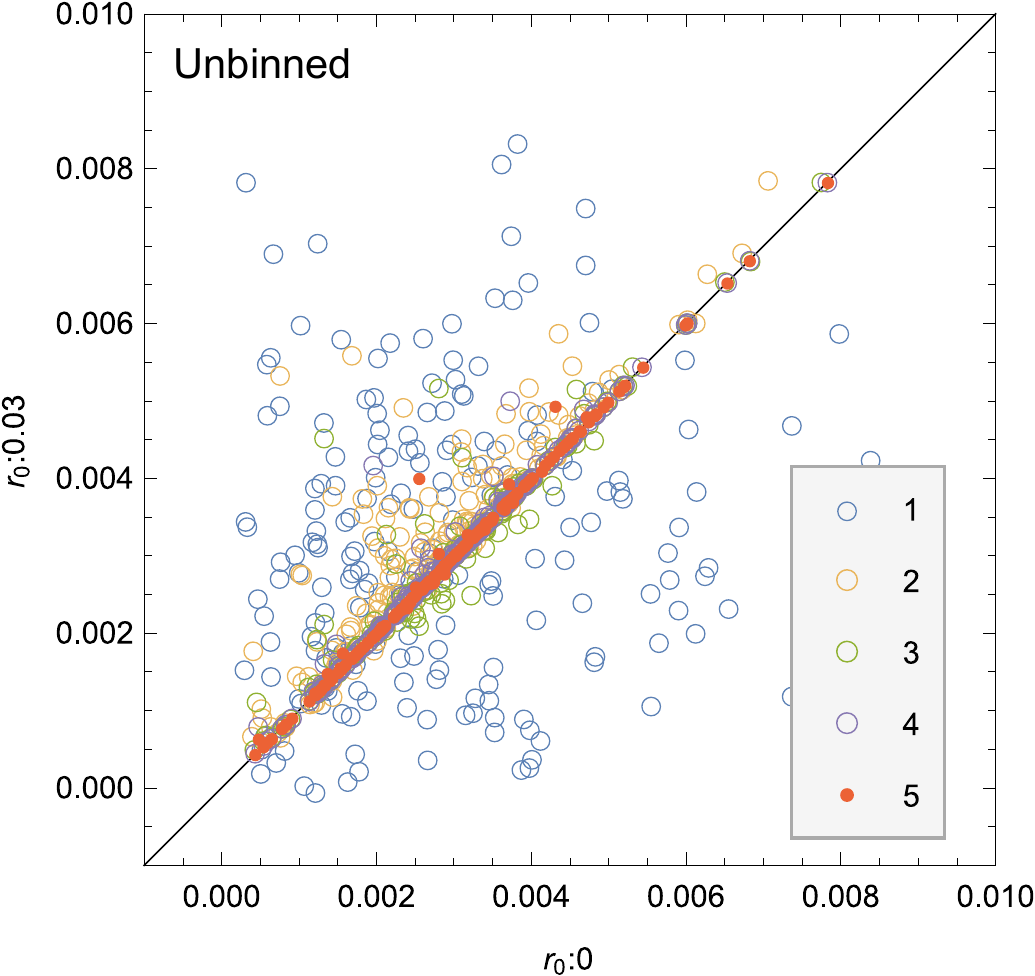}

    \vspace{5 mm}

    \begin{tabular}{lcr}
    \includegraphics[scale=0.65]{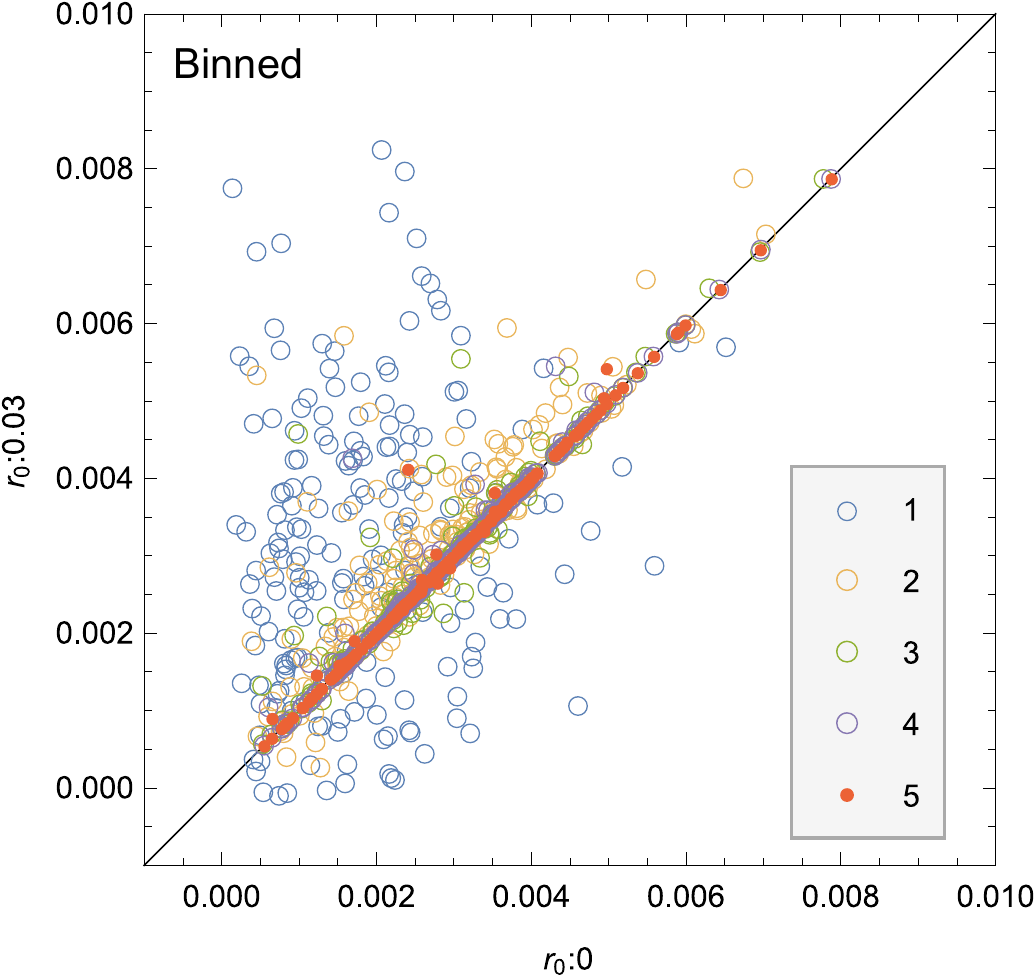} & & \includegraphics[scale=0.65]{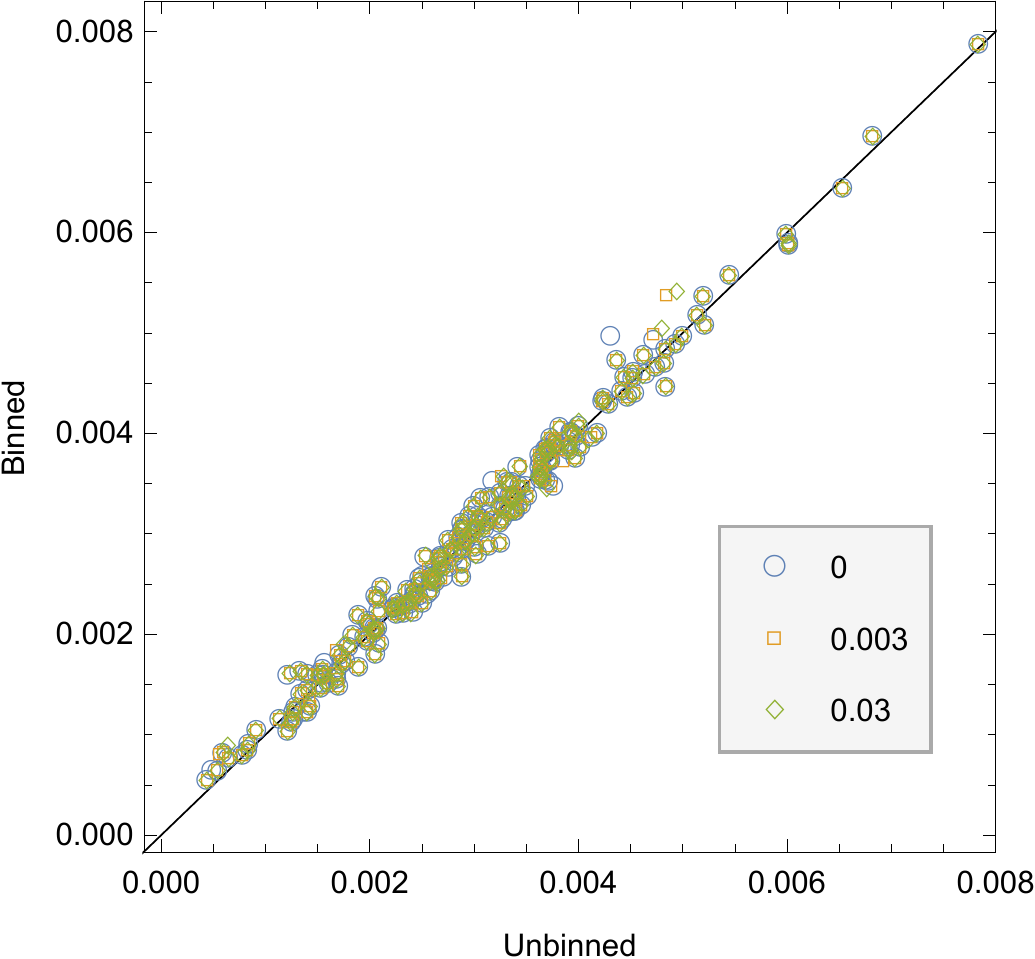}
    \end{tabular}

	\caption{Top: Tensor-to-scalar ratio estimated from the QML spectra starting with a fiducial model with $r=0.03$ (ordinate) plotted against the tensor-to-scalar ratio estimated starting with a fiducial model with $r_0=0$  (abscissa) for different number of iterations. Bottom-left: same plot using the binned spectra. Bottom-right: $\hat{r}$ obtained at the last step of the iteration from the unbinned spectra versus that obtained from binned spectra, for the three different starting points of $r_0$ considered. In all cases, points from 200 simulations are shown.}
    \label{fig:CorrelcacionesIteracion}
  \end{center}
\end{figure}

The good convergence of the iterative QML can also be confirmed by looking at the top and bottom-left panels of figure~\ref{fig:CorrelcacionesIteracion} that show, for each simulation, the estimated value of $r$ for one starting point ($r_0=0$) versus the one estimated for the other initial value ($r_0=0.03)$ for different number of iterations. For both, the unbinned (top panel) and binned (bottom-left panel) QML, we see that individual values of $\hat{r}$ tend to migrate to the diagonal of the plot through the steps of the iteration, showing the good performance of the iterative approach.\footnote{Note that this test is stronger than that of figure~\ref{fig:EvolucionREstimado}, since we impose convergence at each individual simulation, finding that a few values of $\hat{r}$ still deviate from the diagonal at step 5. Indeed, we tested that with more iterations (with around ten steps in total), these values also move to the diagonal} For the binned QML, it also becomes apparent that when no iterating, the estimation of $r$ tends to be lower when starting with $r_0=0$ versus the values obtained with $r_0=0.03$ as reflected in the asymmetric distribution of blue circles around the diagonal. The bottom-right panel of figure~\ref{fig:CorrelcacionesIteracion} shows the estimated values of $r$, obtained at the last step of the iterations, from the binned spectra versus those obtained from the unbinned one, for the three different starting points considered. It is apparent that the two estimations of $r$ (obtained from the binned and unbinned QML) are very similar, clustering around the diagonal. In addition, we see again that the position of the points is independent of the value of $r_0$.

Finally, table~\ref{tab:FinalIteraciones} gives the mean value and standard deviation of $\hat{r}$ in the last iterative step for the three starting points for the unbinned (top) and binned (bottom) spectra, showing an excellent agreement between the different cases. This shows that, at least in the considered case, no information is lost when using the binned version of the QML with respect to the standard implementation and that the iterative approach is robust versus the choice of the initial tensor-to-scalar ratio for the binned and unbinned QML. For comparison, we note that the theoretical errors ($\Delta r$ given by eq.~(\ref{ec:ErrorLikelihood})) for a fiducial model with $r = 0.003$, are $1.16\times 10^{-3}$ and $1.18 \times 10^{-3}$ for the unbinned and binned spectra, respectively, which are somewhat below those found for $\sigma_{\hat{r}}$. This difference is due to the relatively small number of simulations, which only allows estimating the error of $\hat{r}$ with limited precision.

\begin{table}
\begin{center}
\begin{tabular}{|l|ccc|}
\hline
&$r_0$ & $\expec{\hat{r}}  $ & $\sigma_{\hat{r}} $\\
\hline
&0 & $3.00 \times 10^{-3}$ & $1.26 \times 10^{-3}$ \\
Standard QML & 0.003 & $3.02 \times 10^{-3}$ & $1.26 \times 10^{-3}$ \\
& 0.03 & $3.02 \times 10^{-3}$ & $1.26 \times 10^{-3}$ \\
\hline
&0 & $2.98 \times 10^{-3}$ & $1.26 \times 10^{-3}$ \\
Binned QML &0.003 & $2.99 \times 10^{-3}$ & $1.26 \times 10^{-3}$ \\
&0.03 & $2.99 \times 10^{-3}$ & $1.26 \times 10^{-3}$\\
\hline
\end{tabular}

\caption{Mean and standard deviation of $\hat{r}$ from 200 simulated maps (space configuration) at resolution $\ns = 16$ generated with $r_{\mathrm{true}}=0.003$ at the last of the five steps of an iterative scheme for the standard and binned QML. In both cases, values obtained after an iterative process with different starting points are shown.} \label{tab:FinalIteraciones}
\end{center}
\end{table}

\subsection{Only-polarization QML}

Finally, we may also wonder whether the estimation of $\hat{r}$ is degraded when using the only-polarization QML configuration (\mc) versus the full estimator (\ma) described in section~\ref{sec:CasosTEB&EB}. This is important because of the reduction in CPU time achieved when working only with polarization. Therefore, we have applied the configurations \ma and \mc\ (and also \mb for comparison) to 5000 simulations (space configuration, $\ns = 64$, $\ellm=128$) starting from the correct fiducial $r=0.003$ and have estimated the mean value and error for $\hat{r}$.

Table~\ref{tab:EstimacionesR2} shows the mean value, standard deviation and theoretical error bar of $\hat{r}$ for the three considered configurations. As one would expect, since \ma\! includes the full information from the power spectra, it provides a slightly smaller error on $\hat{r}$ than the other two configurations which, at the considered precision, are indistinguishable. However, the differences are very small and, therefore, in practice, it is perfectly acceptable to use the \mc\! configuration in order to save computational resources.

\begin{table}
  \begin{center}
    \begin{tabular}{|cccc|} \hline
      Specification  & $\langle \hat{r} \rangle (\times 10^{-3}) $ &  $\sigma_r (\times 10^{-4}$) & $\Delta r (\times 10^{-4}$) \\ \hline
  \ma   & 3.00 	& 6.51 & 6.50 	\\
  \mb  	& 3.00	& 6.52 & 6.51 	\\
  \mc   & 3.00	& 6.52 & 6.51  \\
 \hline
    \end{tabular}
  \end{center}
  \caption{Results on the estimation of $r$ in the configurations \ma, \mb\! and \mc\! described in section~\ref{sec:CasosTEB&EB}.}
 \label{tab:EstimacionesR2}
\end{table}

\section{Comparison between QML and {\tt NaMaster}}
\label{sec:ComparacionMaster}
The so-called pseudo-spectrum methods (e.g.  \cite{Hivon02}), have become widely used to estimate the CMB power spectra since they require significantly lower computational resources than QML, allowing their computation up to very high multipoles. These methods calculate the spherical harmonic transform in a masked sky and try to deconvolve the effect of this mask through the inverse of the kernel that encodes the coupling in the harmonic space produced by the loss of orthogonality due to the incomplete sky. To reduce the effect of the coupling, masks are usually apodized (e.g.  \cite{Grain_2009}). In the polarization case, due again to the loss of orthogonality, a leakage between E and B modes is also present, which adds an additional complexity to the sought of the very weak primordial B-mode of polarization. The most advanced pseudo-spectrum methods incorporate purification techniques of the E- and B- modes of polarization with the aim to reduce this transfer of power between them  \cite{Elsner_2016}. Although pseudo-spectrum methods are, in practice, of minimum variance for intermediate and high multipoles, they are sub-optimal at large scales, which are particularly relevant for a future determination of the scalar-to-tensor ratio. Therefore, it is interesting to compare the performance of the QML and pseudo-spectrum estimators, to understand the advantages and limitations of each of them (for previous discussions, see e.g.  \cite{Efs04,2014MNRAS.440..957M,Vanneste18}).

In particular, in this section we compare the results from {\tt ECLIPSE}, our QML implementation, to those obtained with {\tt NaMaster}  \cite{Alonso19}, an advanced public implementation of the pseudo-spectrum method,\footnote{{\tt NaMaster} is available at \\ https://github.com/LSSTDESC/NaMASTER} that incorporates different types of apodization as well as the purification technique. For our test, we have applied QML and {\tt NaMaster} to $10000$ simulations in the space configuration at resolution $\ns=64$, with the usual mask (left panel of figure~\ref{fig:Mascaras}) and considering $\ellm=128$. In the case of {\tt NaMaster}, it is possible to tune several parameters (such as type of apodization, apodization scale, to include or not purification) in order to improve the estimated spectra. Although a detailed study of the optimal choice of these parameters is outside the scope of this paper, we have explored several possibilities, prioritizing the recovery of the lowest multipoles for BB. In particular, we find that the \emph{C2}\footnote{In this case, pixels are multiplied by a factor f given by
\begin{equation}
  f  = \left \lbrace  \begin{array}{ll}
  0.5 \left[ 1- \cos{(\pi x)} \right] & {\rm if} \quad x<1 \\
  1 &  {\rm otherwise} \\
  \end{array} \right.,
\end{equation}
where $x=\sqrt{(1- \cos{\theta})/(1- \cos{\theta_*})}$, $\theta_*$ is the apodization scale and $\theta$ is the angular separation between a pixel and the nearest masked pixel. Note that all pixels separated from any masked pixel by more than the apodization scale are left untouched.} option for apodization with a scale of $22^{\circ}$ and the use of B-mode purification is well suited for our purpose when the previous simple Galactic mask is used. However, note that, as it will also be shown, different configurations may produce better results for other components or scales of the spectra as well as for different masks.

\begin{figure}
    \centering
     \includegraphics[scale=0.65]{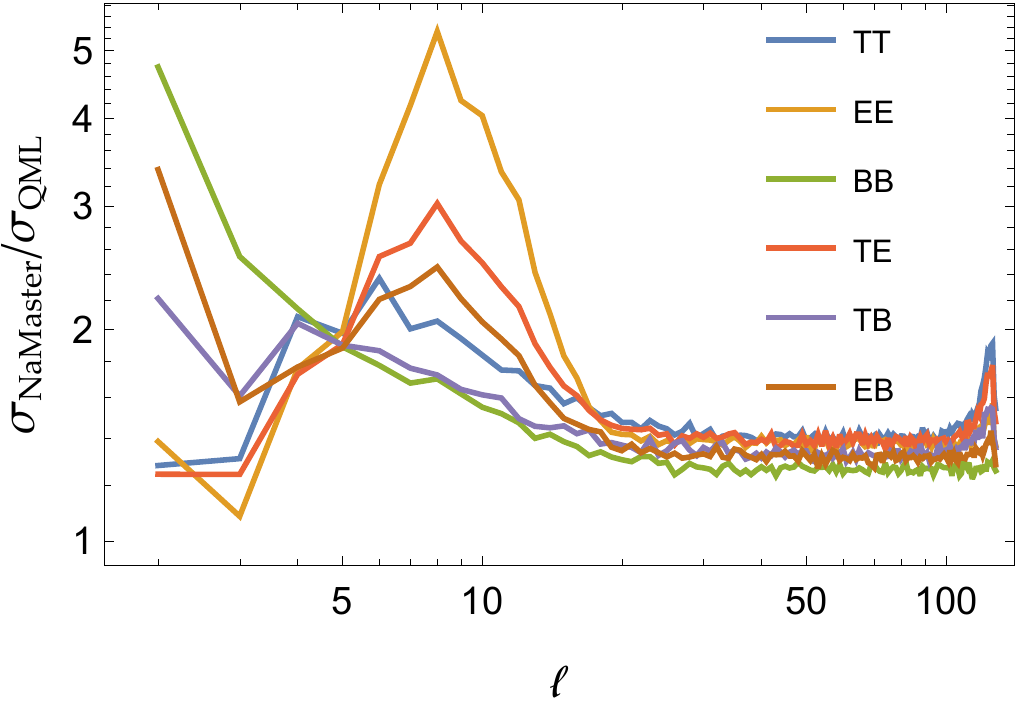}
    \begin{tabular}{ccc}
    & & \\
         \includegraphics[scale=.65]{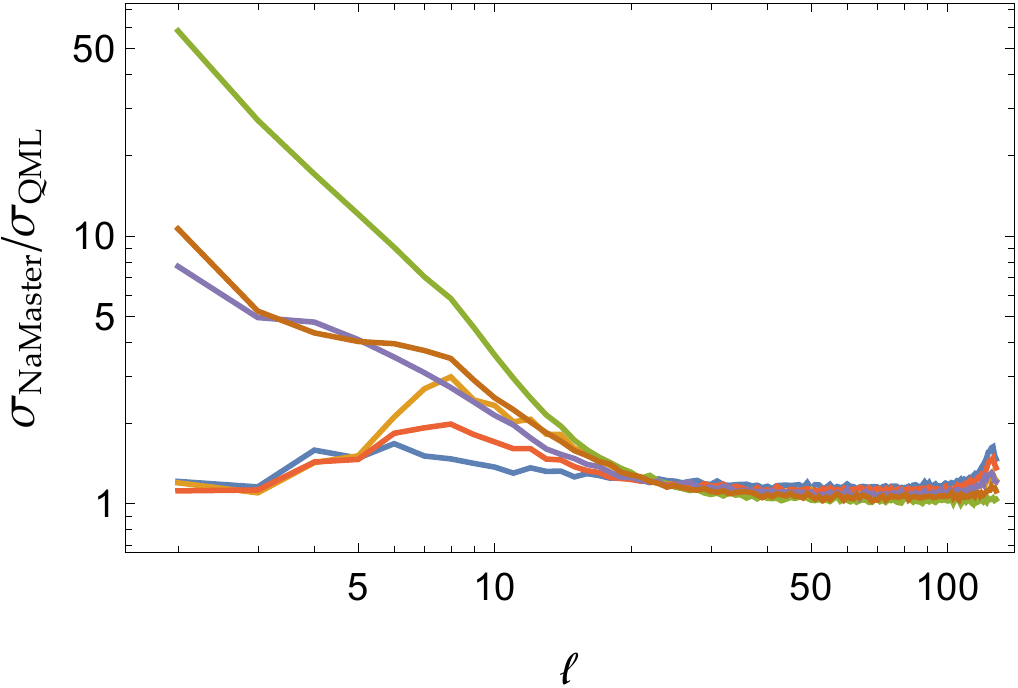} & &
         \includegraphics[scale=0.65]{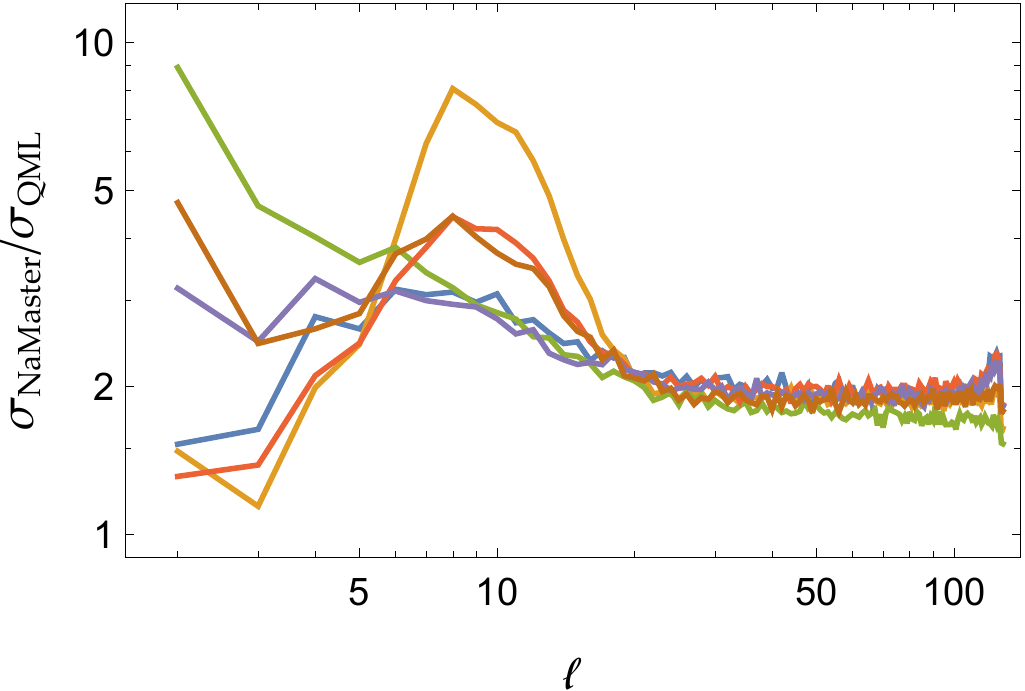} \\
    \end{tabular}

    \caption{Ratio of the errors of the power spectra obtained with {\tt NaMaster} over those estimated with QML. For {\tt NaMaster}, B-mode purification is always used. Top  panel and left-bottom panel, a simple Galactic mask has been considered, whereas for the right-bottom panel, the extended mask from figure~\ref{fig:MascaraProducto} was used. The following apodization options were considered in each case for {\tt NaMaster}: \emph{C2} option with a scale of $22^{\circ}$ (top), \emph{C2} with a scale of $4^{\circ}$ (left-bottom),
    \emph{Smooth} with a scale of 4.5$^{\circ}$ (right-bottom).}
    \label{fig:RatiosNaMASTER}
\end{figure}

As expected, we find that both {\tt ECLIPSE} and {\tt NaMaster} provide unbiased estimations of the different components of the spectra (in the case of NaMaster after subtracting the noise bias). Figure~\ref{fig:RatiosNaMASTER} (top panel) shows the ratio of the estimation errors (obtained from the simulations) achieved with {\tt NaMaster} over those from QML for the considered case. It becomes apparent that important differences are found up to $\ell$ around 20, with maximum values for the ratio of around 5. At the largest multipoles, the ratio is close to 1, but we see that QML still provides better results. This is actually due to our choice of a very large scale of apodization, which in practice reduces the effective available information, degrading the error of the spectra at higher multipoles. By choosing a smaller apodization scale, this ratio tends to unity at these multipoles although at the price of degrading the recovery of BB at large scales very significantly. This is shown in the left-bottom panel of figure~\ref{fig:RatiosNaMASTER} where an apodization scale of $4^\circ$ is instead used. We also note an increase in the estimated error for {\tt NaMaster} at the highest considered multipoles. We found that, at the limit of the resolution of the map, QML also performs better than pseudo-spectrum methods. However, in practice, this is not a real limitation of this technique in comparison to QML, since the pseudo-spectrum method can recover this range of multipoles from maps with higher resolution (where this effect will move to the highest considered resolution, that in any case will not usually be achieved by QML due to computational limitations).

\begin{figure}
    \centering
    \includegraphics[scale=.30]{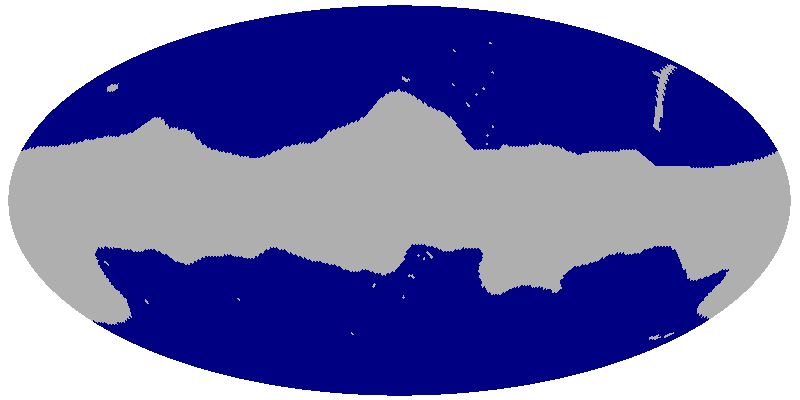}
    \caption{Extended mask that covers the Galactic mask from the top panel of Fig~\ref{fig:Mascaras} plus some additional regions outside the Galactic plane constructed at $\ns=64$. The number of valid pixels is 28824.}
    \label{fig:MascaraProducto}
\end{figure}

The mask that we have considered in this test is well suited for apodization, since it only presents one boundary between the included and excluded regions. However, masks can also exclude regions outside the Galactic plane that, when apodizing, will introduce a further loss of information for pseudo-spectrum methods. This is not the case for QML, where only the pixels discarded by the mask are removed from the analysis. To test this situation, we have repeated the previous exercise considering the mask given in figure~\ref{fig:MascaraProducto}, which is constructed by excluding additional regions outside the Galactic plane, that are present in the Planck common confidence mask for polarization  \cite{Planck_compsep18}. Note that this extended mask allows the use of 58.6 per cent of the sky versus 59.0 allowed by the original mask (left panel of figure~\ref{fig:Mascaras}). For this case, we found that the use of the B purification technique and the \emph{Smooth} apodization option\footnote{In this case all pixels closer than 2.5 times the apodization scale to a masked pixel are initially set to zero. The resulting mask is then smoothed with a Gaussian kernel with standard deviation given by the apodization scale. Finally, all pixels originally masked are put back to zero.} with a scale of $4.5^{\circ}$ were giving better results for the lowest multipoles of BB (again we remark that different configurations could be better suited for other purposes). The bottom panel of figure~\ref{fig:RatiosNaMASTER} shows the ratio between the errors obtained with {\tt NaMaster} versus those of QML. We see that the behavior is qualitatively similar to that found for the original mask (top panel) but the differences between both methods are amplified, confirming that a loss of a small fraction of the sky can degrade the performance of the pseudo-spectrum methods with respect to QML very significantly if the mask is not compact. Indeed, we find that for QML the errors of the estimated spectra increase only slightly (at the subpercent level for all multipoles) with respect to the original mask. However, for {\tt NaMaster}, the estimated errors for the case of the extended mask are between 1.4 to 2.0 times larger than those of the original mask (being the largest scales more affected). A practical way to improve these results for the pseudo-spectrum methods would be to perform some kind of inpainting in the data that allows the use of a more compact mask, although this incorporates an additional complication to the procedure, whose effect should be carefully quantified.

\section{Conclusions}
\label{sec:Conclusiones}

This work presents {\tt ECLIPSE}, a fast implementation of the Quadratic Maximum Likelihood (QML) method written in {\tt FORTRAN} for the estimation of the power spectrum, together with a thorough study of different aspects of the estimator,
including computational improvements,
a binned implementation, tests of robustness, an iterative approach and a comparison of the performance of QML with a pseudo-spectrum technique.

The QML provides an unbiased and minimum variance estimator of the intensity and polarization of the CMB power spectra provided a correct fiducial model is assumed. However, the method is very computationally demanding and, therefore, can be applied only at limited resolution. In this paper, we have developed a new implementation of the QML that reduces considerably the required computationally resources (CPU time and memory). In a direct implementation, one would need to calculate around $6 \times \ellm $ square matrices of $3 \times N_{\rm pix}$ size and, in addition, the product of each of them by the inverse of the covariance matrix. Instead, in our approach, the only operations which are computationally demanding are: the calculation of the covariance matrix and its inverse, the product of the inverse of the covariance matrix by the spherical harmonics matrix and the multiplication of the blocks of the resulting product  by blocks of the spherical harmonic matrix. This yields to a very significant reduction in the number of operations. For instance, we can compute the intensity and polarization power spectra of 1000 simulated maps of 39322 observed pixels at resolution $\ns = 64$ up to $\ellm = 191$ in 90 minutes using 144 cores at the Altamira supercomputer, with a reduction in the number of operations around a factor 1000 with respect to the direct implementation.

In addition, if one is only interested in polarization power spectra, we show that it is possible to use an only-polarization QML implementation reducing further the required computational resources (roughly a factor of 2 in CPU and memory requirements) while obtaining basically the same results as those of the full QML implementation (where intensity and polarization are simultaneously considered).

The QML estimator requires the inversion of the Fisher matrix. However, for small fractions of the sky, this matrix may become singular. To overcome this problem, we have developed a binned version of the QML that provides unbiased and optimal results for the estimation of the bandpowers, making use of the fiducial model. As an example, the method has been shown to perform very well for the configuration of a future ground-based experiment covering 8.4 per cent of the sky.

QML also requires the assumption of a fiducial model. Strictly speaking, the method is only unbiased and of minimum variance if this model corresponds to the underlying true model of the data, which is in general unknown. To overcome this shortcoming, the use of the QML within an iterative scheme that updates the fiducial model at each step has been tested for different scenarios. Our results show that, even when starting with a wrong fiducial model, the estimated power spectra is close to unbiased without iterating, although its error is somewhat larger than when starting with the correct value. Moreover, if we iterate, independently of the starting point, the errors are also consistent with those obtained when starting with the right fiducial model.
The same behavior is found for the estimation of the tensor-to-scalar ratio $r$ from the output power spectra (in a simplified case where all other cosmological parameters are assumed to be known), i.e., its estimation is close to unbiased independently of the assumed initial power spectra but its error can be reduced by iterating. In addition, we have also repeated similar tests for the binned version of the QML, finding that, in certain cases, this may be more sensitive to the assumed initial fiducial model than the standard QML. However, when iterating, both the binned and standard QML estimators provide very similar results.
Therefore, when applying the QML estimator to future CMB data is advisable to iterate at least a few steps in order to check the consistency between the assumed fiducial model and the one estimated with QML, in order to obtain optimal results. Note that the high efficiency of our code allows the use of this iterative approach that could not be easily carried out with previous algorithms due to the high required computational resources.

We have also compared the performance of the QML method with that provided by a pseudo-spectrum estimator (using the {\tt NaMaster} implementation). For the configuration of a future satellite experiment, we find that the errors of the estimated power spectra at low multipoles ($\ell \lesssim 20 $), which are critical for the detection of the tensor-to-scalar ratio, are significantly higher (up to a factor of 5 for a typical Galactic mask) for the pseudo-spectrum method. This method is also found to be much more sensitive than QML to the geometry of the mask, degrading considerably its performance when the mask is not compact, even if only a small fraction of the sky is removed. Also, the use of large scales of apodization, which are useful to improve the recovery of the BB spectra at low multipoles, increases the errors at the smallest scales considered, due to the loss of information. This also illustrates the fact that different tunings of {\tt NaMaster} are needed to obtain the best possible spectra for different components or scale ranges.

In summary, we believe that {\tt ECLIPSE} will be very useful for the community, allowing to reach higher multipoles and to carry out, in practice, analyses that before were prohibitively slow. The code will be made publicly available,\footnote{\href{https://github.com/CosmoTool/ECLIPSE}{https://github.com/CosmoTool/ECLIPSE}} including the possibility of calculating the six spectra, only temperature or only polarization. Other options, as the use of the binned version of QML or obtaining cross-power from two different maps will also be implemented.

\section*{Acknowledgements}
{The authors would like to thank Spanish Agencia Estatal de Investigaci\'on (AEI, MICIU) for the financial support provided under the projects with references PID2019-110610RB-C21, ESP2017-83921-C2-1-R and AYA2017-90675-REDC, co-funded with EU FEDER funds, and also acknowledge the funding from Unidad de Excelencia Mar{\'\i}a de Maeztu (MDM-2017-0765). DH acknowledges partial financial support from the Spanish Ministerio de Ciencia, Innovaci\'on y Universidades project PGC2018-101814-B-I00. We acknowledge J.A. Rubi{\~n}o-Mart{\'\i}n for kindly providing the mask used for the ground-based configuration.
We acknowledge Santander Supercomputacion support group at the University of Cantabria who provided access to the supercomputer Altamira Supercomputer at the Institute of Physics of Cantabria (IFCA-CSIC), member of the Spanish Supercomputing Network, for performing simulations/analyses. The {\tt HEALPix} package  \cite{healpix} was used throughout the data analysis.}

\vspace{0.5cm}


\appendix

\section{Computational approach}\label{ap:AspectosComputacionales}

In this section we describe the fast implementation of the QML method that we have developed for {\tt ECLIPSE}. As we will show, the key element of our approach is the connection between the pixel and harmonic space through matrix operations.

\subsection{Connection between pixel and harmonic spaces}

The expansion in spherical harmonics of the temperature and polarization of a CMB signal map reads \citep[see, for instance,][for details]{Zaldarriaga}
%
\begin{eqnarray}
  T(\hat{n})&=&\sum_{\ell m} a_{T,\ell m} Y_{\ell m}(\hat n), \nonumber \\
  Q(\hat{n})&=&\sum_{\ell m} -a_{E,\ell m} X_{1,\ell m}(\hat n) -i a_{B,\ell m}X_{2,\ell m}(\hat n), \nonumber \\
  U(\hat{n})&=&\sum_{\ell m} i a_{E,\ell m} X_{2,\ell m}(\hat n) - a_{B,\ell m} X_{1,\ell m}(\hat n),
  \label{ec:DefY0}
\end{eqnarray}
where $X_{1,\ell m}$ and $X_{2,\ell m}$ are a combination of the $s=\pm2$ spin-weighted harmonics
\begin{eqnarray}
  X_{1,\ell m}(\hat{n}) &=& (\;_2Y_{\ell m}+\;_{-2}Y_{\ell m})/2, \nonumber \\
  X_{2,\ell m}(\hat{n}) &=& (\;_2Y_{\ell m}-\;_{-2}Y_{\ell m})/2.
  \label{ec:spin}
\end{eqnarray}
Arranging the observed $N_\mathrm{pix}$ pixels of a CMB signal map in a signal vector in the pixel space $\vsp$ of length $3 N_\mathrm{pix}$, being the first $N_\mathrm{pix}$ elements of the vector the values of the map in intensity, the next $N_\mathrm{pix}$ elements the values of Q and finally the values of U, eq.~(\ref{ec:DefY0}) can be expressed in a matrix form
\begin{equation}
  \vsp = \mY \vsa,
  \label{ec:DefY}
\end{equation}
where the vector signal in the harmonic space $\vsa$ is the vector of elements $a_{T,\ell m}$, $a_{E,\ell m}$ and $a_{B,\ell m}$, with dimension 3L (where $L=\sum_{\ell=2,\ellm} (2 \ell +1$)$)$. The matrix $\mY$ has dimensions $3 N_\mathrm{pix} \times 3L$ and its elements are values of the spin-weighted $s=0$ and $s=\pm2$ spherical harmonics, conveniently allocated for the expression~(\ref{ec:DefY}) to be fulfilled. Indeed, the values of these elements can be easily identified by looking at eq.~(\ref{ec:DefY0}) and (\ref{ec:spin}). In particular, this matrix has the structure
\begin{equation}
\left(
\begin{array}{c}
T \\ Q \\ U \\
\end{array}
\right) =
\left(
\begin{array}{ccc}
\mY_{TT} & 0 & 0 \\
0 & \mY_{QE} & \mY_{QB} \\
0 & \mY_{UE} & \mY_{UB} \\
\end{array}
\right)
\left(
\begin{array}{c}
a_{lm}^{T} \\ a_{lm}^{E} \\ a_{lm}^{B} \\
\end{array}
\right).
\label{ec:EstruturaY}
\end{equation}
The covariance matrix of the signal map in pixel space $\mS$ is related to the covariance matrix in harmonic space $\mA$ as
\begin{equation}
  \mS = \mY \mA \mY^{\dag},
  \label{ec:RelacionMatricesC}
\end{equation}
where ${\dag}$ denotes the conjugate transpose of a matrix.
The matrix $\mA$ has dimension 3L $\times$ 3L and it is constituted by $3 \times 3$ diagonal blocks of length $L$ related to the auto and cross-spectra of the intensity and polarization
\begin{equation}
  \mA = \left( \begin{array}{ccc}
  \mA_{TT} & \mA_{TE}  & \mA_{TB}  \\
  \mA_{TE} & \mA_{EE}  & \mA_{EB}  \\
  \mA_{TB} & \mA_{EB}  & \mA_{BB}  \\
  \end{array}  \right).
\end{equation}
For instance, the TT block represents the correlations between all the ($\ell$,m) coefficients for intensity. The off-diagonal elements of each block are zero, since for an isotropic Gaussian field, as expected in the standard cosmological model, we have

\begin{equation}
\left< a_{\ell m} a^*_{\ell' m'} \right> =   C_{\ell}\delta_{\ell \ell'}\delta_{mm'}.
\end{equation}

Note also that since each block runs over all the values of ($\ell$,m), we have $2 \ell+1$ repetitions of the corresponding $C_{\ell}^{XY}$ up to a total of $L$ elements in the diagonal of each block.
In addition, in the standard cosmological model, we expect the TB and EB cross-correlations to vanish. Therefore, the corresponding four blocks of the $\mA$ matrix are also zero.

In section \ref{sec:implementation}, we describe the implementation of the estimator working with the $C_i$ variables. However, the method can also be easily implemented using instead the $D_i$ variables
and also include the beam and pixel window of the experiment by simply redefine the $\mY$ matrix. This is briefly described in section~\ref{sub:YVariablesDl}.

\subsection{The {\tt{ECLIPSE}} QML implementation}
\label{sec:ImplementacionQML}

As shown in section~\ref{sec:implementation}, our implementation of the QML method requires to compute $\mC$, its inverse, $\vx^t \mE_i \vx$, $b_i$ and the Fisher matrix $\mF_{i i'}$. We calculate these quantities sequentially as shown below.

\subsubsection{Computing \boldmath $\mC$ and \boldmath $\mC^{-1}$}

We recall that $\mC$ is the sum of the signal and noise covariance matrices, of dimension $3 N_\mathrm{pix} \times 3 N_\mathrm{pix}$. The signal covariance matrix $\mS$ can be efficiently computed using eq.~(\ref{ec:RelacionMatricesC}).
In particular, $\mS$ has the structure

\begin{equation}
\mS =
\left(
\begin{array}{ccc}
\mS_{TT} & \mS_{TQ} & \mS_{TU} \\
\mS_{QT} & \mS_{QQ} & \mS_{QU} \\
\mS_{UT} & \mS_{UQ} & \mS_{UU} \\
\end{array}
\right).
\label{ec:EstructuraC}
\end{equation}

From eq.~(\ref{ec:RelacionMatricesC}) and assuming $C_{\ell}^{TB}=C_{\ell}^{EB}=0$, six of the blocks are given by
\begin{eqnarray}
\nonumber \mS_{TT} & = &\mY_{TT} \mA_{TT}  \mY_{TT}^{\dag} \\
\nonumber \mS_{TQ} & = & \mY_{TT} \mA_{TE}  \mY_{EQ}^{\dag} \\
\nonumber \mS_{TU} & = &\mY_{TT} \mA_{TE}  \mY_{EU}^{\dag} \\
\nonumber \mS_{QQ} & = & \mY_{QE} \mA_{EE}  \mY_{EQ}^{\dag} + \mY_{QB} \mA_{BB}  \mY_{BQ}^{\dag} \\
\nonumber \mS_{QU} & = & \mY_{QE} \mA_{EE}  \mY_{EU}^{\dag} + \mY_{QB} \mA_{BB}  \mY_{BU}^{\dag} \\
\mS_{UU} & = & \mY_{UE} \mA_{EE}  \mY_{EU}^{\dag} + \mY_{UB} \mA_{BB}  \mY_{BU}^{\dag},
\label{ec:BloquesC}
\end{eqnarray}
where the matrices $\mA_{XY}$ are the diagonal blocks of matrix $\mA$, and the remaining blocks $\mS_{QT}$, $\mS_{UT}$ and $\mS_{UQ}$ are the transpose of their symmetric partners of $\mS$. There are some properties of the mathematical elements of eq.~(\ref{ec:BloquesC}) that allow to reduce the number of computations. Since the matrices $\mA_{XY}$ are diagonal, the products  $\mA_{XY} \mY_{ZY}^{\dag}$ can be computed quickly without resorting to matrix multiplication. In addition, given that the elements of the covariance matrix are real numbers, one can reduce the number of calculations of eq.~(\ref{ec:BloquesC}) by considering only those terms that produce real numbers (since the imaginary part will cancel). Once that the covariance matrix of the signal is computed, the covariance matrix of the noise has to be added.

The next step in the code is to compute the inverse of $\mC$. We accomplish this step using efficient routines for symmetric definite positive matrices of the ScaLAPACK library~\citep{slug}.

\subsubsection{Computing \boldmath $\vx^t \mE_i \vx$}

As mentioned in Section \ref{sec:efficiency}, we can obtain $\vx^t \mE_i \vx$ avoiding the direct calculation of the matrix $\mE_i$ by computing instead the product $\vx^t \mC^{-1} \mP_i \mC^{-1} \vx$. In particular, the matrices $\mP_i$ can be computed from eq.~(\ref{ec:RelacionMatricesC}) introducing a matrix basis $\mD_i$ such that
\begin{equation}
  \mA = \sum_{i} C_i \mD_i.
  \label{ec:DefDi}
\end{equation}
The $\mD_i$ are matrices of dimension $3L \times 3L$ with a structure of nine blocks (corresponding to the different combinations for the spectra of T,E,B) and constituted mostly by zeroes except for some selected elements of value 1 at the adequate positions. In particular, for an $i$ index corresponding to a case of auto-spectra and to a multipole $\ell$, we have $2\ell + 1$ non-null elements in the diagonal of the corresponding block, which are related to the m-elements of the considered multipole. For the case of cross-spectra, there are again $2\ell + 1$ non-null elements but in two of the blocks corresponding to the considered cross-spectrum.

To illustrate better the structure of these matrices, let us consider as a toy model a map with signal only at multipoles $\ell=0$ and $\ell=1$.\footnote{Although in a realistic case $\ell_{min}=2$ is usually assumed (in fact, the polarization signal is not even defined for $\ell < 2$), for the sake of simplicity, we will present some examples assuming that those multipoles actually exist in order to illustrate the calculations in a simple case. Whenever necessary, we will also indicate general results for the case $\ell_{min}=2$.} In this case, the index $i$ of eq.~(\ref{ec:DefDi}) runs from 1 to 12, corresponding to $\ell=0,1$ for each of the 6 possible power spectra (TT, EE, BB, TE, TB and EB). The ordering of the indices is such that i=(1, 2, 3, ..., 12) $\rightarrow$ ($\ell=0$, TT; $\ell=1$, TT; $\ell=0$, EE; ... ; $\ell=1$, EB).\footnote{To avoid confusion, we remark that, just by chance, in our simple example, the $\mD_i$ matrices have dimension $12\times 12$ (since this is fixed by the quantity $3L$ and we have L=4) and there are also 12 values of the $i$ index, which is given by the number of considered multipoles times the number of different power spectra (i.e., $2\times 6=12$). However, in a general case, these two numbers will be different.}
As an example the $\mD_i$ matrix for $i=8$, corresponding to $\ell=1$ and the TE cross-spectrum is given by

\begin{equation}
\mD_{8} =
\left(
\begin{array}{cccc|cccc|cccc}
 0 & 0 & 0 & 0 & 0 & 0 & 0 & 0 & 0 & 0 & 0 & 0 \\
 0 & 0 & 0 & 0 & 0 & 1 & 0 & 0 & 0 & 0 & 0 & 0 \\
 0 & 0 & 0 & 0 & 0 & 0 & 1 & 0 & 0 & 0 & 0 & 0 \\
 0 & 0 & 0 & 0 & 0 & 0 & 0 & 1 & 0 & 0 & 0 & 0 \\ \hline
 0 & 0 & 0 & 0 & 0 & 0 & 0 & 0 & 0 & 0 & 0 & 0 \\
 0 & 1 & 0 & 0 & 0 & 0 & 0 & 0 & 0 & 0 & 0 & 0 \\
 0 & 0 & 1 & 0 & 0 & 0 & 0 & 0 & 0 & 0 & 0 & 0 \\
 0 & 0 & 0 & 1 & 0 & 0 & 0 & 0 & 0 & 0 & 0 & 0 \\ \hline
 0 & 0 & 0 & 0 & 0 & 0 & 0 & 0 & 0 & 0 & 0 & 0 \\
 0 & 0 & 0 & 0 & 0 & 0 & 0 & 0 & 0 & 0 & 0 & 0 \\
 0 & 0 & 0 & 0 & 0 & 0 & 0 & 0 & 0 & 0 & 0 & 0 \\
 0 & 0 & 0 & 0 & 0 & 0 & 0 & 0 & 0 & 0 & 0 & 0 \\
\end{array}
\right).
\label{ec:MatrizPTE1}
\end{equation}

Equation~(\ref{ec:DefDi}) is the analogous in the harmonic space to $\mS = \sum_{i} C_i \mP_i$ in the pixel space. Combining eq.~(\ref{ec:DefDi}) and (\ref{ec:RelacionMatricesC}), we have

\begin{equation}
\mS = \mY \mA \mY^{\dag} = \mY \left( \sum_i C_i \mD_i \right) \mY^{\dag} = \sum_i C_i \mY  \mD_i \mY^{\dag}.
\label{ec:CalculoPi}
\end{equation}
Where we can identify
\begin{equation}
  \mP_i = \mY \mD_i \mY^{\dag}.
\label{ec:DefPi}
\end{equation}
Thus
\begin{equation}
  \vx^t \mC^{-1} \mP_i \mC^{-1} \vx = \vx^t \mC^{-1} \mY \mD_i \mY^{\dag} \mC^{-1} \vx.
  \label{ec:ProductoxEx}
\end{equation}
To get the value of the last expression, we compute first the vector  $\mY^{\dag} ( \mC^{-1} \vx)$.
The last operation transforms information in the pixel space to information in harmonic space. The next step is to collect it adequately taking into account the effect of multiplying
by the $\mD_i$ matrices.
When $\mD_i$ is one of the matrices associated to the cases TT, EE and BB, the vector product $\mD_{i}  \mY^{\dag} (\mC^{-1} \vx)$ is a vector of zeroes, except the values of $\mY^{\dag} \mC^{-1} \vx$ at the positions of the 1's in the diagonal of $\mD_i$. Therefore the product of eq.~(\ref{ec:ProductoxEx}) is directly the sum of the square of the real part plus the square of the imaginary part of the values of $ \mY^{\dag} (\mC^{-1} \vx)$ at the adequate positions, what allows for a fast computation of these elements.

When $\mD_i$ is one of the matrices associated to the cases TE, TB and EB, the vector product $\mD_{i}  \mY^{\dag} (\mC^{-1} \vx)$ is also a vector of zeroes, except for some values of $ \mY^{\dag} (\mC^{-1} \vx)$ moved to other positions (since the non-null values of $\mD_i$ are not in the diagonal in these cases). To illustrate this, let us consider again the toy model of a map with signal only at multipoles $\ell=0$ and $\ell=1$. In this case the vector has 12 elements grouped in three blocks of four elements each block; one of T, one of E and another of B. Written as a row vector
\begin{equation}
   \mY^{\dag} ( \mC^{-1} \vx) =  (T_{0},\,T_{1}, \,T_{2}, \,T_{3}, \,E_{0}, \,E_{1}, E_{2}, \,E_{3}, \,B_{0}, \,B_{1}, \,B_{2}, \,B_{3}),
\end{equation}
where the sub-indices code the pairs $(\ell, m)$: $(\ell=0, m=0) \rightarrow 0$, \ldots\, and $(\ell=1, m=1) \rightarrow 3$.
If we wanted to obtain the value of eq.~(\ref{ec:ProductoxEx}) for the cross-power TE and $\ell=1$, the product $\mD_{8} \mY^{\dag} ( \mC^{-1} \vx)$ would be
\begin{equation}
  \mD_{8}  \mY^{\dag} ( \mC^{-1} \vx) = (0,\, E_{1},\, E_{2},\, E_{3},\, 0,\, T_{1},\, T_{2}, T_{3},\, 0,\, 0,\, 0,\, 0)
\end{equation}
and, finally, to obtain the product of eq.~(\ref{ec:ProductoxEx}) for the value of the index $i=8$ all we have to do is
\begin{equation}
  \vx^t \mC^{-1} \mY \mD_8 \mY^{\dag} \mC^{-1} \vx =  \sum_{k=1}^{3} [  T_{k} E_{k}^* +  T_{k}^* E_{k}].
\end{equation}
This establishes a rule also for a fast computation of all the elements associated to the cross-power correlation and therefore the expression (\ref{ec:ProductoxEx}) can be efficiently computed.

The previous paragraphs show that by including the matrix $\mY$, we go from working with the matrices $\mP_i$, which implied carry out very demanding computations, to operate with some kind of matrices of selection $\mD_i$. What makes this approach more efficient is that, in practice, we do not need to calculate these matrices, and their action on vectors (on matrices in the next sections) is just to select (or select and move the position of) certain elements of the vector. Therefore, with our approach, we significantly simplify this part of the QML calculation.

Note that it is also possible to work with several maps at the same time. In this case, one just need to replace in $\mY^{\dag} \mC^{-1} \vx$ the vector $\vx$ by a matrix whose columns are the maps, transform the matrix-vector multiplications into matrix-matrix multiplications and adequately collect and combine the values of that matrix product.

\subsubsection{Computing \boldmath $\vb_i$}

The next step is to compute the noise contribution in harmonic space. To do this, we need to compute just one matrix multiplication, $\mC^{-1} \mY$, that will also be used to compute the Fisher matrix. Introducing $\mD_i$ in eq.~(\ref{ec:PotenciaRuido}), we get
\begin{equation}
  \vb_i =\frac{1}{2} \tr \left[ \mN \mC^{-1} \mY \mD_i \mY^{\dag} \mC^{-1} \right].
  \label{ec:DetalleBi}
\end{equation}
Let us show how to compute this quantity step by step.
The matrix product $\mC^{-1} \mY$ has as many rows as the vector map (i.e. 3$N_{\mathrm pix}$) and as many columns as the matrix $\mY$ (i.e. 3L). For instance, for our toy model with $\ell \in \{0, 1\}$ and only two pixels in the map, this is a matrix of six rows and twelve columns of structure
\begin{equation}
		\mC^{-1} \mY = \left(
		\begin{array}{ccccccccc}
			TT_{10} & TT_{11} & \dots & TB_{13} \\
			TT_{20} & TT_{21} & \dots & TB_{23} \\
			QT_{10} & QT_{11} & \dots & QB_{13} \\
			QT_{20} & QT_{21} & \dots & QB_{23} \\
			UT_{10} & UT_{11} & \dots & UB_{13} \\
			UT_{20} & UT_{21} & \dots & UB_{23} \\
		\end{array}
		\right),
		\label{ec:ProductoICY}
\end{equation}
where, for example, $QB_{23}$ means that this element corresponds to a row associated to the second pixel of the $Q$ component and to the $a_{X,\ell m}$ with $X=E$, $\ell=1$ and $m=1$.

In our model the noise is assumed to be isotropic and uncorrelated, thus the noise matrix $\mN$ is diagonal with values $\sigma_T^2$ and $\sigma_Q^2 = \sigma_U^2$. In this case, the matrix obtained by the operation $\mN\mC^{-1} \mY$ is simply the matrix given in eq.~(\ref{ec:ProductoICY}) with each $j$ row multiplied by the diagonal $j$ element of the noise matrix. Note that this is also valid if the noise is anisotropic and uncorrelated, since in this case the noise matrix is also diagonal.

In order to calculate the trace of eq.~(\ref{ec:DetalleBi}), let us recall that the trace of the product of two matrices $\maa$ and $\mB$ is
\begin{equation}
  \tr \maa \mB = \sum_{ij} \maa_{ij} \mB_{ji},
  \label{ec:TrazaProducto}
\end{equation}
i.e., we can obtain the trace of the product of a matrix without calculating the actual matrix multiplication. In this way, the trace can be written as
\begin{equation}
  \vb_i =\frac{1}{2} \sum_{\alpha \beta} (\mN \mC^{-1} \mY \mD_i)_{\alpha \beta}  (\mC^{-1} \mY)^*_{\alpha \beta}.
  \label{ec:DetalleBiTraza}
\end{equation}
Given that the matrices $\mD_i$ are sparse, the first matrix product in the previous equation has a large number of null elements what, as we will see, simplifies the calculation of $\vb_i$.

Let us show one example, again for our toy model with $\ell = \{0,1\}$ and two pixels, calculated for $i=2$ (what corresponds to $\ell=1$ and the component $TT$). In this case, the product $\mN \mC^{-1} \mY \mD_{2}$ has dimensions $6 \times 12$, and has the structure given by

\begin{equation}
  \mN \mC^{-1} \mY \mD_{2} =
  \left(
  \begin{array}{llllllllll}
		0 & \sigma_T^2 TT_{11} & \sigma_T^2 TT_{12} & \sigma_T^2 TT_{13} & 0 & \dots & 0 \vspace{1mm} \\
		0 & \sigma_T^2 TT_{21} & \sigma_T^2 TT_{22} & \sigma_T^2 TT_{23} & 0 & \dots & 0 \vspace{1mm} \\
		0 & \sigma_T^2 QT_{11} & \sigma_Q^2 QT_{12} & \sigma_Q^2 QT_{13} & 0 & \dots & 0 \vspace{1mm} \\
		0 & \sigma_T^2 QT_{21} & \sigma_Q^2 QT_{22} & \sigma_Q^2 QT_{23} & 0 & \dots & 0 \vspace{1mm} \\
		0 & \sigma_T^2 UT_{11} & \sigma_U^2 UT_{12} & \sigma_U^2 UT_{13} & 0 & \dots & 0 \vspace{1mm} \\
		0 & \sigma_T^2 UT_{21} & \sigma_U^2 UT_{22} & \sigma_U^2 UT_{23} & 0 & \dots & 0 \vspace{1mm} \\
  \end{array}
  \right),
  \label{ec:MatrizParaBi}
\end{equation}
where the first index indicates the pixel number and the second index runs over the different ${(\ell,m)}$ pairs. Therefore, this matrix has $2\ell+1$ columns with non-null elements. Taking this into account, to compute the required trace, we only need to multiply element by element, the second, third and fourth columns of the matrix of eq.~(\ref{ec:MatrizParaBi}) by the complex conjugates of the elements of the second, third and fourth columns of $\mC^{-1} \mY$, and then sum the results.

From the previous calculations, it becomes apparent that the noise bias of $y_{\ell}^{TT}$ depends explicitly not only on the temperature noise but also on the noise of the polarization components. This is not surprising because the $y_i$ quantities are a combination of the different power spectra, and not the power spectrum itself. So the noise bias is also a combination of the noise of the different components.

When computing the terms $b_i$ associated to the cross power components, the effect of multiplying by matrices $\mD_i$ is to select and reorder columns of $\mN \mC^{-1} \mY$. For example, for $i=7$ in our simple model, which corresponds to $\ell=0$ and the TE component, we have

\begin{equation}
  \mN \mC^{-1} \mY \mD_{7} =
  \left(
  \begin{array}{lllllllllllll}
    \sigma_T^2 TE_{10} & 0 & 0 & 0 & \sigma_T^2 TT_{10} & 0 & \dots & 0 \vspace{1mm} \\
    \sigma_T^2 TE_{20} & 0 & 0 & 0 & \sigma_T^2 TT_{20} & 0 & \dots & 0 \vspace{1mm} \\
    \sigma_Q^2 QE_{10} & 0 & 0 & 0 & \sigma_Q^2 QT_{10} & 0 & \dots & 0 \vspace{1mm} \\
    \sigma_Q^2 QE_{20} & 0 & 0 & 0 & \sigma_Q^2 QT_{20} & 0 & \dots & 0 \vspace{1mm} \\
    \sigma_U^2 UE_{10} & 0 & 0 & 0 & \sigma_U^2 UT_{10} & 0 & \dots & 0 \vspace{1mm} \\
    \sigma_U^2 UE_{20} & 0 & 0 & 0 & \sigma_U^2 UT_{20} & 0 & \dots & 0 \vspace{1mm} \\
  \end{array}
  \right).
  \label{ec:MatrizParaBi2}
\end{equation}
In this case, the number of columns with non-null elements is $2(2\ell+1)$.

Therefore, looking at eq.~(\ref{ec:ProductoICY})--(\ref{ec:MatrizParaBi}), we can infer that to compute an element $\vb_i$ corresponding to a given multipole $\ell$ and an auto spectra XX component, we have to multiply the elements of the appropriate $2 \ell + 1$ columns of the matrix of eq.~(\ref{ec:ProductoICY}) by their complex conjugates. The next step would be to multiply these terms by the appropriate noise variance but, since this quantity is the same for all the elements of a given row, it is more convenient to construct first a vector with one column, where each element is given by the sum of the $2\ell+1$ products corresponding to a fixed row. We then calculate the dot product of this vector by the vector constructed with the diagonal of the noise matrix $\mN$ and, finally, the result is divided by two. A similar procedure can be inferred to calculate $\vb_i$ for the cross-spectra components ($X \neq Y$) taking into account eq.~(\ref{ec:MatrizParaBi2}).

Thus, in this section we have shown that the only demanding operation needed to obtain $\vb_i$ is the matrix multiplication $\mC^{-1} \mY$. Its computational cost can be reduced by taking into account that, as shown in eq.~(\ref{ec:EstruturaY}), $\mY$ is a matrix constituted by two diagonal blocks. Thus a good strategy to calculate this product is to split the matrix $\mC^{-1}$ in four blocks that fit to the structure of $\mY$ and to compute four multiplications of matrices of smaller dimensions.

\subsubsection{Computing the Fisher matrix}

Writing eq.~(\ref{ec:MatrizFisher}) in terms of $\mY$ and the matrices $\mD$ and taking into account that the trace is invariant under cyclic permutations, we have
\begin{equation}
  \mF_{ii'} = \frac{1}{2} \tr \left[  \mY^{\dag} \mC^{-1} \mY  \mD_{i}  \mY^{\dag} \mC^{-1} \mY \mD_{i'} \right] .
  \label{ec:MatrizFisherDi}
\end{equation}
Since we have already computed $\mC^{-1} \mY$, the next step is to multiply this matrix by $\mY^{\dag}$, another strong consuming operation. The product $\mY^{\dag} \mC^{-1} \mY$ is an Hermitian matrix\footnote{Note that in practice this reduces the computational time, since only six out of the nine blocks needs to be computed.} structured on nine squared blocks of length $L$ of the form
\begin{eqnarray}
  \mY^{\dag} \mC^{-1} \mY =
  \left(
  \begin{array}{ccc}
    TT & TE & TB \\
    ET & EE & EB \\
    BT & BE & BB \\
  \end{array}
  \right).
  \label{ec:DefProductofinal}
\end{eqnarray}
Let us define a notation to refer to the elements of the last matrix.  For example, in the framework of our toy model, the element $TE_{02}$ is located in the $TE$ block in row number 1, corresponding to the harmonic of index 0 ($\ell=0$ and $m=0$) and in column number 3 of the block, corresponding to the harmonic of index 2 ($\ell=1$ and $m=0$). Note that this corresponds to the element in row number 1 and column number $L+3$ of the full matrix.

According to eq.~(\ref{ec:MatrizFisher}), each element of the Fisher matrix involves a pair of multipoles. Since the power spectrum is composed of the six modes TT, EE, BB, TE, TB and EB, $\mF$ is a symmetric matrix of $6^2 (\ellm -1)^2$ elements.\footnote{For this number we are assuming $\ell_{min} = 2$ and, therefore, we have $(\ellm -1)$ different multipoles.} For convenience, let us arrange them in a matrix conformed by 36 squared blocks of $(\ellm-1)^2$ elements each one
\begin{eqnarray}
  \mF =
  \left(
  \begin{array}{cccccc}
    TTTT  & TTEE & TTBB & TTTE & TTTB & TTEB \\
    EETT  & EEEE & EEBB & EETE & EETB & EEEB \\
    BBTT  & BBEE & BBBB & BBTE & BBTB & BBEB \\
    TETT  & TEEE & TEBB & TETE & TETB & TEEB \\
    TBTT  & TBEE & TBBB & TBTE & TBTB & TBEB \\
    EBTT  & EBEE & EBBB & EBTE & EBTB & EBEB \\
  \end{array}
  \right),
  \label{ec:EstructuraMF}
\end{eqnarray}
where, for example, the block TTTE contains the cross terms between the multipoles $C_{\ell}^{TT}$ and $C_{\ell'}^{TE}$.

Taking into account eq.~(\ref{ec:TrazaProducto}), we can obtain the Fisher matrix without actually calculating the direct matrix multiplication of eq.~(\ref{ec:MatrizFisherDi}). Moreover, once the product $\mY^{\dag} \mC^{-1} \mY$ is computed, to obtain a given element of the Fisher matrix $\mF_{ii'}$, we just have to localise the values and positions of the non null elements of $ \mY^{\dag} \mC^{-1} \mY \mD_i$ and of the transpose of $  \mY^{\dag} \mC^{-1} \mY \mD_{i'}$ (characterised by the positions of the 1's in $\mD_i$ and $\mD_{i'}$), multiply element by element those pairs formed by two non null elements and sum the result. In the following, we will show in detail how this technique leads to a fast method to compute $\mF$.

Let us show the procedure by showing how to calculate the elements of the first block (TTTT) of the Fisher matrix. For example, let us consider the element $\mF_{i=1,i'=2}$ corresponding to $C_1 = C^{TT}_0$ and $C_{2}=C^{TT}_1$  of eq.~(\ref{ec:MatrizFisher}). In this case, the matrix $ \mY^{\dag} \mC^{-1} \mY \mD^{TT}_0$, that we will call M0, reads
\begin{eqnarray}
  \mathrm{M0} \equiv \left(
  \begin{array}{cccccccccccc}
   TT_{00} 	& 0 	 	& \dots 	& 0 \\
   TT_{10}  & 0		 	& \dots		& 0 \\
   TT_{20}  & 0		 	& \dots		& 0 \\
   TT_{30}  & 0		 	& \dots		& 0 \\
   ET_{00}  & 0		 	& \dots		& 0 \\
   \vdots	& \vdots	& \ddots	& \vdots \\
   BT_{30}	& 0 		& \dots		& 0 \\
  \end{array}
  \right).
  \label{ec:MP1}
\end{eqnarray}
Analogously, the product $ \mY^{\dag} \mC^{-1} \mY \mD^{TT}_1 $ is given by
\begin{eqnarray}
  \mathrm{M1} \equiv \left(
  \begin{array}{cccccccccccc}
0 &   TT_{01} 	& TT_{02} 	& TT_{03} 	& 0 	& 0 	& \dots		& 0 \\
0 &   TT_{11} 	& TT_{12} 	& TT_{13} 	& 0 	& 0 	& \dots 	& 0 \\
0 &   TT_{21} 	& TT_{22} 	& TT_{23} 	& 0 	& 0 	& \dots 	& 0 \\
0 &   TT_{31} 	& TT_{32} 	& TT_{33} 	& 0 	& 0 	& \dots 	& 0 \\
0 &   ET_{01} 	& ET_{02} 	& ET_{03} 	& 0 	& 0 	& \dots 	& 0 \\
0 &   \vdots	& \vdots	& \vdots	& \vdots	& \vdots	& \ddots	& \vdots \\
0 &   BT_{31} 	& BT_{32} 	& BT_{33} 	& 0 	& 0 	& \dots 	& 0 \\
  \end{array}
  \right).
  \label{ec:MP2}
\end{eqnarray}
The next step is to calculate the matrix whose elements are the multiplication of the elements of M0 and the transpose of M1
\begin{eqnarray}
  \mathrm{R01} \equiv \left(
  \begin{array}{cccccccccccc}
    0                   & 0 &  \dots    & 0 \\
    TT_{10} TT_{01} 	& 0 &  \dots 	& 0 \\
    TT_{20} TT_{02} 	& 0 &  \dots 	& 0 \\
    TT_{30} TT_{03} 	& 0 &  \dots 	& 0 \\
    0                   & 0 &  \dots    & 0 \\
    \vdots              & \vdots & \ddots & \vdots \\

    0                   & 0 & \dots         & 0 \\
  \end{array}
  \right).
  \label{ec:FinalMF1}
\end{eqnarray}

The sum of the elements of R01 is twice the element $\mF_{12}$ of the Fisher matrix.

A more complete analysis shows that, in fact, it is possible to compute, almost simultaneously, all the elements of each one of the 36 blocks of eq.~(\ref{ec:EstructuraMF}) by means of adequate sums of the product of elements of certain blocks of $\mY^{\dag} \mC^{-1} \mY$ by elements of the transpose of certain blocks of the matrix $\mY^{\dag} \mC^{-1} \mY$. To illustrate this, let us show the matrices analogous to R01 of eq.~(\ref{ec:FinalMF1}) that emerge when computing with this technique another elements of the block TTTT of the Fisher matrix. For example, for $\ell = 0$ and $\ell' = 0$, R00 reads
\begin{eqnarray}
  \mathrm{R00} = \left(
  \begin{array}{cccccccccccc}
    TT_{00} TT_{00} & 0         & \dots     & 0 \\
    0               & 0         & \dots     & 0 \\
    \vdots          & \vdots    & \ddots    & 0 \\
    0               & 0         & \dots     & 0 \\
  \end{array}
  \right),
  \label{ec:FinalMF2}
\end{eqnarray}
while for $\ell = 1$ and $\ell' = 0$, we have
\begin{eqnarray}
  \mathrm{R10} \equiv \left(
  \begin{array}{cccccccccccc}
    0 & TT_{01} TT_{10} 	& TT_{02} TT_{20} & TT_{03} TT_{30} & 0 		& \dots 	& 0 \\
    0 & 0                   & 0                 &  0            & 0 & \dots & 0 \\
   \vdots					& \vdots		& \vdots		& \vdots	& \vdots	& \ddots & 0 \\
    0 & 0 					& 0 			& 0 			& 0 & \dots & 0 \\
  \end{array}
  \right),
  \label{ec:FinalMF3}
\end{eqnarray}
and for $\ell = 1$ and $\ell' = 1$, R11 is
\begin{eqnarray}
  \mathrm{R11} = \left(
  \begin{array}{cccccccccccc}
    0 & 0 & 0 & 0 & 0 & \dots \\
    0 & TT_{11} TT_{11} & TT_{12} TT_{21} & TT_{13} TT_{31} & 0 & \dots \\
    0 & TT_{21} TT_{12} & TT_{22} TT_{22} & TT_{23} TT_{32} & 0 & \dots \\
    0 & TT_{31} TT_{13} & TT_{32} TT_{23} & TT_{33} TT_{33} & 0 & \dots \\
    0 & 0 & 0 & 0 & 0 & \dots \\
    \vdots & \vdots & \vdots & \vdots & \vdots & \ddots \\
  \end{array}
  \right).
  \label{ec:FinalMF4}
\end{eqnarray}

In each case, the different elements of the first block (TTTT) of the Fisher matrix can be computed as the sum of the elements of the corresponding $\mathrm{R}$ matrices, divided by two. Moreover, one can infer that these elements of the Fisher matrix can also be computed by calculating the product, element by element, of the block TT of $\mY^{\dag} \mC^{-1} \mY$ times its transpose (which would give rise to a dense matrix filled by the non-null elements of eq.~(\ref{ec:FinalMF1} -- \ref{ec:FinalMF4})) and then sum over the appropriate elements, selected taking into account the value of $\ell$ and $\ell'$, divided by two.
That is, the element of the Fisher matrix associated to $C_{\ell}^{TT}$ and $C_{\ell'}^{TT}$ is given by
\begin{equation}
    \mF^{TTTT}_{\ell \ell'} = \frac{1}{2}  \sum_{k_{\ell} k_{\ell'}} TT_{k_{\ell'} k_{\ell}} TT_{k_{\ell} k_{\ell'}},
    \label{ec:DetalleFisherTTTT}
\end{equation}
where $TT$ on the right hand side refers to the first block of $\mY^{\dag} \mC^{-1} \mY$, the index $k_{\ell}$ runs over $2 \ell +1$ values and $k_{\ell'}$ over $2 \ell' + 1$. If we denote $k$ as the index that runs from 0 to $L-1$ (L is the size of the TT block in eq.~(\ref{ec:DefProductofinal})), for $\ell_{min}=2$, we have that $k=0$ corresponds to the pair $(\ell,m)=(2,-2)$. It can be shown that, for a given $\ell$, $k_{\ell}$ runs from $\ell^2-4$ to $2 \ell + \ell^2 -4$, what defines the elements of the block that must be selected to construct the corresponding Fisher element.

For convenience, let us define the operator $\{\, \}$ to symbolize the computation of the block TTTT of the Fisher matrix eq.~(\ref{ec:EstructuraMF}) for all the pairs $(\ell, \ell')$ using the expression~(\ref{ec:DetalleFisherTTTT}) as
\begin{eqnarray}
  \mathbf{TTTT} & = & \{TT,TT\},
  \label{ec:FinMF1}
\end{eqnarray}
where the first block TT in $\{TT, TT\}$ enters as it is and the second one is transposed.

To compute the elements of the rest of the blocks of $\mF$, eq.~(\ref{ec:EstructuraMF}), one has to apply the same technique used for the case TTTT but varying the matrices $\mD_i$. Following a similar procedure as before, one can infer that there are three different cases

\begin{enumerate}
  \item The blocks of $\mF$ where the two considered power spectra correspond to autocorrelations (TT, EE, BB), that is, the blocks TTTT, TTEE, TTBB, EEEE, EEBB, BBBB.
  \item The blocks of $\mF$ that mix one autocorrelation and one cross-correlation, i.e., TTTE, TTTB, TTEB, EETE, EETB, EEEB, BBTE, BBTE, BBBE.
  \item The blocks that involve two cross-correlations, i.e., TETE, TETB, TEEB, TBTB, TBEB, EBEB.
\end{enumerate}

For instance, if one derives the equations equivalent to (\ref{ec:MP1}) - (\ref{ec:DetalleFisherTTTT}) for the block TTEE (case i), it is found that one has to multiply element by element the block ET of eq.~(\ref{ec:DefProductofinal}) and the transpose of the block TE, sum the sub blocks adequately, and divide by two. That is
\begin{eqnarray}
  \mathbf{TTEE} & = & \{ET, TE\}.
  \label{ec:FinMF2}
\end{eqnarray}
The rest of the blocks of the case (i) are obtained as
\begin{eqnarray}
  \mathbf{TTBB} & = & \{BT, TB\} \nonumber \\
  \mathbf{EEEE} & = & \{EE, EE\} \nonumber \\
  \mathbf{EEBB} & = & \{BE, EB\} \nonumber \\
  \mathbf{BBBB} & = & \{BB, BB\}.
  \label{ec:FinMF3}
\end{eqnarray}

Let us show an example to find the rule to compute the blocks of the case (ii), in particular, we will consider the element of $\ell=0$ and $\ell'=1$ of the block TTTE of our toy model, that is, the element $\mF_{ii'}$ of $C_i = C^{TT}_0$ and $C_{i'}=C^{TE}_1$  of eq.~(\ref{ec:MatrizFisher}). In this case, one has to compute $ \mY^{\dag} \mC^{-1} \mY \mD^{TT}_0$  and multiply element by element by the transpose of $\mY^{\dag} \mC^{-1} \mY \mD^{TE}_1$. Given that the $\mD^{TE}_1$ matrix has a total of six elements with the value of 1, grouped in two sets of three, outside the main diagonal (see eq.~(\ref{ec:MatrizPTE1})),\footnote{Note that in the notation used in this subsection $\mD^{TE}_1 \equiv \mD_8$} this gives rise to two blocks of three non-null elements in $\mathrm{R01}$.

In particular, the result reads
\begin{eqnarray}
  \mathrm{R01} = \left(
  \begin{array}{ccccccc}
   0 &               0 & \dots & 0 \\
   TT_{10} TE_{01} & 0 & \dots & 0 \\
   TT_{20} TE_{02} & 0 & \dots & 0 \\
   TT_{30} TE_{03} & 0 & \dots & 0 \\
   0               & 0 & \dots & 0 \\
   0               & 0 & \dots & 0 \\
   0               & 0 & \dots & 0 \\
   ET_{10} TT_{01} & 0 & \dots & 0 \\
   ET_{20} TT_{02} & 0 & \dots & 0 \\
   ET_{30} TT_{03} & 0 & \dots & 0 \\
   0               & 0 & \dots & 0 \\
  \vdots & \vdots & \ddots &  0 \\
  0 & 0           & \dots & 0 \\
  \end{array}
  \right).
  \label{ec:FinalMF1X}
\end{eqnarray}
Therefore, following the same reasoning as in previous cases, one can write
\begin{eqnarray}
  \mathbf{TTTE} & = &\{TT, TE\} + \{ET, TT\}.
  \label{ec:CasoIIA}
\end{eqnarray}
The imaginary part of $\{TT, TE\} + \{ET,TT\}$ is zero and the real part of $\{TT, TE\}$ is equal to the real part of $\{ET, TT\}$, therefore
\begin{eqnarray}
  \mathbf{TTTE} & = & 2 \{TT, TE\}.
  \label{ec:FinMFTTTE}
\end{eqnarray}
For the sake of clarity, let us write explicitly how to calculate an element of the Fisher matrix in the block TTTE
\begin{equation}
    \mF^{TTTE}_{\ell \ell'} =  \frac{1}{2}  \times 2 \sum_{k_{\ell} k_{\ell'}} TT_{k_{\ell'} k_{\ell}} TE_{k_{\ell} k_{\ell'}}.
    \label{ec:DetalleFisherTTTE}
\end{equation}
Applying the technique to the rest of the blocks of the Fisher matrix of case (ii), we get
\begin{equation}
\begin{array}{lcccccc}
  \mathbf{TTTB} & = & \{TT, TB\} & + & \{BT, TT\}  & = & 2 \{TT, TB\}  \\
  \mathbf{TTEB} & = & \{ET, TB\} & + & \{BT, TE\}  & = & 2 \{ET, TB\} \\
  \mathbf{EETE} & = & \{TE, EE\} & + & \{EE, ET\}  & = & 2 \{TE, EE\}  \\
  \mathbf{EETB} & = & \{TE, EB\} & + & \{BE, ET\}  & = & 2 \{TE, EB\}  \\
  \mathbf{EEEB} & = & \{EE, EB\} & + & \{BE, EE\}  & = & 2 \{EE, EB\}  \\
  \mathbf{BBTE} & = & \{TB, BE\} & + & \{EB, BT\}  & = & 2 \{TB, BE\}  \\
  \mathbf{BBTB} & = & \{TB, BB\} & + & \{BB, BT\}  & = & 2 \{TB, BB\}  \\
  \mathbf{BBEB} & = & \{EB, BB\} & + & \{BB, BE\}  & = & 2  \{EB, BB\}. \\
\end{array}
\label{ec:FinMF5}
\end{equation}

Finally, for the blocks corresponding to case (iii), we find
\begin{equation}
\begin{array}{lcccccccccccc}
	\mathbf{TETE} & = & \{TE, TE\} & + & \{TT, EE\}  & + & \{EE, TT\} & + & \{ET, ET\} \\ & = & 2  \{TE, TE\} & + & 2 \{TT, EE\} \\
	\mathbf{TETB} & = & \{TE, TB\} & + & \{TT, EB\}  & + & \{BE, TT\} & + & \{BT, ET\} \\ & = & 2  \{TE, TB\} & + & 2 \{TT, EB\} \\
	\mathbf{TEEB} & = & \{EE, TB\} & + & \{ET, EB\} & + & \{BE, TE\} & + & \{BT, EE\} \\ & = & 2  \{EE, TB\} & + & 2 \{ET, EB\} \\
	\mathbf{TBTB} & = & \{TB, TB\} & + & \{TT, BB\} & + & \{BB, TT\} & + & \{BT, BT\} \\ & = & 2  \{TB, TB\} & + & 2 \{TT, BB\} \\
	\mathbf{TBEB} & = & \{EB, TB\} & + & \{ET, BB\} & + & \{BB, TE\} & + & \{BT, BE\} \\ &  = & 2  \{EB, TB\} & + & 2 \{ET, BB\} \\
	\mathbf{EBEB} & = & \{EB, EB\} & + & \{EE, BB\}  & + & \{BB, EE\} & + & \{BE, BE\} \\
& = & 2  \{EB, EB\} & + & 2 \{EE, BB\}. \\
\end{array}
  \label{ec:FinMF6}
\end{equation}

\subsection{Working with variables $D_i$, beam and pixel window}
\label{sub:YVariablesDl}

To implement the QML in terms of the variables $D_i$ and to take into account the effect of the beam of the experiment and the pixel window, one just needs to define adequately the matrix $\mY$.

Writing an element of the matrix $\mS$ according to the expression~(\ref{ec:RelacionMatricesC}), we have
\begin{equation}
    \mS_{ij} = \sum_{kk'} \mY_{ik} \mA_{kk'} \mY^{\dag}_{k'j} = \sum_{kk'} \mY_{ik} C_{kk'} \mY^{\dag}_{k'j},
\end{equation}
where, taking into account the structure of $\mA$, $C_{kk'}$ is some of the  $C_\ell^{TT}$, $C_\ell^{EE}$, $C_\ell^{BB}$, when $k=k'$, while for $k \neq k'$ corresponds to $C_\ell^{TE}$ in some cases or to zero otherwise.

Introducing the beam and pixel window functions through  $B_k$ and the $D_{k k'}$ variables instead of $C_{k k'}$, we have
\begin{equation}
    \mS_{ij} =  \sum_{kk'} \mY_{ik} \frac{2 \pi B_k B_{k'}} {\ell (\ell +1)} D_{kk'} \mY^{\dag}_{k'j}.
\end{equation}
We can rewrite the previous equation as
\begin{equation}
    \mS_{ij} =   \sum_{kk'} \left( \sqrt{\frac{2 \pi}{\ell (\ell +1)}} B_k \mY_{ik} \right)  D_{kk'} \left( \sqrt{ \frac{2 \pi }{\ell (\ell +1)}} B_{k'} \mY^{\dag}_{k'j} \right).
\end{equation}
From the last expression we can define a new matrix $\check{\mY}$ that satisfies
\begin{equation}
  \mS_{ij} = \sum_{kk'}  \check{\mY}_{ik} D_{kk'} \check{\mY}^{\dag}_{k'j}.
\end{equation}
That is
\begin{equation}
  \check{\mY}_{ik} = \sqrt{\frac{2 \pi}{\ell (\ell +1)}} B_k \mY_{ik}.
\end{equation}
Hence the columns of $\check{\mY}$ are just the columns of $\mY$ multiplied by the appropriate factor  $\sqrt{\frac{2 \pi}{\ell (\ell +1)}} B_k$.

\subsection{About the parallelized implementation}

Our {\tt ECLIPSE} code is parallelized, and the matrix operations, such as inversion and multiplication, are computed by subroutines of the libraries BLACS and ScaLAPACK. In this context, the matrices are block-cyclically distributed through the grid of processors. The most efficient implementation that we have found is oriented to minimize the interchange of blocks of matrices between processors. For example, to compute the block TTTT of the Fisher matrix, each processor calculates, for all the $\ell, \ell'$, the sum of the terms of eq.~(\ref{ec:DetalleFisherTTTT}) to which it has direct access in its memory, i.e., it obtains a partial sum for each $\mF^{TTTT}_{\ell \ell'}$. These partial results of each processor are then added together to obtain the final value of each element of the block.


\section{Effect of sampling variance on the power spectrum error}\label{ap:fsky}
\label{sec:fsky}

It is well known that, due to sampling variance, the error of the estimated power spectrum increases as the observed sky fraction ($f_{\rm sky}$) is reduced. Its effect can be roughly approximated by an increment in the variance by a factor $1/f_{\rm sky}$  \cite{Sco94}. However, it is also well known that, in practice, the specific geometry of the considered mask will also affect this error.
Although for our purpose (constructing the optimal binning) this approximation is sufficient, it is still interesting to perform some simple tests in order to understand the limitations of this assumption.

\begin{figure}
  \begin{center}
   \includegraphics[scale=1]{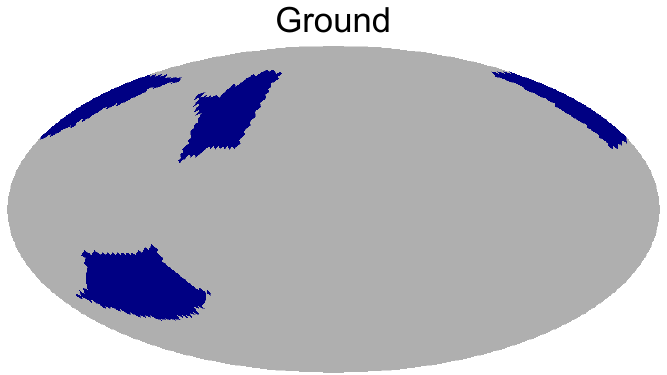}
  	\includegraphics[scale=1]{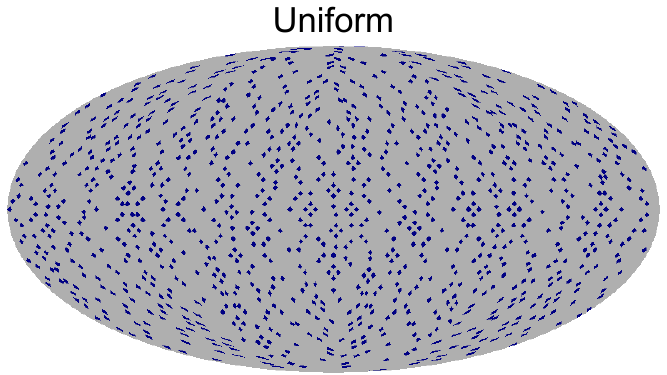}
  		\includegraphics[scale=1]{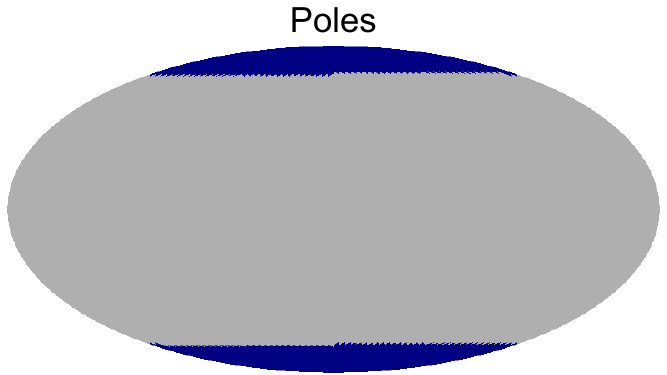}
  		\includegraphics[scale=1]{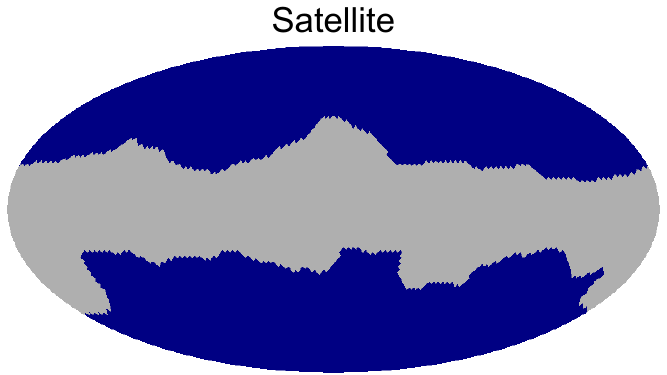}
    \caption{Masks used to test how the error of the estimated power spectra depends on the geometry and fraction of the observed sky. All masks are shown at $\ns=32$. Top-left: ground-based mask. Top-right: toy-model mask  with pixels distributed uniformly over the sky. Bottom-left: toy-model mask with the observed pixels distributed on the poles. Bottom-right: mask for a satellite-based experiment. The first three masks have a total of 1028 valid pixels, corresponding to a sky fraction of around 8.4 per cent, while $f_\mathrm{Sky}$=0.58 for the last mask. Note that the ground and satellite masks correspond to those of figure~\ref{fig:Mascaras}, but at a resolution of $\ns=32$, and are repeated here for easiness of reading.}
    \label{fig:MascaraAlterna}
  \end{center}
\end{figure}

As an illustration, we have studied how the errors degrade with the considered sky fraction in different scenarios, in particular for the four masks shown in figure~\ref{fig:MascaraAlterna}.
The errors have been obtained from the Fisher matrix. A resolution of $\ns=32$
and a binning of $\Delta_\ell = 3$ have been considered in all the cases. Noise has also been included according to the ground-based or space configurations levels, as it corresponds.
Three of the considered masks correspond to three
different geometries but with the same sky coverage as that of the ground-based scenario, that allows for an observed sky fraction of around 8.4 per cent (1028 pixels at $\ns=32$ resolution). Specifically, the first mask (top left panel of figure~\ref{fig:MascaraAlterna}) corresponds to that of the ground-based experiment used along the paper. A second toy-model mask contains the observed pixels distributed uniformly over the sky (top right panel of figure~\ref{fig:MascaraAlterna}), and we will refer to it as the uniform mask. A third mask with pixels distributed in the poles (bottom left panel of figure~\ref{fig:MascaraAlterna}) has also been added, referred to as poles mask; although this mask is not expected for a ground-based experiment, it allows us to check the geometrical effect of an (extreme) Galactic mask and to compare it with the other configurations. Finally, we have considered the case of the satellite-based experiment (bottom right panel of figure~\ref{fig:MascaraAlterna}) at a resolution of $\ns=32$
($f_\mathrm{sky}$=0.58).

\begin{figure}
\begin{center}
    \includegraphics[scale=1]{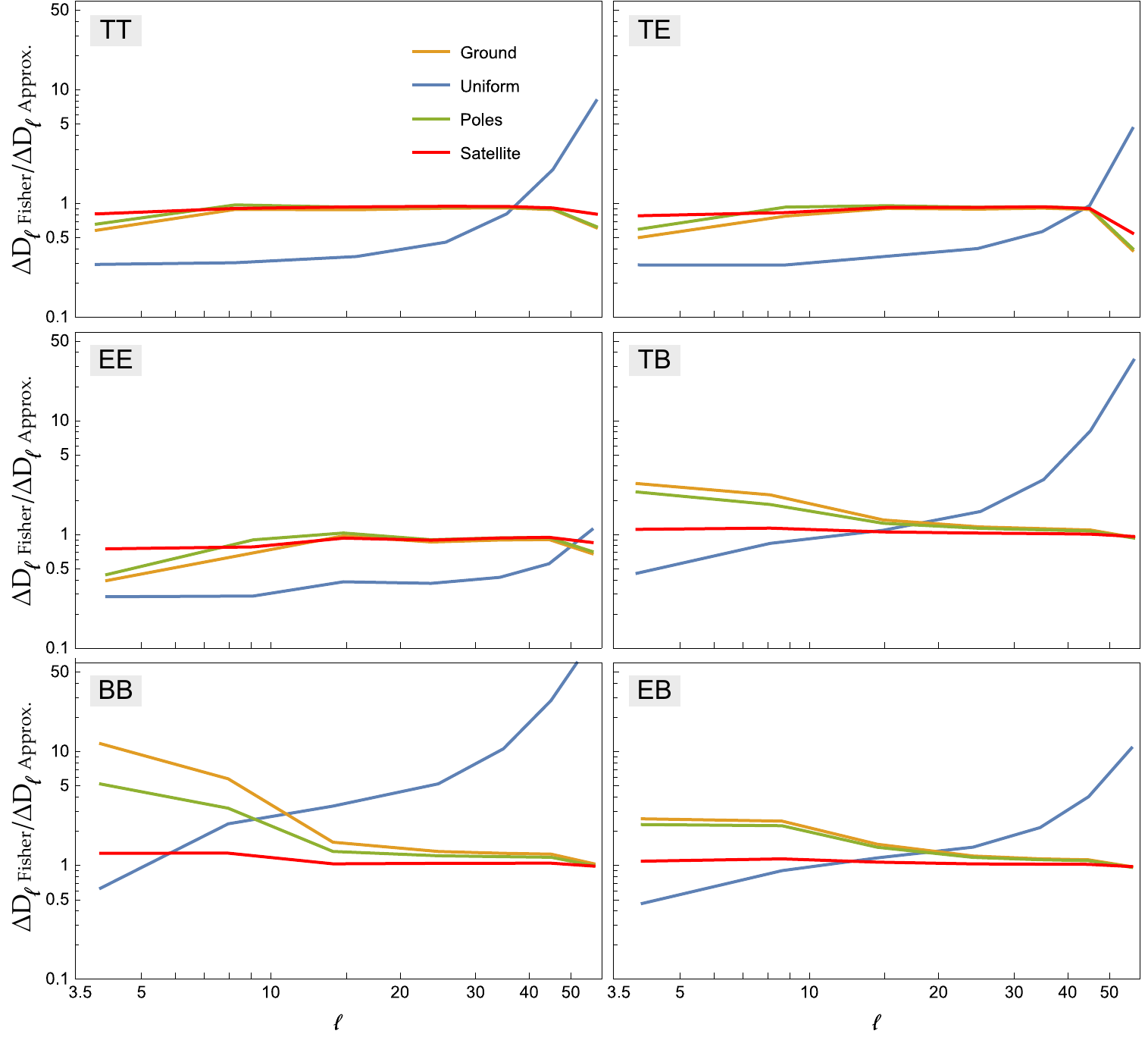}
    \caption{Ratio between the error of the estimated power spectrum obtained with QML from the Fisher matrix, $\Delta D_{\ell\, \mathrm{Fisher}}$, over that of the theoretical approximation~\cite{Eis99}, $\Delta D_{\ell \, \mathrm{Approx.}}$, for the different masks shown in figure~\ref{fig:MascaraAlterna}.}
    \label{fig:ErroresSegunDistribucionPuntos}
\end{center}
\end{figure}

Figure~\ref{fig:ErroresSegunDistribucionPuntos} shows the ratio of the error on the estimation of the power spectrum for the four masks using the Fisher matrix over the theoretical approximation (according to eqs.~(4-11) from \cite{Eis99}). Therefore values close to one indicate that the theoretical expression is a good approximation, whereas larger (smaller) values indicate an under- (over-) estimation of the error. As seen, for the uniform case (blue line), the estimated errors are smaller than those given by the theoretical expectation at large scales, while they explode at larger multipoles. This is expected since to estimate the power spectra at the largest scales, it is only necessary to provide data on a limited number of pixels but conveniently distributed, while this configuration can not provide the required information at small scales. It is also worth noting that the range of scales where this mask provides better results than the theoretical ones also varies significantly with the considered component of the power spectrum, working worse especially for BB. The ground (orange line) and poles (green line) masks have a qualitatively similar behavior. For TT, EE and TE, the estimated error is better than the theoretical expectation at low multipoles, while at small scales the error is underestimated by the naive $f_{sky}$ scaling. The minimum observed at TE at the smallest scales is due to the fact that the TE spectrum becomes zero at that multipole.
For BB, TB and EB, the error estimated by the Fisher matrix is larger than the approximation at all scales, with a particular large deviation for the large scales of BB. The theoretical approximation is working better for the Galactic mask in the space configuration (red line), finding only relatively small deviations between both estimations.

Therefore, if a precise estimation of the error introduced by the sky fraction is needed, it is advisable to carry out a full analysis that takes into account the effect of the geometry of the mask, especially when a small fraction of the sky is considered, as is the case in most ground-based experiments.

\section{Smoothing function for the iterative QML}
\label{ap:Suavizado}

Although the power spectra given by a physical model will follow a smooth curve, a particular realization of the spectra and its corresponding estimation will in general be noisy. In fact, if we take directly the power spectra estimated with QML as the fiducial model for a subsequent iteration, this process may be unstable and does not necessarily lead to convergence. Therefore, it becomes necessary to apply some kind of smoothing to the estimated power spectra in order to provide a suitable fiducial model to the next iteration.
An obvious choice would be to fit the estimated spectra to a cosmological model, and use that as our initial guess for the next step. This has been the approach used in section~\ref{sec:robust_r}, where we have considered a simplified case where all cosmological parameters are assumed to be known except the tensor-to-scalar ratio r. However, in a more general case, it could be computationally very costly to fit all the cosmological parameters and a simpler approach would be more convenient.
With this aim we have constructed a smoothed version $q_{\ell}$ of each component of the estimated power spectra $D_{\ell}$ by minimizing the following function:
\begin{equation}
    \label{Sua:Suavizado}
    \delta = \sum_{\ell=2}^{\ell_\mathrm{max}-1} L_{\ell} M_{\ell} \, ,
\end{equation}
where $L_{\ell}$ and $M_{\ell}$ are given by
\begin{eqnarray}
    \label{Sua:FlMl}
    L_{\ell} &=& \frac{(q_{\ell+1}-q_\ell)^2}{\Delta D_{\ell+1}^2 + \Delta D_{\ell}^2} + 1 \, , \\
     M_{\ell} &= &w_{\ell} + \frac{1}{2}\left[\frac{\left(q_{\ell+1} - D_{\ell+1}\right)^2}{\Delta D_{\ell+1}^2} + \frac{\left(q_{\ell} - D_{\ell}\right)^2}{\Delta D_{\ell}^2}\right] \, ,
\end{eqnarray}
and $\Delta D_{\ell}$ is the error associated to the $D_{\ell}$ estimation.
The aim behind this approach is to minimize the sum of the length of the segments that join a pair of consecutive points (controlled by the $L_{\ell}$ term) but penalizing large differences between the smoothed and estimated values of the power spectra (encoded in the $M_{\ell}$ factor).
Note that the +1 term in the first factor avoids that the minimization defaults to a constant straight line (i.e. $q_\ell$ = constant). $w_{\ell}$ are weights that control the relative importance between both effects, such that larger values of the weights would tend to produce smoother curves and vice versa, allowing us to modify the level of required smoothing. In particular $w_{\ell}=0$ would lead to the solution $q_\ell = D_\ell$, i.e., no smoothing. Note that a particular choice of the weights could still leave to an unstable iterative process for a given map in which case the smoothing process to provide a fiducial model for the next step would need to be repeated with different weights, such that the required number of iterations could be achieved. For simplicity, in our case we have simply discarded from the analysis those simulations that do not converge.

\begin{figure}
  \begin{center}
  	\includegraphics[scale=.8]{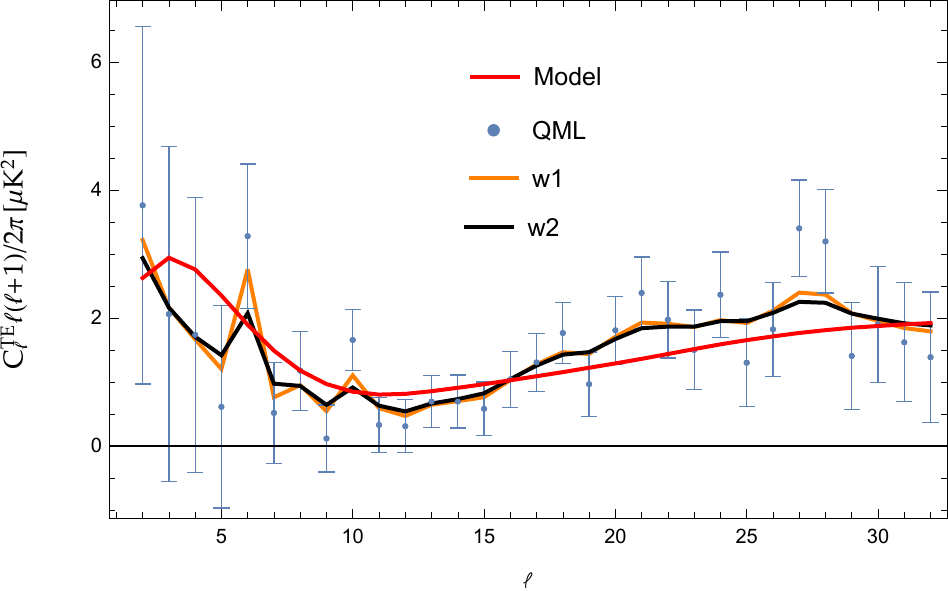}
    \caption{TE power spectra estimated from a simulation (blue points) with its corresponding error. Two smoothed versions of the spectrum are plotted obtained with weights ${w_1}_\ell$ (orange line) and ${w_2}_\ell=2{w_1}_\ell$ (black line). For comparison, the fiducial model used to generate the simulation is also given (red line).}
    \label{fig:SuavizadoApendice}
  \end{center}
\end{figure}

For our results in section~\ref{sec:iterativeQML}, we have chosen weights in the range 0.5 to 7, depending on the considered component of the power spectrum and on the multipole (larger weights are assigned to higher multipoles). Although the choice of the weights is somewhat arbitrary, we do not expect that their specific values affect to our conclusions, since they are used only to provide an initial guess at each step of the iterative process, facilitating convergence. As an illustration, figure~\ref{fig:SuavizadoApendice} shows the TE power spectrum estimated for one simulation (blue points) and its smoothed version using two different sets of weights ${w_1}_\ell$ (orange line) and ${w_2}_\ell=2{w_1}_\ell$ (black line). As expected, smaller weights lead to a smoother curve. Although our results are presented with the ${w_1}_\ell$ set of weights, we have tested that very similar results are achieved when using instead ${w_2}_\ell$.

\section{Estimator of the tensor-to-scalar ratio}
\label{ap:r}

In order to evaluate the performance of the QML technique in some of the considered cases, it is useful to estimate the value of the tensor-to-scalar ratio from the recovered power spectra. This will be done in a simplified case, where we assume that all cosmological parameters are known but $r$. In this way, we can write the fiducial power spectra as a function of r
\begin{equation}
  \label{ec:DefModelo}
  \vc(r) = \vc_{S}+ r \vc_T,
\end{equation}
where  $\vc$ is the vector containing the six different components (temperature and polarization) of the power spectra. $\vc_S$ and $ \vc_T$ are the scalar and tensor contributions, including lensing effects, which are fixed for a given fiducial model except for the value of $r$.

For a given map, let us assume that we have obtained with the QML an estimation of the power spectra ($\vct$) and of the corresponding Fisher matrix ($\mF$). If we assume that the variables $\vc(r)$ follow a Gaussian distribution,\footnote{It is well known that the $C_{\ell}$'s actually follow a $\chi^2$ distribution with $2\ell+1$ degrees of freedom and, therefore, the Gaussian approximation improves as $\ell$ increases. However, in our case, this approximation is sufficient to estimate $r$ and, therefore, enough for our purpose.} the probability density function of $\vct$ can be written as a multivariate normal distribution of mean $\vc(r)$ and covariance matrix $\mF^{-1}$. Therefore, we can obtain $\hat{r}$ by maximizing the likelihood
\begin{equation}
  \label{ec:Likelihood1}
  \log L = -\frac{1}{2} (\vct - \vc(r))^t \mF (\vct - \vc(r)) + \log ((2 \pi)^{6 (1-\ell_{\mathrm{max}})} |\mF|^{1/2}).
\end{equation}
Since we are fitting $r$ to the estimated power spectrum, our goal is to minimize the quantity
\begin{equation}
  \label{ec:Likelihood2}
  \chi^2 = (\vct - \vc(r))^t \mF (\vct - \vc(r)).
\end{equation}
This $\chi^2$ variable is a polynomial of degree two, thus it can be maximized analytically to obtain
\begin{equation}
    \hat{r} = \dfrac{\vc_T^t \mF (\vct - \vc_S)}{\vc_T^t \mF \vc_T}.
    \label{ec:EstimacionR_DesdeEspectro}
\end{equation}
The analytical expression of the error on the estimation is
\begin{equation}
  \label{ec:ErrorLikelihood}
  (\Delta r)^2= - \left(\frac{d^{2} \ln L}{dr^{2}} \right)^{-1} = (\vc_{T}^t \mF \vc_{T})^{-1}.
\end{equation}
It is interesting to note that since $\vct = \mF^{-1} \vy$, $r$ can be expressed directly in terms of $\vy$, that is
\begin{equation}
    \hat{r} = \dfrac{\vc_T^t \vy - \vc_T^t \mF \vc_S }{\vc_T^t \mF \vc_T}.
    \label{ec:EstimacionR_DesdeYl}
\end{equation}
This is useful when the fraction of the sky covered is small and the Fisher matrix becomes singular, since it would allow the estimation of $r$ without inverting the Fisher matrix. This also shows that all the cosmological information is actually encoded into the coupled power spectrum vector $\vy$.

As shown in section~\ref{sec:robust_r}, if the value of the estimated tensor-to-scalar ratio that minimizes $\chi^2$ differs significantly from the value used for the fiducial model (to compute $\mC$ and $\mF$, and to estimate $\vct$), one can obtain an updated power spectrum from the estimated value $\hat{r}$, that can be used as input for a new QML step in an iterative scheme until convergence is achieved.

This method can also be used with the binned estimator described in section~\ref{sec:EstimadorBineado}.
In particular, one would need to replace the quantities $\mF$ and $\vct$ in eq.~(\ref{ec:EstimacionR_DesdeEspectro})  by their analogous binned versions, described in eq.~(\ref{ec:MatrizDeFisher}) and (\ref{ec:FinalEstimador}), respectively. The fiducial power spectra $\vc_S$ and $\vc_T$ should also be binned in the same way.


\bibliographystyle{JHEP}
\bibliography{library}

\end{document}